\documentclass[twocolumn]{aastex7}

\usepackage{siunitx}
\usepackage[colorinlistoftodos]{todonotes}
\usepackage{appendix}
\usepackage{tikz}
\usetikzlibrary{shapes.geometric, arrows}

\shorttitle{Galaxy Morphologies and GOODS AGN}
\shortauthors{Jarvis et al.}

\graphicspath{{./}{Figures/}}

\begin{document}

\title{Obscured AGN at z $<$ 1.5: X-ray to Far-Infrared SEDs and Host Galaxy Morphologies in the GOODS Fields}

\correspondingauthor{William W.H. Jarvis}

\author[0000-0001-5916-5641]{William W.H. Jarvis}
\affiliation{Department of Astronomy, University of Massachusetts, Amherst, MA 01003, USA}
\affiliation{Astronomy Department, University of Wisconsin - Madison, Sterling Hall, 2535, 475 N Charter St, Madison, WI 53706}
\email[show]{wjarvis@umass.edu}

\author[0000-0002-5504-8752]{Connor Auge}
\affiliation{Eureka Scientific, 2452 Delmer Street, Suite 100, Oakland, CA 94602-3017, USA}
\affiliation{Institute for Astronomy, University of Hawai`i, 2680 Woodlawn Drive, Honolulu, HI 96822, USA}
\email{cauge@gmail.com}

\author[0000-0002-1233-9998]{David Sanders}
\affiliation{Institute for Astronomy, University of Hawai`i, 2680 Woodlawn Drive, Honolulu, HI 96822, USA}
\email{sanders@ifa.hawaii.edu}

\author[0000-0001-8917-2148]{Xuheng Ding}
\affiliation{School of Physics and Technology, Wuhan University, Wuhan 430072, China}
\email{dingxh@whu.edu.cn}

\author{Jeana Kim-Bolt}
\affiliation{Northwestern University}
\email{jeanakim-bolt2029@u.northwestern.edu}

\author[0000-0002-0745-9792]{C. Megan Urry}
\affiliation{Physics Department and Yale Center for Astronomy and Astrophysics, PO Box 208120, New Haven, CT 06520-8120, USA}
\email{meg.urry@yale.edu}

\author[0000-0003-0713-3300]{Eric Hooper}
\affiliation{Astronomy Department, University of Wisconsin - Madison, Sterling Hall, 2535, 475 N Charter St, Madison, WI 53706}
\email{ehooper@astro.wisc.edu}

\author[0000-0003-2196-3298]{Alessandro Peca}
\affiliation{Eureka Scientific, 2452 Delmer Street, Suite 100, Oakland, CA 94602-3017, USA}
\affiliation{Department of Physics, Yale University, P.O. Box 208120, New Haven, CT 06520, USA}
\email{alessandro.peca@yale.edu}

\author[0000-0002-2525-9647]{Aritra Ghosh}
\affiliation{DiRAC Institute and the Department of Astronomy, University of Washington, Seattle, WA, U.S.A}
\email[]{aritrag@uw.edu}

\author[0000-0003-4056-7071]{Chuan Tian}
\affiliation{Department of Physics, Yale University, New Haven, CT, USA}
\email{ufsccosmos@gmail.com}

\author[0000-0001-8211-3807]{Tonima T. Ananna}
\affiliation{Department of Physics and Astronomy, Wayne State University, Detroit, MI 48202, USA}
\email{tonima@wayne.edu}

\author[orcid=0009-0007-1608-923X]{Md Mahmudunnobe}
\affiliation{Department of Physics and Astronomy, Wayne State University, Detroit, MI, USA}
\affiliation{Center for Astronomy, Space Science and Astrophysics, Independent University, Bangladesh}
\email{mahmud.nobe@wayne.edu}

\begin{abstract}
We present an analysis of spectral energy distributions (SEDs), galaxy light profiles, and visual morphological classifications for 194 X-ray luminous AGN (intrinsic absorption-corrected $\log (L_{\mathrm{X, 0.5-7\,keV}}) > 42.5 \ (\mathrm{maximum} \ 45.2) \ \mathrm{erg \ s^{-1}}$) with $\mathrm{z} < 1.5$ in the GOODS fields. We generate X-ray to far-infrared SEDs normalized at $1 \ \mathrm{\mu m}$ for all AGN and sort them according to their emission slopes in the ultraviolet and infrared. We visually classify their host galaxies’ morphologies and compute their bulge/total light ratios using the software Galaxy Shapes of Light (\texttt{galight}). Most (94\%) GOODS AGN exhibit obscured (i.e. diminished UV and/or MIR emission) SEDs. Only 6\% show unobscured, quasar-like SEDs. Secular processes appear to play a large role in stimulating AGN emission, as only $\sim 1/3$ of galaxies are undergoing interactions. We also describe the morphological identification of a population of suspected post-merger spheroid galaxies with obscured UV/IR SEDs and distinguish them from the host galaxies of AGN with less obscuration in the UV or IR.
\end{abstract}



\section{Introduction} 
\label{sec:intro}

When a supermassive black hole (SMBH) at the center a galaxy accretes large amounts of new material, it becomes an active galactic nucleus (AGN). When such accretion occurs, a disk of material forms around the SMBH. Moving further away from the SMBH, temperatures decrease, and the disk radiates thermal energy across the electromagnetic spectrum \citep{shakura1973}. Active accretion generates emission from X-ray to optical wavelengths and is often greatest in the ultraviolet (UV) \citep{Koratkar1999}. In turn, dust and gas farther away from the SMBH and the accretion disk absorb this radiation and reprocess it to the infrared (IR) \citep{Sanders1989, urrypadovani}. The strength of the multi-wavelength emission can be measured through a spectral energy distribution (SED). SED features such as the Big Blue Bump (BBB) in the UV and optical or increased infrared radiation are dependent on the accretion and circum-black hole environment \citep{Sanders1989, Elvis1995, Richards2006, Suh2019COSMOS}.

Studying the links between SMBHs and their galaxies is vital to understanding both their evolutions \citep{Kormendy2013BHGal}. Numerous correlations link SMBHs to their host galaxies, such  as the correlations between SMBH mass and total mass of the host galaxy \citep{bandara2009}, SMBH mass and host galaxy velocity dispersion \citep{gebhardt_2000_msigma, Merritt_Ferrarese_01_msigma}, and SMBH and bulge mass \citep{haring_smbh_bulge_mass}. Identifying further relationships can constrain evolutionary pathways for both AGN and their host galaxies across cosmic time.

Studying the morphological properties of galaxies allows us to place them along an evolutionary pathway. Structural properties such as bulge-to-total ratios, merger signatures, and morphology can be used to place a galaxy along an evolutionary pathway \citep{conselice_2014_gal_structure}.

Mergers have been considered a vital component in AGN and galaxy evolution, and are believed to play a role in AGN evolution, either by triggering or quenching AGN feeding \citep[e.g.,][]{sanders_1988_ulirgs, hopkins_2006_agn_model, sanders_agn_ulirg_connection}. However, more recent studies seem to cast doubt on mergers as the primary drivers for AGN, with secular processes being hypothesized as the most common triggers for all but the most powerful AGN \citep[e.g.,][]{cisternas2011_weak_agn_merger_link, treister_2012_majormerger_agn, povic2012_weak_agn_merger_link, villforth_2014_merger_candels, baldassare_2020_low_mass_m_sigma, greene_2020_imbh,smethurst_non_merger_2024}. Resolving this tension requires combining AGN emission properties with sensitive morphological analyses to test whether and how often galaxy interactions coincide with accretion events.

Many different methods have been utilized to study galaxy morphologies and how they correlate with various galaxy properties, from machine learning to citizen science visual classification efforts and more \citep[e.g.,][]{galzoo, huertascompany, candels_visual, simmons_2017_galzoo_CANDELS, Ghosh2023, Ghosh2024}. Our work contributes to the field of AGN/galaxy coevolution by combining visual morphological classifications with an analysis of galaxy light profiles, most notably using the bulge-to-total light ratio (B/T). We focus on faint features indicative of merger activity, either past or present, and attempt to use these to better explain the coevolutionary path of AGN and their host galaxies.

Analyzing aggregated data from numerous observatories is much more straightforward due to online repositories and widely available catalogs, opening many new opportunities for analysis \citep{Xue2016CDFN}. Recent studies have used this wealth of data to perform in-depth analyses of the SEDs for AGN  (e.g., \citealp{hao_2013_quasar_seds, hao2014_type1_seds, Hickox2017, Suh2019COSMOS, suh_2020_no_z_relation_evolution, Li2020CT, coleman_2022_aha_host_galaxy, Auge2023}). By utilizing this wealth of observational data, we can learn much more about the growth and evolution of both AGN and their host galaxies across cosmic time. 

This paper contributes to the Accretion History of AGN (AHA) project\footnote{For more information on AHA, reference the AHA website: \url{https://project.ifa.hawaii.edu/aha}.}. AHA utilizes a wedding cake style survey with three observational fields, each with a different survey areas and flux limits, therefore covering distinct luminosities and redshift ranges. This wedding cake is comprised of sources from four principal fields sorted from deepest to most shallow: Great Observatories Origins Deep Survey-North (GOODS-N) and GOODS-South (GOODS-S), COSMOS, and Stripe 82x \citep{Giavalisco2004GOODS, Scoville2007cosmos, Grogin2011GOODSN, Koekemoer2011CANDELS, GUO2013GOODSS, LaMassa2013stripe}. The widest fields are necessary to understand rare high-luminosity sources, while the narrower fields allow us to probe both high-$z$ and faint AGN. 

We focus on the smallest and deepest of these fields, GOODS-N and GOODS-S. These fields were designed to search for high-redshift galaxies and provide an incredible trove of data for low-$z$ sources \citep{Giavalisco2004GOODS, Grogin2011GOODSN, GUO2013GOODSS}. The overlapping Chandra Deep Fields detects many faint Compton-thick ($\mathrm{N_H} > 1.5 \times 10^{24}~\mathrm{cm}^{-2}$) AGN that are undetected in larger surveys (for CDF, see \citealp{Xue2016CDFN, Luo2017CDFS}; for wide-area surveys and their benefits and limitations, see \citealp{Lamassa2016s82x, peca_2023_xray_s82x, peca_2024_s82xl}). Combined with GOODS-N/S's deep multi-wavelength observations, we can interrogate the properties of low-luminosity sources while revealing more exquisite details about the host galaxies' environments. 


In 2022 and 2023, the JWST Advanced Deep Extragalactic Survey (JADES) collaboration observed both GOODS fields, providing publicly available near and mid-infrared observations of the fields \citep{rieke_jades_goods}. These new observations, combined with existing HST observations, provide us with the opportunity to study the morphologies of low-luminosity AGN host galaxies across nearly 10 billion years of cosmic time.

In this paper, we combine morphological classifications and light-profile analyses with AGN SEDs to test the connection between SMBH activity and host galaxy structure. By focusing on faint signatures of merger activity across nearly 10 billion years of cosmic time, we aim to clarify whether mergers or secular processes most commonly drive AGN fueling. In \S\ref{sec:data}, we describe our sample selection and datasets. \S\ref{sec: methodology} outlines our SED construction and morphological analyses. Results are presented in \S\ref{sec: analysis}, followed by a discussion of their implications for AGN/galaxy coevolution in \S\ref{sec: discussion}.


\section{Data} \label{sec:data}

The GOODS fields combine data from the Hubble Space Telescope, Chandra X-ray Observatory, Spitzer Space Telescope (SST), XMM-Newton, and Herschel Space Telescope, as well as many ground facilities in order to provide deep, multi-wavelength observations on the twin 10\arcmin $\times16$\arcmin\ GOODS-N and GOODS-S fields \citep{Giavalisco2004GOODS}. With the addition of JADES, we now have access to high-resolution and extremely sensitive infrared imaging for $\sim 1/3$ of the fields.

\subsection{Chandra Deep Fields}

Both the GOODS-N and GOODS-S fields overlap with the Chandra Deep Fields (CDF). These fields include 0.5 to 7\,keV X-ray data from the 2Ms Chandra Deep Field-North (CDF-N, \citealp{Xue2016CDFN}) and the 7Ms Chandra Deep Field-South (CDF-S, \citealp{Luo2017CDFS}) and detect 683 and 1008 sources, respectively. Deep X-ray observations allow us to identify faint Compton-thick AGN missed in shallower large area surveys. Most AGN are not bright quasars, but by virtue of their brightness, quasars are the easiest AGN to locate in large area sky surveys. Deep drilling fields like CDF-N and CDF-S provide the perfect opportunity to dig into the low-luminosity AGN which constitute the majority of AGN in the universe \citep{ueda_2014_xlf, fotopoulou_2016_5-10_xlf, tonima_aha_2019_pop_synthesis}. Studying these heavily obscured AGN may be informative for understanding how the majority of SMBHs accrete material throughout their lifetimes, as unobscured quasar-mode accretion constitutes only a small portion of the SMBH's existence of order $10^7$ or $10^8$Myr \citep{Haehnelt_quasar_duty_cycle}. 

\subsection{GOODS - HST}

The Hubble Space Telescope observed the CDF fields as both part of the Great Observatories Origins Deep Survey (GOODS, \citealt{Giavalisco2004GOODS}) and as part of the Cosmic Assembly Near-infrared Deep Extragalactic Legacy Survey (CANDELS, \citealt{Grogin2011GOODSN, Koekemoer2011CANDELS}). The high Galactic Latitude locations of these fields made them ideal for studying the deep extragalactic universe. The original observations taken by the Advanced Camera for Surveys (ACS) filters from \citet{Giavalisco2004GOODS} were supplemented by Wide Field Camera 3 (WFC3) observations taken after the 2009 Hubble servicing mission as part of CANDELS \citep{Koekemoer2011CANDELS}. Combined, the GOODS-N/S fields are approximately $330 \ \mathrm{arcmin^2}$. In this paper, we rely upon the \citet{Koekemoer2011CANDELS} observations, specifically using the F814W and F125W filters. \citet{GUO2013GOODSS} presented a catalog of multi-wavelength observations building upon CANDELS that ``includes data from UV (U band from both CTIO/MOSAIC and VLT/VIMOS), optical (HST/ACS F435W, F606W, F775W, F814W, and F850LP), and infrared (HST WFC3 F098M, VLT ISAAC Ks, VLT HAWK-I Ks, and Spitzer IRAC 3.6, 4.5, 5.8, 8.0 $\ \mathrm{\mu m}$) observations." It also includes Spitzer 24 and 70 $\ \mathrm{\mu m}$ and Herschel Space Observatory (100, 160, 250, 350, and 500 $\ \mathrm{\mu m}$) far-infrared observations from \citet{elbaz_2011_goods_herschel}.

Similarly, our GOODS-N data comes from a catalog presented by \citet{barro2019_goodsn_cat}. From their catalog, we utilized KPNO and LBC data (U-band), HST ACS (F814W), HST WFC3 (F125W), Spitzer IRAC (3.0, 4.5, 5.8, $8.0 \ \mathrm{\mu m}$), Spitzer {MIPS} (24, $70 \ \mathrm{\mu m}$), Herschel PACS (100, $160\ \mathrm{\mu m}$) and  SPIRE (250, 350, $500 \ \mathrm{\mu m}$). 

We acquired all the aforementioned photometric data using the Rainbow Cosmological Surveys Database, which in turn utilizes the sources above. We created a custom catalog combining X-ray data from \citet{Xue2016CDFN} and \citet{Luo2017CDFS} with the Rainbow Navigator\footnote{http://arcoirix.cab.inta-csic.es/Rainbow\_navigator\_public/}. 

To ensure precise associations between the X-ray and other catalogs, we required detections to be within 2\arcsec \ for all data points and compared their redshifts. As noted by \citet{Auge2023}, thanks to numerous efforts to identify mismatches between the catalogs (e.g., \citealt{weaver_2022_cosmos2020_cat} - careful aperture extraction; \citealt{tonima_max_likelihood_2017} - maximum likelihood matching), we can reliably trust the counterpart identifications between catalogs.

\subsection{JADES}
The JADES collaboration observed the GOODS fields throughout 2022 and 2023 and has published image catalogs from their observations \citep{rieke_jades_goods, eisenstein2023_jades_goodss}. We utilize the F115W and F444W images due to their large area coverage. These filters contained the most overlap with our HST-selected sample, making them ideal for our work. JADES includes both high and medium depth fields. The high-depth fields constitute $\sim 45 \ \mathrm{arcmin^2}$, while the medium-depth fields combine for $\sim 175\ \mathrm{arcmin^2}$ of coverage as of the JADES second data release \citep{eisenstein2023_jades_goodss, eisenstein_jades_2023_overview}. This represents 53\% of the total GOODS fields. 

While JADES does not cover the full fields, it provides a valuable resource when studying the morphologies of AGN host galaxies. The high resolution images produced by the JADES team give us a valuable comparison tool so that we can check the accuracy of our HST-based visual classifications.

\subsection{Sample Selection} \label{subsec: sample selection}

\begin{deluxetable*}{cccccccc}
\tablenum{1}
\tablecaption{Sample Selection Cuts \label{tab: sample_selection}}
\tablehead{\colhead{} & \colhead{$L_{\rm X} > 10^{42.5}$} & \colhead{$z < 1.5$} & \colhead{\textbf{Imaging (Sample)}} & \colhead{F814W} & \colhead{F125W} & \colhead{Both} & \colhead{F115W}
}
\startdata
GOODS-N & 401 & 98 & \textbf{68} & 64 & 63 & 61 & 16 \\
GOODS-S & 517 & 155 & \textbf{126} & 126 & 61 & 61 & 23 \\
Total   & 918 & 253 & \textbf{194} & 190 & 124 & 122 & 39 \\
\enddata
\tablecomments{(1) Number of X-ray sources with $L_{\rm X} > 10^{42.5}$~erg~s$^{-1}$ in each GOODS field. (2) Subset with $z < 1.5$. (3) Sources with either HST/F814W or F125W imaging and at least 15 photometric points (the final sample). (4) F814W coverage. (5) F125W coverage. (6) Coverage in both filters. (7) JWST F115W coverage.}
\end{deluxetable*}

\begin{figure}
    \centering
    \includegraphics[width=\linewidth]{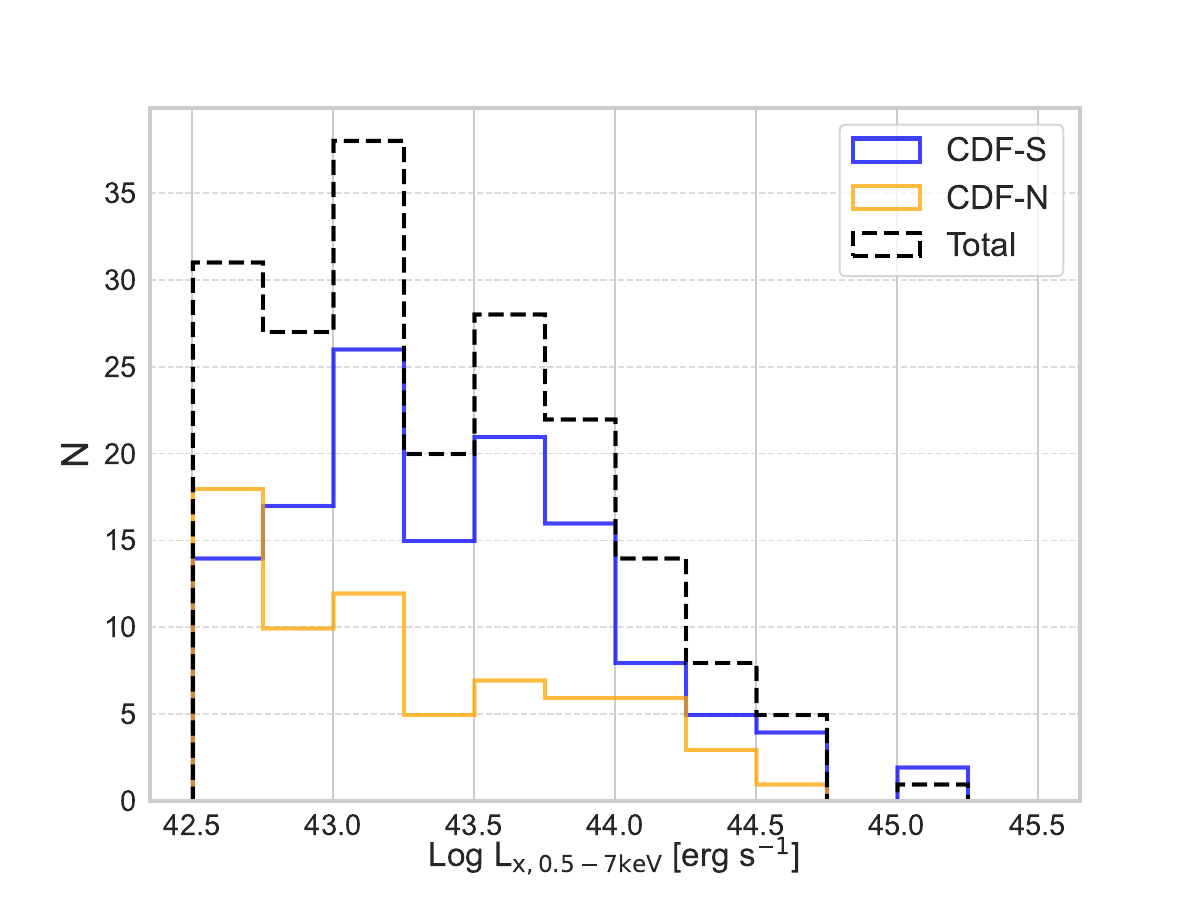}
    \caption{Histogram of intrinsic 0.5-7\,keV X-ray luminosities in units of log [$\mathrm{erg \ s^{-1}}$]. GOODS is unique in that its deep Chandra observations allow us to reliably detect faint, obscured AGN that are selected against when studying fields such as Stripe82X or COSMOS. In studies such as \citet{Auge2023}, these low luminosity ($10^{42.5} < \mathrm{L_{\rm X}} < 10^{43} \ \mathrm{erg \ s^{-1}}$) are removed due to potential contamination from luminous starburst galaxies present in larger area surveys. However, our sample constitutes entirely AGN, as verified through the construction of spectral energy distributions later in this paper.}
    \label{fig: xlum_histo}
\end{figure}

Before implementing physical criteria (i.e. based on the actual physical parameters of the galaxy, such as redshift and X-ray luminosity), we implemented both a cut based on the available photometric data and an imaging check. The Rainbow catalogs have differing numbers of filters available for each field: 24 for GOODS-North and 26 for GOODS-South. We required detections in at least 15 different filters with $\lambda_{\mathrm{rest}} > 0.1\ \  \mathrm{\mu m}$ in order to ensure high quality SEDs. This number is somewhat arbitrary, but it does make certain that we are sufficiently sampling regions of the SED important for classification into a given SED shape as discussed in \S \ref{sec: methodology_seds}. Sources were also required to be detected within $\pm 0.5\ \ \mathrm{\mu m}$ of rest-frame $1\ \mu \mathrm{m}$ with fractional flux error in each filter less than 25\% in order to normalize the SEDs as described in \S \ref{sec: methodology_seds}. AGN host galaxies were also required to have images available in either the Hubble F814W or F125W filter. These filters were selected as the most complete for both GOODS-N and GOODS-S. While not necessary for our selection, if both filters are available, they compliment each other by capturing different features, namely young stars in F814W and old stars and dust in F125W. These filters also allow us to probe the rest-frame B-band to $\mathrm{z \sim 1.5}$, allowing us to compare classifications with previous literature work \citep[e.g.,][]{simmons2011_obscured_goods_agn}. Even if a source met all other criteria discussed below, if the galaxy was blank in both of those filters, it was removed from our sample. 

We selected sources based on their 0.5-7\,keV X-ray luminosity and spectroscopic redshift from the CDF-N and CDF-S fields \citep{Xue2016CDFN, Luo2017CDFS}. The full data set included 683 X-ray sources in the CDF-N and 1008 in the CDF-S. CDF-S shows more sources, particularly at lower-luminosities, due to the deeper 7 Ms field compared to the shallower 2 Ms CDF-N observations. \citet{Xue2016CDFN} and \citet{Luo2017CDFS} calculated intrinsic hydrogen column densities based on a fixed photon index of 1.8 and used the column densities to estimate the rest-frame absorption-corrected X-ray luminosities. We use these intrinsic luminosities throughout the remainder of the paper.

To ensure that we select only AGN, we introduce an X-ray luminosity threshold. Starburst galaxies rarely produce X-ray luminosities above $10^{42} \ \mathrm{erg \; s^{-1}}$ while AGN can produce X-ray luminosities up to $10^{46} \  \mathrm{erg \; s^{-1}}$ \citep{fabbiano_1989_normal_xrays,renalli_2003_lx_sfr, PersicS2004SFR}. This allows us to easily identify the AGN and reduce the risk of contamination by stellar activity. Thus, we set a minimum 0.5-7\,keV X-ray luminosity threshold of $10^{42.5} \ \mathrm{erg \; s^{-1}}$. X-ray selection is one of many methods of identifying AGN, but we do note that X-ray selection methods are likely to miss a fraction of the full AGN population, especially for the most heavily-obscured sources \citep{treister_2004_obscuredagn, lyu_2022_completeness}.

\begin{figure*}[t]
    \centering
    \includegraphics[width=\linewidth]{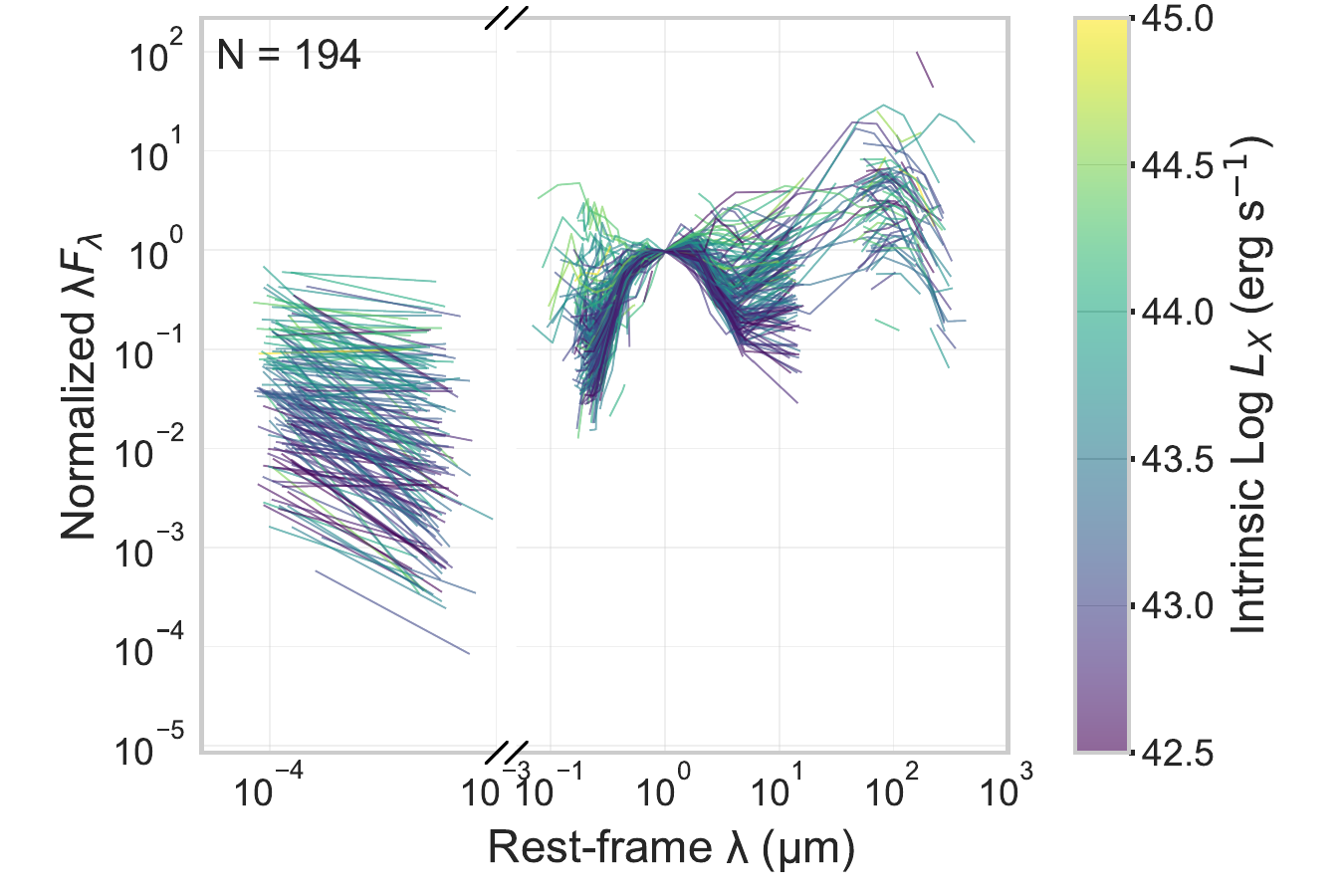}
    \caption{SEDs for the 194 GOODS-N and GOODS-S sources in our sample. 
    Fluxes are normalized at $1\mu$m and are colored based on intrinsic X-ray luminosity. Note the break between $10^{-3} \ \mathrm{\mu m}$ and $0.5 \times 10^{-2} \ \mathrm{\mu m}$ where there is no data.}
    \label{fig: goods_seds}
\end{figure*}

We set an upper redshift limit of $z=1.5$. At higher redshifts, key emission bands shift to longer wavelengths and leave our photometric coverage. Importantly, the rest-frame MIR emission shifts beyond our observations. Without it, we cannot construct our spectral energy distributions as described in \S \ref{sec: methodology_seds}. The upper limit also attempts to reduce classification errors, as \citet{simmons_urry_serc} have shown that at higher redshifts, galaxies are more likely to be misclassified as spheroids. There are several potential reasons for misclassification. Unresolved AGN components may dominate the bulge components of and any extended components---possibly including evidence of mergers---will experience surface brightness dimming and susbsequent loss of signal to noise ratio at higher redshifts \citep{simmons_urry_serc}.

Ultimately, there are 194 galaxies that have available X-ray luminosities, photometric data to construct SEDs, are within our redshift range, and have either F814W or F125W HST observations available. Most (122 out of 194) galaxies have coverage in both F814W and F125W; only 63 of 194 were captured by JWST in either the F115W or F444W filters. However, this is consistent with the JWST coverage available at the time of the analysis, which covers $\sim 1/3$ of the fields. Put differently, where JWST has coverage, every sample member detected with HST is also detected by JWST. For our primary metrics---namely visual classifications and bulge/total (B/T) light ratios---we compared the results between HST and JWST before our analysis and verified that they tell a consistent picture. We discuss this in greater depth in \S \ref{sec: morphology_methods}. 

Table \ref{tab: sample_selection} shows the breakdown of sources according to different selection criteria, culminating in our final sample. 

Likewise, in Figure \ref{fig: xlum_histo}, we note AGN emitting from $10^{42.5}$ $\mathrm{erg\ s^{-1}}$ to $10^{45.2} \ \mathrm{erg\ s^{-1}}$, only $15\%$ (N=28) show X-ray luminosities greater than $10^{44} \ \mathrm{erg\ s^{-1}}$. As high X-ray luminosity AGN are rare, we expect few in the small GOODS fields. By contrast, \citet{peca_2023_xray_s82x} analyzed the X-ray AGN population the 31 deg$^2$ Stripe 82X survey. Their Figure 6 shows that the median X-ray AGN has an intrinsic X-ray luminosity of $\sim 10^{44} \ \mathrm{erg \ s^{-1}}$ with the vast majority of sources above $\sim 10^{43.5} \ \mathrm{erg \ s^{-1}}$. Meanwhile, $\sim60\%$ of our sample has $\mathrm{L_{\rm X}} < 10^{43.5} \ \mathrm{erg \ s^{-1}}$. This highlights one of the benefits of our study: we focus on lower-luminosity sources which are missed by or comprise only a small percentage of wide-field surveys.

\section{Methodology} \label{sec: methodology}
\subsection{SEDs}
\label{sec: methodology_seds}
To generate the SEDs, we interpolated between the available photometric data points discussed in \S \ref{subsec: sample selection}. We did not utilize any modeling software in the creation of the SEDs. While detailed modeling of the host galaxy properties presents a wealth of opportunities for analysis, this work is specifically focused on correlating the host galaxy morphology with the slope of UV and IR emission. 

Sources were normalized to the $1\ \mu$m flux to emphasize AGN emission, especially in the UV and IR. AGN emission is also minimized here, and the $1\ \mu$m emission should be dominated by the stellar population of the host galaxy for all but the most luminous unobscured AGN \citep{neugebauer_quasar_energy, sanders_quasar_continuum}.

Figure \ref{fig: goods_seds} shows X-ray to MIR SEDs for the 194 sources with $z < 1.5$ normalized at $1\ \mu$m. We also note FIR (100\ \ $\mathrm{\mu m}$) detections for 76/194 sources. Note the wide range of X-ray luminosities present, as well as a significant variance in the shapes of the SEDs in the UV and MIR. An AGN's SED encodes significant information about the system it resides inside of \citep{ciesla_2015_agn_seds, hickox_alexander_obscuredagn_review2018, Auge2023}. Obscuration impacts both the optical/UV and IR, and different shapes of the SEDs tell us about the levels of obscuration in different regions around the AGN such as the accretion disk and dusty torus. 

\subsubsection{Normalized SED Shapes}
\label{normalized_sed_shapes}
\begin{deluxetable*}{cccc|c}
\tablenum{2}
\tablecaption{SED Shape Classification Criteria \label{tab: panel_lims}}
\tablehead{\colhead{} & \colhead{(0.15--1.0$\mu$m)} & \colhead{(1.0--6.5$\mu$m)} & \colhead{(6.5--10$\mu$m)} & \colhead{\# Sources} }
\startdata
Shape 1--2 & $\alpha < -0.3$ & $-0.4 < \alpha$ & --- & 11 \\
Shape 3 & $-0.3 < \alpha < 0.2$ & $-0.4 < \alpha$ & --- & 94 \\
Shape 4 & $0.2 < \alpha$ & $\alpha < -0.4$ & $0.0 < \alpha$ & 47 \\
Shape 5 & $0.2 < \alpha$ & $\alpha < -0.4$ & $\alpha < 0.0$ & 41 \\
\enddata
\tablecomments{
Criteria used to separate SEDs into the four SED shapes shown in Figure~\ref{fig: goods5p}, following \citet{Auge2023}.  
Shapes~1 and~2 are combined due to low counts in each.  
Each column corresponds to a wavelength region used to determine the slope~$\alpha$.  
A dash (---) indicates that no criterion applies in that region.  
The criteria are designed to emphasize AGN contributions to each galaxy's normalized SED.}
\end{deluxetable*}


\citet{Auge2023} developed a technique of sorting SEDs according to the shape of their SEDs normalized at $1 \ \mathrm{\mu m}$. The AGN dominates the SED in the UV and MIR, so by normalizing at $1 \ \mathrm{\mu m}$, we emphasize the contribution of the AGN. \citet{Auge2023} breaks these normalized SEDs into five distinct shapes that trace varying levels of UV and/or MIR obscuration and X-ray luminosity. These shapes capture a range of features of the X-ray luminous AGN found within the full AHA sample, including the GOODS fields.

\begin{figure} [!tp]
    \centering
    \includegraphics[height=15cm, keepaspectratio]{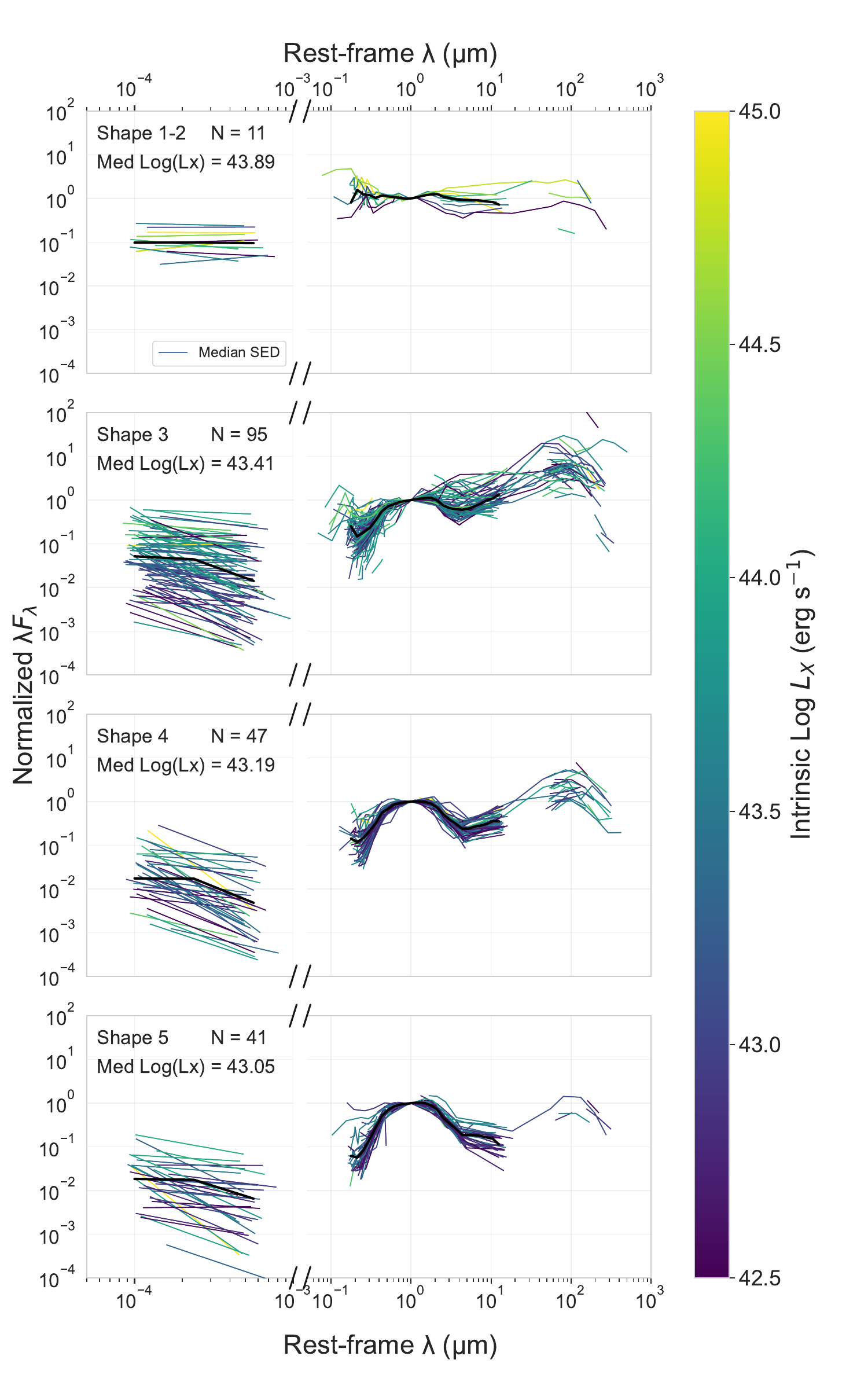}
    \caption{Four shapes of SEDs sorted according to the criteria in Table \ref{tab: panel_lims}. The SEDs are normalized to 1\,$\mathrm{\mu m}$ to emphasize the emission of the AGN. Note the break between $10^{-3} \ \mathrm{\mu m}$ and $0.5 \times 10^{-2} \ \mathrm{\mu m}$ where there is no data. For each shape, the bold black line shows the median SED. Almost half the AGN ($94/194$) are in Shape~3, as in \citet{Auge2023} (note: \citet{Auge2023} breaks the SEDs into five panels. Our Shape~1-2 is a combination of their Shape~1 and Shape~2 due to the dearth of unobscured AGN in our sample). 51\% of Shape~3 galaxies have far-infrared (100\,$\mathrm{\mu m}$ detections, higher than both Shape~4 (45\%) and Shape~5 (12\%). }
    \label{fig: goods5p}
\end{figure}

Table \ref{tab: panel_lims} shows how SEDs are sorted according to their slope $\alpha$. Due to only 11 sources in shapes 1 and 2, we combine them into a single shape that catches all UV/MIR unobscured AGN. Figure \ref{fig: goods5p} shows the four SED shapes. For the remainder of this paper, we denote the combination of shapes 1 and 2 as ``Shape~1-2."

Shape~1-2 sources are most akin to quasars, i.e., bright, unobscured AGN with high X-ray luminosities radiating across the entire electromagnetic spectrum. The AGN dominate their host galaxies' emissions. 

Shape~3 includes sources with lower UV emission but high IR, likely due to reprocessing of light from the accretion disk or from contamination due to star formation. It also shows a broad range of X-ray luminosities, including the brightest X-ray sources in our sample, with log $\mathrm{L_{\rm X}} \sim 45.2 \ \mathrm{\ erg \ s^{-1}}$. Almost half (48.7\%) of all galaxies in our sample fall into SED Shape~3. 

Shape~4 SEDs have a much steeper drop in the 1-6.5\,$\mathrm{\mu m}$ IR, while Shape~5 is differentiated by IR emission decreasing out to 10\,$\mu$m and X-ray luminosities that are on average the lowest of the sample. 

The GOODS-N/S samples are distinguished by having a high fraction (94.3\%) of shapes 3, 4, and 5 sources. The Accretion History of AGN Collaboration also studies the Stripe82X and COSMOS fields, sees only $73\%$ of sources in shapes 3, 4, and 5 \citep{Auge2023}. The deep GOODS imaging in small fields probes the less luminous, more common AGN, whereas large surveys sample the extremely luminous end of the AGN luminosity function.

Shape~3 has the highest fraction of FIR detections (51\%). Shape~4 follows close behind (by ratio, though not by number) with 45\% of sources showing a FIR detection. Shape~5 shows only 12\% with FIR detections. 

FIR emission tells us much about a galaxy's physical properties. Cold dust, the primary emitter of FIR photons, thus appears to be missing in unobscured quasar-like galaxies and in potential fading AGN as seen in shape~5, pointing to later evolutionary stages for many of those AGN host galaxies.

The overall picture presented here is consistent with earlier results showing that most AGN growth occurs in the obscured phase shown in our shapes 3, 4, and 5 \citep{treister_2004_obscuredagn, ananna_2019_aha_seds, Auge2023}. Unobscured quasar-like emission occurs only at specific points in an AGN's lifecycle and has at most a few tens of millions of years of activity before fading \citep{Haehnelt_quasar_duty_cycle, khrykin+2021_quasar_lifetime}.

\subsection{Galaxy Profile Decomposition}
\label{sec: galight_description}

\begin{figure*}
    \centering
    \includegraphics[width=\linewidth]{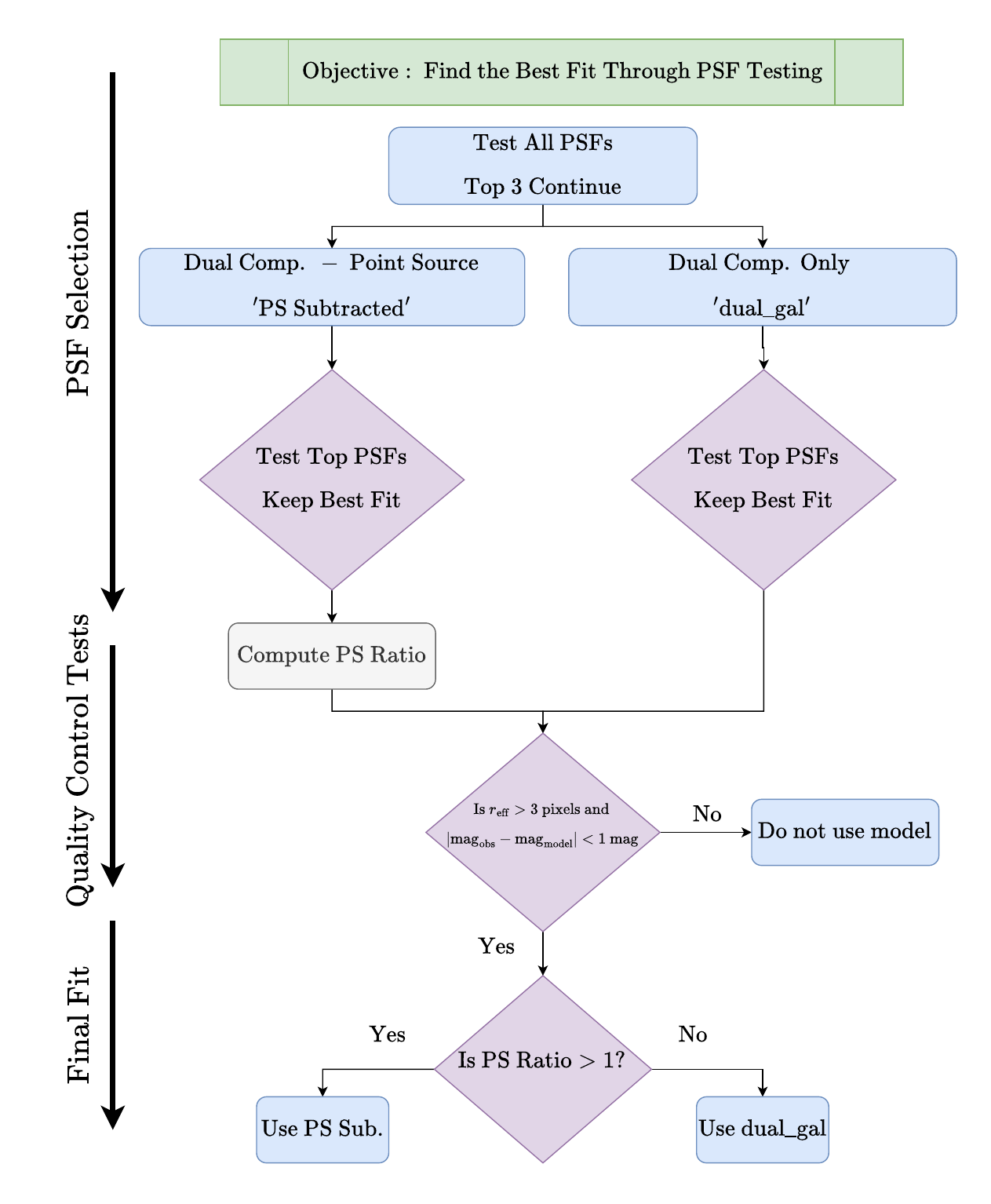}
    \caption{Flow chart showing the process for iteratively choosing the best fit for a given galaxy. Through this process we test multiple PSFs, choose the best PSF for a given source, and decide whether it is necessary to subtract a point source based on the ratio of light from the point source to light from the host galaxy. Colors correspond to different steps in the process. Blue denotes fitting steps, purple denotes selection steps, and beige computational steps.}
    \label{fig: flow_chart}
\end{figure*}

\begin{figure}
    \centering
    \includegraphics[width=\linewidth]{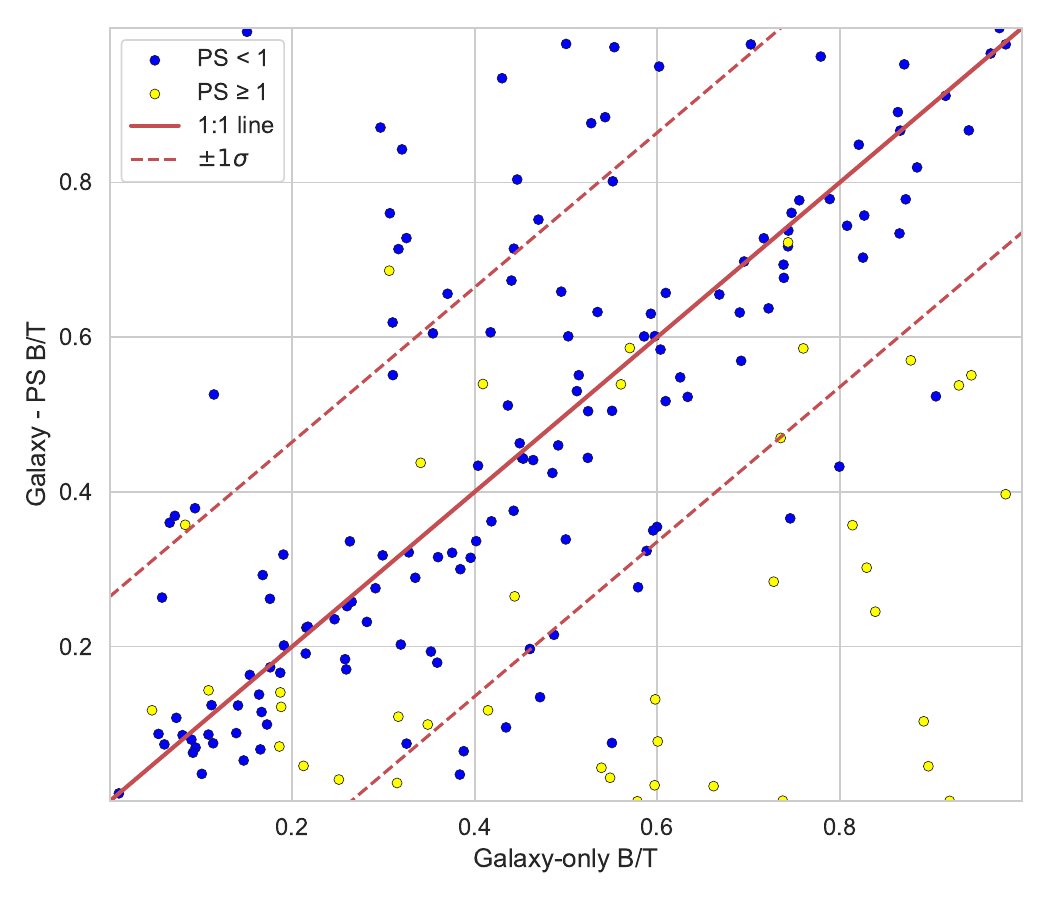}
    \caption{Galaxy-only bulge/total (B/T) ratio  versus galaxy minus point source B/T ratio for 187 GOODS X-ray AGN colored by the ratio of light in the point source to the host galaxy (PS Ratio). Note that the majority of fits with $\mathrm{PS \ Ratio} > 1$ show large diversions between the fit results. In some of these cases, the two-component only fit performs comparably to those with a point source subtracted, but for many sources with a large point source contribution, the B/T ratio appears to be overestimated if a point source is not subtracted. Sources above the $+1\sigma$ line show, except for three galaxies, no improvement and even some evidence for over-fitting if a point source is included.}
    \label{fig: agn_frac_deviation}
\end{figure}

\begin{figure*}[t!]
    \centering
    \includegraphics[width=\textwidth]{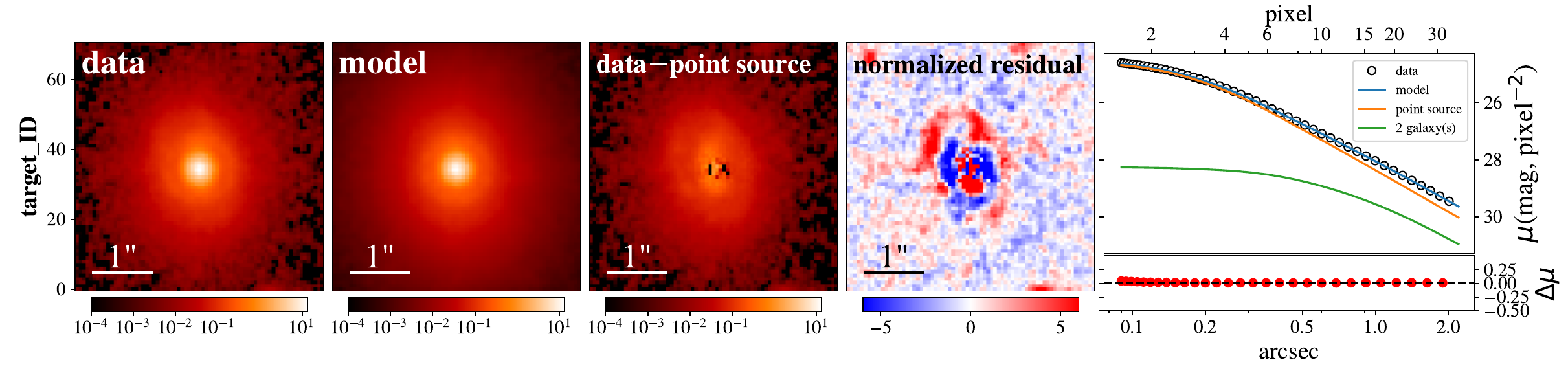}
    \caption{Example HST F125W \texttt{galight} results for GOODSN-ID 459 (visual classification: Irregular). The left panel shows the raw F125W data. The second panel shows the model of the source created by \texttt{galight} using the best-fit PSF. The third shows the data with a point source subtracted away, revealing the galaxy underneath. The fourth shows a normalized residual after subtracting the model from the data. The fifth shows the azimuthally-averaged radial profiles for (a) the observed data, (b) the full \texttt{galight} model, which is the sum of (c) the point source and (d) bulge and disk component radial profiles (listed as ``galaxy(s)" in the figure) showing that the model fits the data quite well, except for an excess from the underlying host galaxy. There is strong evidence for spiral or ring-like structure under the point source, though the spiral arms are irregular and asymmetric.}
    \label{fig: galight_ex}
\end{figure*}

\begin{figure}
    \centering
    \includegraphics[width=\linewidth]{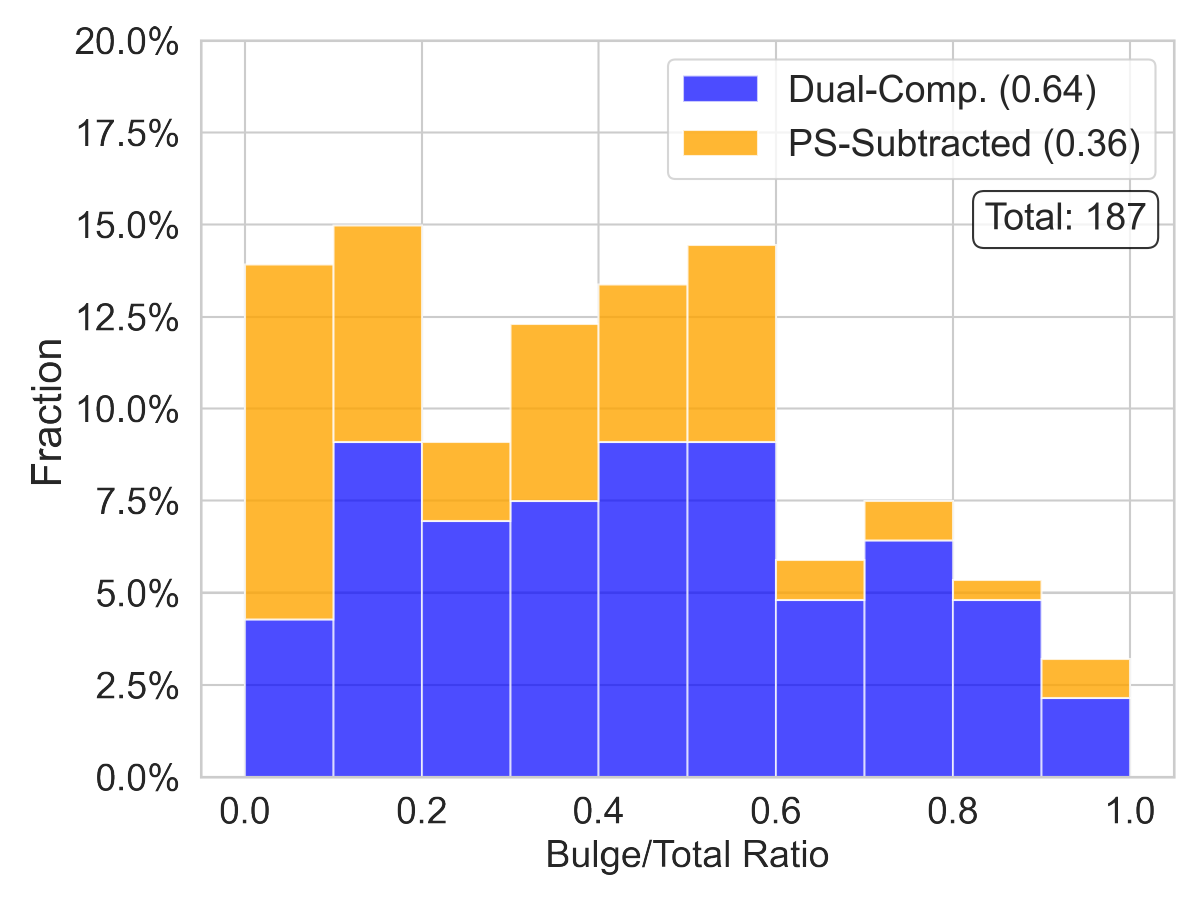}
    \caption{Stacked histogram B/T ratio distribution for the 187 sources with successful point source subtracted JWST and HST fits. The stacked histograms are colored according to the ratio of light from the point source to the host galaxy. A point source is subtracted if the PS ratio, the ratio of light from the point source to the host galaxy, is greater than one. 36\% of galaxies have had a point source subtracted.}
    \label{fig: all_bt}
\end{figure}

Analyzing the light profiles of galaxies can provide important insights into their morphologies. \citet{vanderwel_2012_candels} included \texttt{GALFIT} Sérsic Indices for each galaxy within the CANDELS survey, including GOODS. However, these indices were not calibrated for AGN and did not include point source subtraction, thus potentially skewing the Sérsic index distribution. To conduct our own fits properly accounting for the prevalence of point sources, we utilized the Galaxy Shapes of Light (\texttt{galight}) software from \citet{galight_ding}. \texttt{galight} is a Python-based alternative to \texttt{GALFIT}, offering a more automated and streamlined approach utilizing the modeling functionality of \texttt{lenstronomy}\footnote{https://github.com/lenstronomy/lenstronomy} \citep{birrer_lenstronomy_2018, birrer_lenstronomy_ii_2021}. \texttt{galight} includes automated masking, PSF generation, and a simple interface.

\texttt{galight}, as a dependent of \texttt{lenstronomy}, uses particle swarm optimization (PSO) to test various solutions and understand the posterior distribution within the provided parameter space. The PSO utilizes a number of semi-independent particles to sample the given input parameters. Users input the target image, a noise map for that image, optionally a weights map, and a point spread function which together are used by the PSO to understand the posterior distribution and minimize the returned $\chi^2$ value for the fit. The user may also constrain the fits by setting thresholds for parameters like the effective radius or Sersic index. The fitting depth (i.e., the number of particles in the swarm and the number of iterations or movements of said particles) is controlled by the `$\mathrm{fitting\_level}$' or `depth' parameter. Shallow modes use fewer particles and less iterations, making their results susceptible to temporary biases and local maxima and/or minima, as well as small number statistics. Deeper modes can produce better results by allowing the swarm to settle on a consistent solution, but can lead to overfitting. In this paper, we use the `Shallow' mode for our initial testing fits, and then ``extra-deep" for our final results due to the complexity of decomposing multiple components for our galaxies as well as modeling nearby galaxies to prevent overlapping galaxy profiles from impacting one another. We elucidate further details and rationale for both below.

PSF selection is essential to ensuring successful fits. The ideal PSF for a particular galaxy may not match well for a different part of the field. HST in particular is known to have an unstable PSF varying as the telescope moves in and out of the Earth's shadow \citep{lallo_2005_hst_orbital_breathing, ding_2023_galight_nature}. Over many orbits, these effects will prove additive over different parts of the field, and accounting for it becomes a priority. JWST's PSF is considerably more stable due to its location at the Sun-Earth L2 point, but the best PSF will still vary across the observation tiles \citep{lajoie_2023_jwst_psf}. Figure \ref{fig: flow_chart} gives a brief overview of the PSF and fit selection process for an individual galaxy, described in detail in the following paragraphs.

For each filter and field (e.g., HST F814W GOODS-S), we constructed a unique PSF library using \texttt{galight}. We selected large numbers of valid PSF stars for each library (minimum: 27 for JWST GOODS-S F115W; maximum: 127 for HST GOODS-S F814W). The number of PSFs varied due to differences in the number of potential PSF stars in each filter, which are determined by the resolution, sensitivity, and filter observation area. 

We then tested the goodness-of-fit for all PSFs by conducting a ``shallow"-mode single component 2D Sérsic profile minus QSO fit for all galaxies and potential PSFs in our sample. The ``shallow" fit utilizes less particles and iterations for each fit, resulting in much faster but less precise fitting results, completely sufficient for determining which PSFs will perform well for a given galaxy. We acknowledge that not all host galaxies in our sample contain a bright central point source but emphasize that these are not the final fits. These first fits were designed to quickly identify well-performing PSFs for each individual galaxy and streamline our later dual-component fits. Once we identified the top three performing PSFs for each source, we mean-stacked them to create an averaged PSF. Thus, for every source, we have four possible PSFs: three field PSFs plus one averaged PSF.

To evaluate the robustness of the goodness of fit to PSF choice, we asked the following question: for a given source, how much does the expected goodness of fit vary across the top three available field PSFs? For each source, filter, and fitting mode, we computed the fractional standard deviation of the goodness-of-fit metric (BIC or reduced $\chi^2$) across the three non-reference PSFs, normalized by their mean value.

For both HST F814W and F125W, we find fractional standard deviations of approximately $5\%$ for both fitting modes, with the \texttt{PS - subtracted} fits exhibiting marginally lower scatter (by $\lesssim 1\%$) than \texttt{dual\_gal}. For JWST F115W, the fractional standard deviation remains low for \texttt{PS - subtracted} fits ($\sim 4\%$) but increases substantially for \texttt{dual\_gal} fits ($\sim 10\%$).

This behavior is expected and reflects the increased sensitivity of models lacking an explicit PSF component to field-dependent PSF mismatches. In the absence of a PSF component, point source contributions must be absorbed by the bulge model, causing variations in PSF structure to propagate directly into the extended components and thus the goodness of fit. In contrast, for \texttt{PS - subtracted} fits, differences in PSF structure are largely absorbed by the PSF component itself, primarily affecting the inferred QSO flux (and point source ratio) while leaving the extended components comparatively stable.

As a consistency check, we checked whether the PSF yielding the minimum reduced $\chi^2$ also yielded the minimum BIC. We find excellent agreement between the two metrics: $100\%$ for JWST F115W in both modes, $98\%$ (\texttt{dual\_gal}) and $100\%$ (\texttt{PS - subtracted}) for HST F125W, and $\geq 98\%$ for HST F814W in both modes. Although $\chi^2$ and BIC are not independent diagnostics within \texttt{galight}, this high level of agreement is reassuring. 

We therefore conclude that, for any given source, the top three PSFs generally yield very similar goodness-of-fit values, with increased PSF-sensitivity appearing only for JWST \texttt{dual\_gal} fits where PSF mismatches cannot be explicitly modeled.

For our final fits, we utilized a two-component host galaxy profile in a particle swarm-optimized mode, run in ``extra-deep" mode, a custom mode created by the author to allow high processing depths. Test cases found a Bayesian Information Criterion (BIC) improvement of $\sim 5-10\%$ for challenging sources which hit the standard ``deep"-mode's maximum processing depth without converging on a fit. This change did not impact sources that were satisfactorily fit in shallower processing modes, since \texttt{galight} will naturally halt once it has converged on a solution.

For the initial fit, we instituted a prior that the bulge should be less than 15\% of the disk size and the disk should have a higher ellipticity than the bulge. This prevents erroneous B/T ratio values. We fixed the bulge component Sérsic index at 4 and the disk component at 1. For each source, we tested all four best-performing PSFs. We first performed a two-component (bulge+disk, or `dual\_gal') fit and then a two-component minus point source (bulge+disk-PS, or `Point Source (PS) Subtracted') fit. For the PS subtracted fits, \texttt{galight}'s automated point source-subtraction feature determined the point source location and subtracted the point source. \texttt{galight} automatically computed the flux from the point source, bulge, and disk components and used them to compute the B/T ratio and the point source ratio (PS ratio).

To determine the best fit from both modes, we used the Bayesian Information Criterion (BIC, \cite{Schwartz_BIC_1978}). The BIC attempts to account for goodness-of-fit based on the available degrees of freedom, in our case determined by the resolution of the image. A lower BIC corresponds to a better fit. Equation \ref{eq: bic} shows the computation of BIC, where k is the number of free parameters, n the number of data points, and $\mathrm{\hat{L}}$ is the maximum likelihood of the model. 

\begin{equation}
\label{eq: bic}
BIC = k \ln(n) - 2 \ln(\hat{L})
\end{equation}

Thus, for each fitting mode, we select the source with the lowest BIC and take it as the best fit for that mode. We also calculated the Akaike Information Criterion (AIC) and found that they agreed more than 90\% of the time, with exceptions only for galaxies with exceptionally bright point sources, which tended in general to have higher BICs. For those point sources, we visually inspected the outputs to ensure the best fits were used. 

Next, we must decide whether to use the point source-subtracted model. The naive assumption is that as all galaxies in our sample host AGN, all should have a point source subtracted. However, as discussed, the majority of AGN in our sample are obscured in the UV into the optical. Thus, it is not immediately clear whether point source subtraction is appropriate.

During the point source subtracted fits, we saved the ratio of light contained within each galaxy's point source to the light contained in the host galaxy, hereafter identified as the ``Point Source Ratio (PS Ratio or PSF/Host Ratio)." This ranges from near zero for sources without a bright central point source to over 1000 for quasar-like AGN. We expect the PS ratio to be related to the AGN's properties, such as the intrinsic X-ray luminosity. For these powerful AGN, the point source may contribute to the bulge light, creating a larger bulge-to-total ratio than would otherwise be expected for a given galaxy. However, this is complicated by the column densities of the host galaxies. High column densities of dust around the AGN can suppress optical emission, even for powerful AGN. Thus, we found that motivating point source subtraction based on the X-ray luminosity was not feasible. 

Instead, we searched for galaxies with large offsets between their dual-component only and dual-component minus point source fits. Figure \ref{fig: agn_frac_deviation} shows a scatter plot of galaxy-only B/T and galaxy minus point source B/T ratio. We found that sources with a point source ratio above one were likely to be modeled with a high B/T ratio below the $-1\sigma$ line in the figure. Thus, we subtract a point source from a galaxy if the point source ratio is greater than one, or the point source contributes as much or more light than its host galaxy. 

For those sources with a higher B/T ratio in the PS-subtracted mode, we visually inspected all residuals. We found that three galaxies - GOODS-S xIDs 89, 888, and 959 - showed notable improvements in their residuals when a point source was included even though their PS ratios were less than one. For the remainder, the inclusion of the point source either induced overfitting or provided no measurable improvement to the model. This demonstrates an important benefit of our methodology: we only remove point sources where their contribution is necessary for the best fit, reducing potential errors induced by the inclusion of unnecessary additional components. 

It is important to note that BIC cannot be used to distinguish between the two models, as including the point source makes the models incomparable in a Bayesian sense. Comparing the dual-component and \texttt{PS - subtracted} fits is, as far as a Bayesian statistician is concerned, comparing apples and oranges. The models are fundamentally distinct due to the inclusion of an additional element in the point source models. The BIC can be used to choose the best of a series of different fits for the same model, but not for choosing between models. 

Thus, for the final fit, we use in all cases (except for IDs 89, 888, and 959, as discussed earlier) the point source subtracted result if the point source ratio is greater than one. In most figures showing the B/T ratios, we show not only the distribution of ratios, but also the fraction that have had a point source subtracted. 

\begin{deluxetable*}{cccccccccccc}
\tablenum{3}
\tablecaption{X-ray Selected GOODS-AGN \label{tab: machine_readable}}
\tablehead{\colhead{xID} & \colhead{xRA} & \colhead{xDec} & \colhead{$z_{\mathrm{spec}}$} & \colhead{$L_{X,\ \mathrm{0.5-7}}$} & \colhead{\texttt{gal.} Filter} & \colhead{B/T Ratio} & \colhead{Red. $\chi^2$} & \colhead{BIC} & \colhead{PS Ratio} & \colhead{Class.} & \colhead{Sub-Class.}}
\startdata
11 & 52.937 & $-27.761$ & 0.731 & 42.871 & F814W & 0.685 & 0.811 & 11225.636 & 3.005 & 1 & 0 \\
20 & 52.948 & $-27.808$ & 1.37 & 43.469 & F814W & 0.212 & 0.791 & 7559.701 & 58.993 & 2 & 0 \\
28 & 52.956 & $-27.721$ & 1.32 & 44.276 & F814W & 0.301 & 0.598 & 3204.908 & 14.440 & 2 & 1 \\
\enddata
\tablecomments{
Summary of classifications and properties. (1) Chandra Deep Field (CDF) ID. (2) CDF Right Ascension. (3) CDF Declination. (4) Spectroscopic Redshift. (5) 0.5–7\,keV Intrinsic Absorption-Corrected X-ray Luminosity. (6) Filter used for \texttt{galight} fitting. (7) Best-performing B/T Light Ratio. (8) Associated Reduced $\chi^2$. (9) Associated Bayesian Information Criterion. (10) Ratio of point source light to host galaxy light. (11) Visual Morphology Classification. 1=Disk, 2=Disk-Spheroid, 3=Irregular, 4=Spheroid, 5=Point Source. (12) Morphology Sub-Classification. 0=No Classification, 1=Disturbed, 2=Major Merger, 3=Ring.}
\end{deluxetable*}

Once we choose the preferred fit for each source for each filter, we implement both physical and quality-control checks on the resultant fits. 

As redshift increases, the rest-frame wavelength probed by a given filter also increases. We can see direct evidence of this in our fits. The median B/T ratio of the F125W and F115W filters is nearly 0.1 higher than F814W, consistent with probing an older stellar population more densely concentrated in the bulge, as well as the impact of resolution losses at longer wavelengths. This emphasizes the importance of utilizing a standard rest-frame filter to minimize the effects of observing different stellar populations.

For our purposes, we wish to compare rest-frame B-band morphologies. This is consistent with previous work in the GOODS fields (e.g., \citealt{simmons2011_obscured_goods_agn}, hereafter S+2011). Thus, using the spectroscopic redshift for a given galaxy, we compute the rest-frame wavelength observed using each available filter. For sources with $z < 1$, this generally means the HST/ACS F814W filter. For sources with $z > 1$, we generally use JWST/NIRCAM F115W if it is available, otherwise HST/ACS F125W, falling back to F814W if neither of the redder filters are available.

We also must check the quality of our models. We adapt a modified methodology from S+2011, checking that the effective radius of the galaxy \( r_{\mathrm{eff}} \) be greater than three pixels and the difference between observed and modeled apparent magnitude \( \Delta m \) must be less than 1~mag. This ensures that our fits exclude outliers or unphysically small, bright, or dim results. 

As a final test before analyzing our results, we compared our final B/T ratios with those from S+2011. There are 66 galaxies in common between our datasets. For those galaxies, a 1:1 line was the best fit for the available data, but with substantial scatter, especially at low B/T ratios ($\mathrm{Pearson \ R \ coefficient} \sim 0.53$; $\mathrm{Mean \ Absolute \ Deviation}= 0.2371$). Simmons tends to be skewed towards higher B/T ratios. For the 66 matching galaxies, 36 are above the 1:1 line, 30 below. Interestingly, the average offset from the 1:1 line for sources above the 1:1 line is 0.28 units of B/T ratio. For sources below the 1:1 line, it is 0.12. This may appear somewhat surprising since S+2011 subtracted a point source from all galaxies. However, differences in the selection of the point spread function may explain this difference. They created an “analytical PSF based on analysis of dozens of real stars in the GOODS fields, created independently for each band using the IRAF package \texttt{daophot},” (page 3, below Fig. 2; S+2011). As discussed in section \ref{sec: galight_description}, an averaged PSF does not provide the best fit for many of the sources in our shared sample, with an individual PSF often fitting just as well or better than the averaged result. We suggest that our detailed PSF selection methodology is sufficient to explain the slight discrepancies between our works. Overall, we are pleased with the agreement between our studies, and take what disagreements exist as further reinforcement of our methods.

Figure \ref{fig: all_bt} shows the B/T ratio distribution for our full sample. We were able to successfully fit 187/194 galaxies using HST or JWST data. The distribution is broadly flat out to $\mathrm{B/T} = 0.6$ where the fraction drops precipitously. Note that many of the sources with point source contributions greater than or equal to that of the host galaxy tend to have lower B/T ratios. Only $\sim 21\%$ of B/T ratios are greater than 0.6. This is slightly lower than the fraction of bulge-dominated sources seen in Figure \ref{fig: all_morphs}, but it is not surprising considering that many visually bulge-dominated galaxies have a non-negligible point source contribution impacting the visual classification. 

\subsection{Host Galaxy Morphologies}
\label{sec: morphology_methods}
\begin{figure*}
    \centering
    \includegraphics[width=\linewidth]{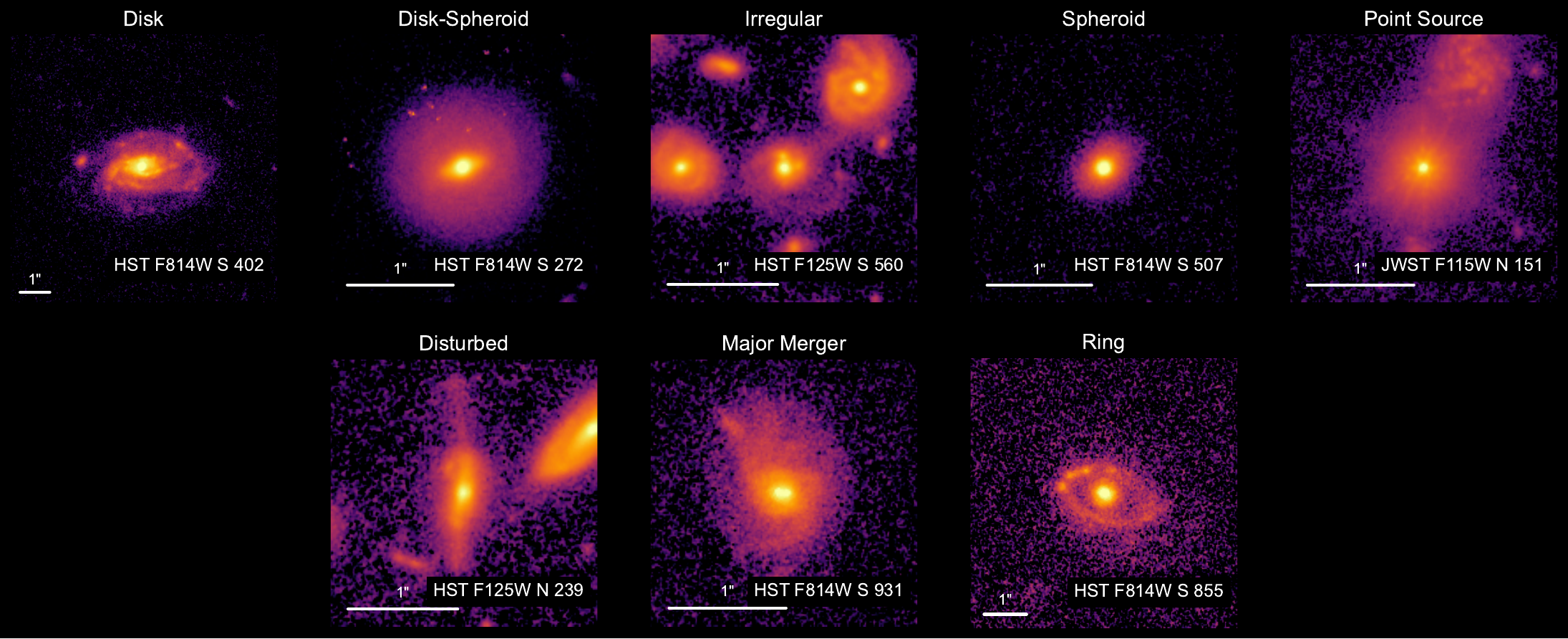}
    \caption{Top row: examples of primary visual classifications. Bottom row: examples of secondary classifications. The telescope, filter, field, and Chandra Deep Field source ID are shown for each source. Every galaxy in our sample was given a primary classification; secondary classifications were awarded as necessary to provide additional context about interacting galaxies. Disturbed is the lowest-priority secondary classification. For example, if a galaxy is clearly in a major merger (as indicated, for example, by the dual nucleus shown in the Figure), then the major merger secondary classification is given in lieu of disturbed. Nearly all major mergers show disturbances (nuclear asymmetry, tidal features, etc.), but not all disturbances are indicative of a major merger.}
    \label{fig: visual_morph_examples}
\end{figure*}

\begin{figure*}
    \centering
    \includegraphics[height = 14cm, keepaspectratio]{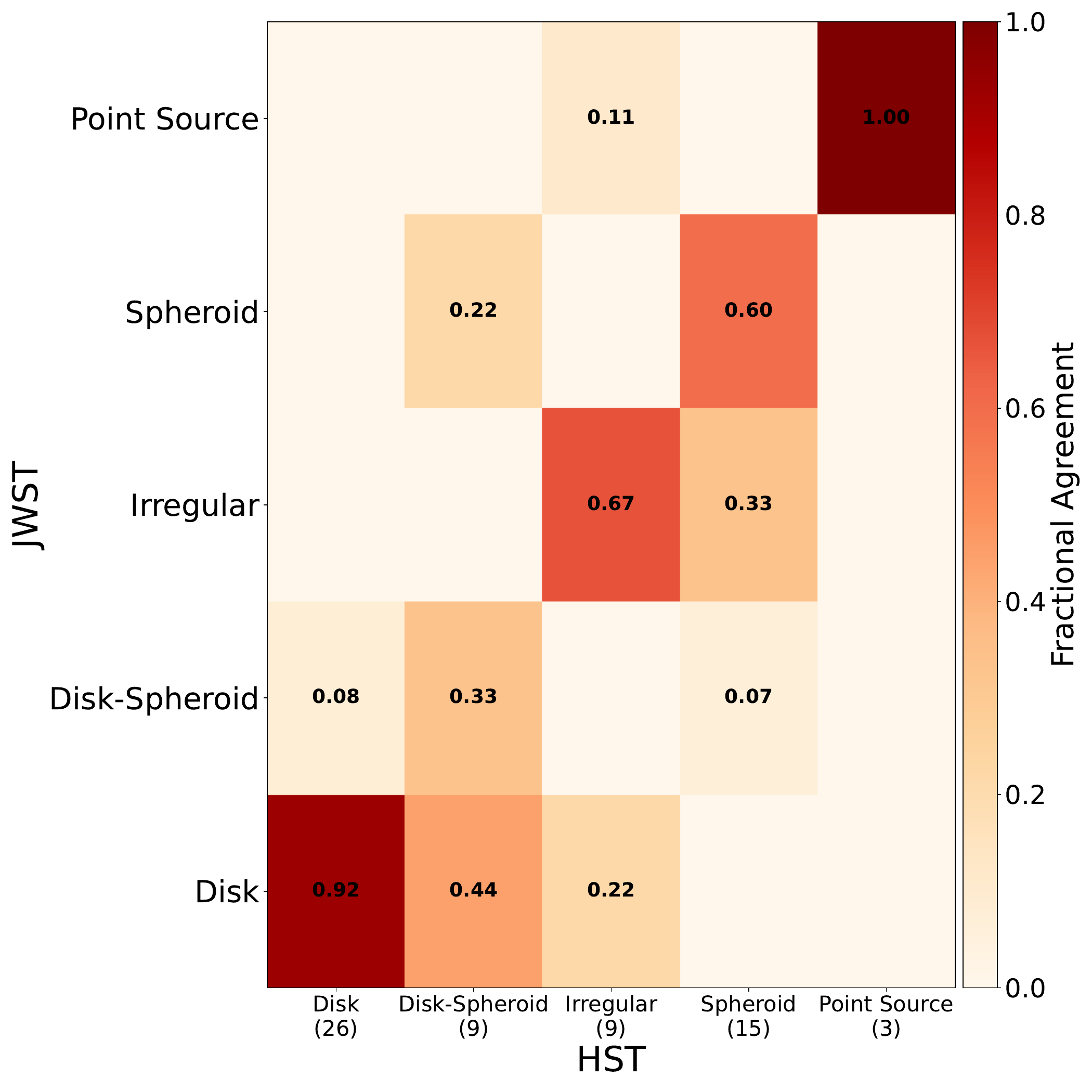}
    \caption{2D Heat Map showing the agreement between HST and JWST sources. Numbers in parentheses denote the number of sources in the column. The results typically agree whether a source is disk-dominated or spheroidal, as noted by the 92\% agreement on disks and 60\% for spheroids. Disk-spheroids show the largest disagreement, with $4/9$ HST disk-spheroids being classified as disks using JWST. This difference is likely due to the increased resolution and sensitivity of JWST which more clearly resolved the disks from any nuclear components.}
    \label{fig: jeana_agreement}
\end{figure*}

\begin{figure}[t]
    \centering
    \includegraphics[width=\linewidth]{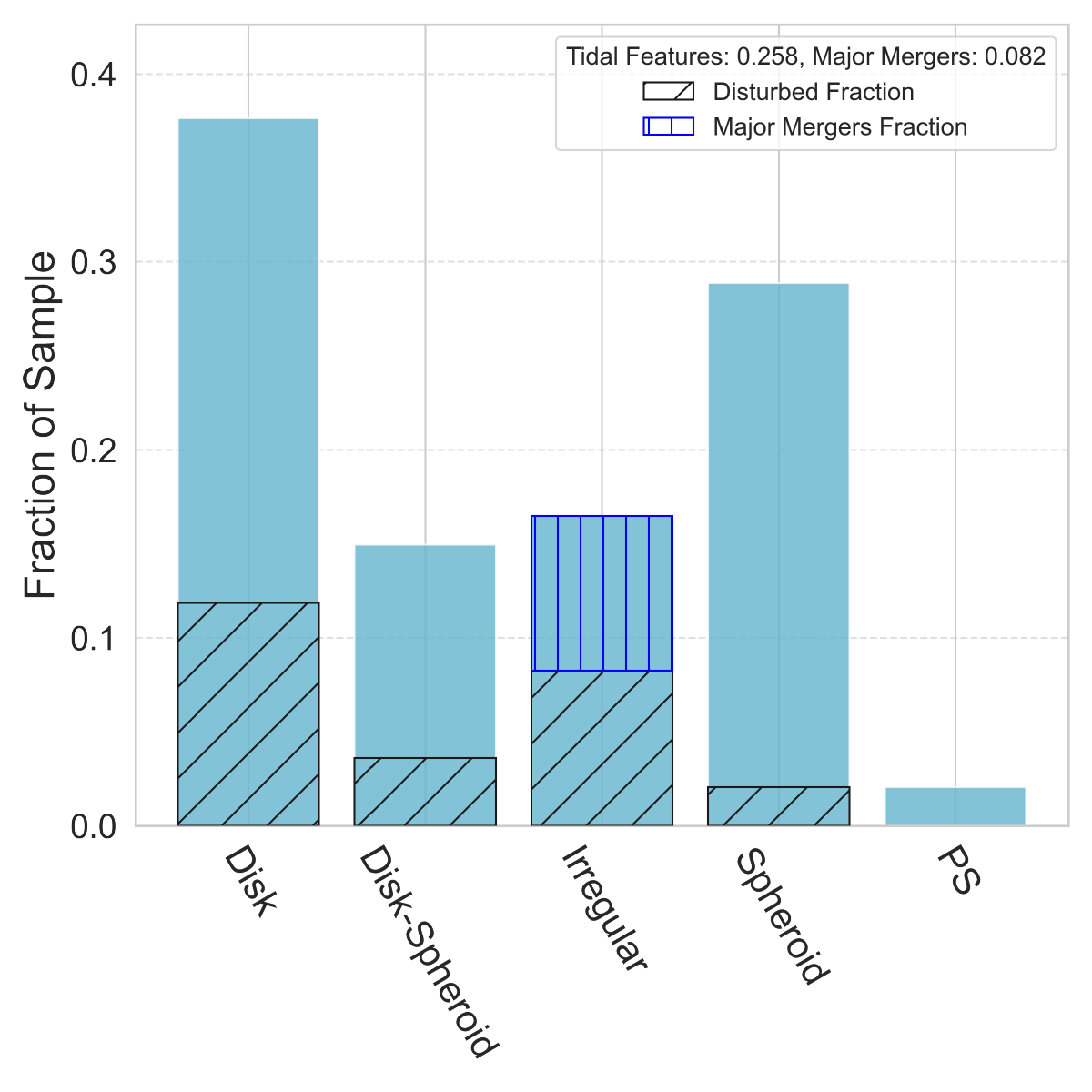}
    \caption{Morphologies for all 194 sample galaxies. disk-dominated sources (disk or disk-spheroid) comprise over half ($\sim 52\%$) the galaxies in our sample; irregulars comprise $\sim 16.5\%$, spheroids $\sim 29.3\%$, and point sources $\sim 2\%$. Ring galaxies are not included due to the small number in our sample.}
    \label{fig: all_morphs}
\end{figure}

\label{sec: morph_intro}
We created 200\,kpc diameter cutouts of each galaxy using the available data from the HST F814W and F125W filters and the JWST F115W and F444W filters and classified the AGN host galaxy morphologies using these cutouts. These filters were selected because they have the largest available spatial coverage across the survey as discussed in \S \ref{subsec: sample selection}.

We used SAO Image DS9 to view each galaxy \citep{sao_ds9}. DS9 offers numerous different contrasts and color schemes to help identify galaxy emission. One of this project's goals was to identify faint features indicative of galaxy interactions. However, the imaging depth and sensitivity varies across the fields and filters \citep{Grogin2011GOODSN, Koekemoer2011CANDELS}. The best stretch for one galaxy may be inappropriate for another. Thus, we did not utilize a standard stretch for our sources, instead relying on classifiers to individually stretch each image within DS9. 

For sources with both HST images, we opened both images simultaneously using DS9. As each filter will emphasize different features at different redshifts, having multiple filters open simultaneously can yield a more complete view of a galaxy's morphology. Ultimately, we make a classification based on viewing all available filters. For JWST data, we follow the same procedure for F115W and F444W data. We describe the classifications below.

Each galaxy has a primary classification and a flag for a secondary classification when applicable. The primary classifications are Disk, Disk-Spheroid, Irregular, Spheroid, and Point Source. The secondary flags are Disturbed, Major Merger, and Ring, and are exclusive from each other. Disturbances include features such as bent galactic disks, tidal streams, and off-center nuclei. Major mergers show multiple galaxies in the process of merging, including dual-AGN. Ring galaxies include a clearly defined ring without necessarily showing tidal features or other interacting companions. 

All secondary classifications are exclusive from the other secondary classifications. There is an implicit hierarchy of secondary classifications. Clearly, major mergers will share features with disturbed galaxies, but the classification of major merger provides us with more information than simply marking it as disturbed. The flags allow us to deal with special circumstances that provide the necessary context to the environments and surroundings of the host galaxies.

Figure \ref{fig: visual_morph_examples} shows examples of all primary and secondary visual classifications.

Disk-spheroids are differentiated from disks by either (1) the presence of a large central bulge or (2) a large stellar halo (indicative of, for example, a lenticular galaxy). For example, the disk-spheroid in Figure \ref{fig: visual_morph_examples} shows a central bar, some faint evidence of spiral arms, and a large halo. Spheroidal galaxies can be either compact or show a large halo. We differentiate spheroids from point sources in that point sources must show no distinct morphological features other than the point source \ref{fig: visual_morph_examples}. It is important to note that we did not use point source subtracted results from \texttt{galight}, instead choosing to characterize the galaxy morphologies distinctly from the \texttt{galight} results.

Before continuing, we note that we have three ring galaxies present in our sample. There are no commonalities in their AGN properties and we do not have sufficient information to comment on their possible formation mechanisms. One is visually classified as a disk and the other two as irregular. It is possible that 2/3 rings are the result of merger activity, but we cannot conclusively prove that with the available data. 

After classification, we compared the HST and JWST morphologies in Figure \ref{fig: jeana_agreement}. 62/194 galaxies in the sample could be fit with both HST and JWST. They generally agree on whether a source is spheroidal, with $73\%$ of HST spheroids being identified as spheroids or point sources by JWST. Spheroids and point sources both share round morphologies with few asymmetries, so they are relatively similar-looking. Despite the higher resolving power of JWST (HST F814W FWHM = $0\farcs08$, HST F125W FWHM = $0\farcs12$, JWST F115W $0\farcs03 < \mathrm{FWHM} < 0\farcs05$, depending on region of field and choice of PSFs), morphologies are remarkably consistent across different telescopes. The greatest disagreement is between HST Disk-spheroids/JWST Disks. 4/9 HST disk-spheroids are classified as pure disks using JWST. 

We also see great agreement in the tiles relating disks and spheroids. $92\%$ of disks and 60\% of spheroids have identical classifications in both telescopes.  It is important to note that the fraction of disks and spheroids does not meaningfully evolve with redshift (see \S \ref{sec: z_evolution} for a larger discussion mostly focused on their B/T ratios). 

Noting good agreement, we combined the results into one master catalog by replacing any HST classifications or other morphological measurements with their JWST counterpart, if available. 

The lead author visually classified all 194 galaxies in our sample. Figure \ref{fig: all_morphs} shows these classifications broken down by galaxy morphology and Table \ref{tab: morphology_counts} displays the raw numbers of each morphology and sub-classification. The distribution is dominated by disks and disk-spheroids with $\sim 37\%$ and $\sim 15\%$, respectively. Irregulars contribute $\sim 17\%$, spheroids $\sim 29\%$, and point sources $\sim 2\%$. $52\%$ of all galaxies show disk-dominated morphologies compared to $31\%$ bulge-dominated (spheroid or point source). The bulge-dominated fraction may be a slight overestimate, as disk components may be lost at high redshifts \citep{simmons_urry_serc}. We will discuss this in greater depth in \S \ref{sec: z_evolution}.

It is unsurprising that we would see a high fraction of disk-dominated sources based on the prevalence of low-luminosity X-ray sources as shown in Figure \ref{fig: xlum_histo}. This highlights one of the strengths of studying the GOODS fields: GOODS allows us to identify a sample of X-ray luminous AGN fainter than those in large surveys that capture some of the earlier phases of galaxy and AGN evolution.

\begin{deluxetable*}{lccccc}
\tablenum{4}
\tablecaption{Summary of Visual Classifications \label{tab: morphology_counts}}
\tablehead{\colhead{Classification} & \colhead{Disks} & \colhead{Disk-Spheroids} & \colhead{Irregulars} & \colhead{Spheroids} & \colhead{Point Sources}}
\startdata
Primary Class. & 73 & 29 & 32 & 56 & 4 \\
Disturbed & 23 & 7 & 16 & 4 & 0 \\
Major Mergers & 0 & 0 & 16 & 0 & 0 \\
Rings & 4 & 0 & 0 & 0 & 0 \\
\enddata
\tablecomments{
Primary classifications for 194 AGN host galaxies in the GOODS fields and their associated secondary flags. Only one secondary flag may be associated with a given galaxy.}
\end{deluxetable*}

Table \ref{tab: morphology_counts} shows the overall number of sources classified into each morphology and the three subclassifications. 

\subsection{Validating Visual Classifications}

To assess whether our classifications are representative of what a larger group of classifiers would identify, we compared them against two literature sources: the CANDELS collaboration visual classifications from \citet{Kartaltepe2015morph} (hereafter K+2015b) and the machine-learning based “visual-like” classifications from \citet{huertascompany} (hereafter HC+2015).

K+2015b visually classified every detected galaxy with an H-band magnitude upper limit 24.5 and $\mathrm{z} \ < \ 4$. This classification scheme had 3-5 classifiers per galaxy and attempted to categorize many important galaxy features including bars, spiral arms, merger features, and bulge or disk dominance. It is important to note that they did not specifically analyze AGN host galaxies as part of their study. 

They sorted galaxies into main categories including Disk, Spheroid, Irregular, Disk-Spheroid, Point Source, Spheroid-Irregular, Disk-Spheroid-Irregular, and Unclear. They also took a host of data about specific features. However, for this simple cross-check, we utilized only the main categories except for three specific morphological flags: Disk-dominated, Merger, and Bulge-dominated. These three flags provide additional information to clarify their classification scheme. For disks and spheroids, respectively, if a source was disk or bulge-dominated, they were automatically assigned to that classification. If a source had a merger score of 0.8 or better, it was automatically classified as irregular. 

The HC+2015 machine learning model was trained on the classifications from K+2015b. This dependency is inherent in the data, so it should not be surprising that the results are similar. In Appendix \ref{app: additional visual classifications}, we discuss independently training a group of classifiers from the UW Madison Astronomy Club and local high schools to classify the galaxy morphologies. We also find good agreement with their results.

For both catalogs, we needed to convert fractional likelihoods of different morphologies into discrete classifications. HC+2015 does include a mapping from fractional to firm classifications. We adapted the HC+2015 thresholds, except for a modified irregular flag designed to mimic our visual classifications which tend to have a broader definition of irregular than HC+2015. We considered irregular galaxies to be those that showed more than 10\% irregularity and did not fall into another morphological category. 

We note that HC+2015 calibrated these thresholds using ``visual inspection." They provide example image stamps of their morphological classes within their paper.

\begin{figure*}
    \centering
    \includegraphics[width=\linewidth]{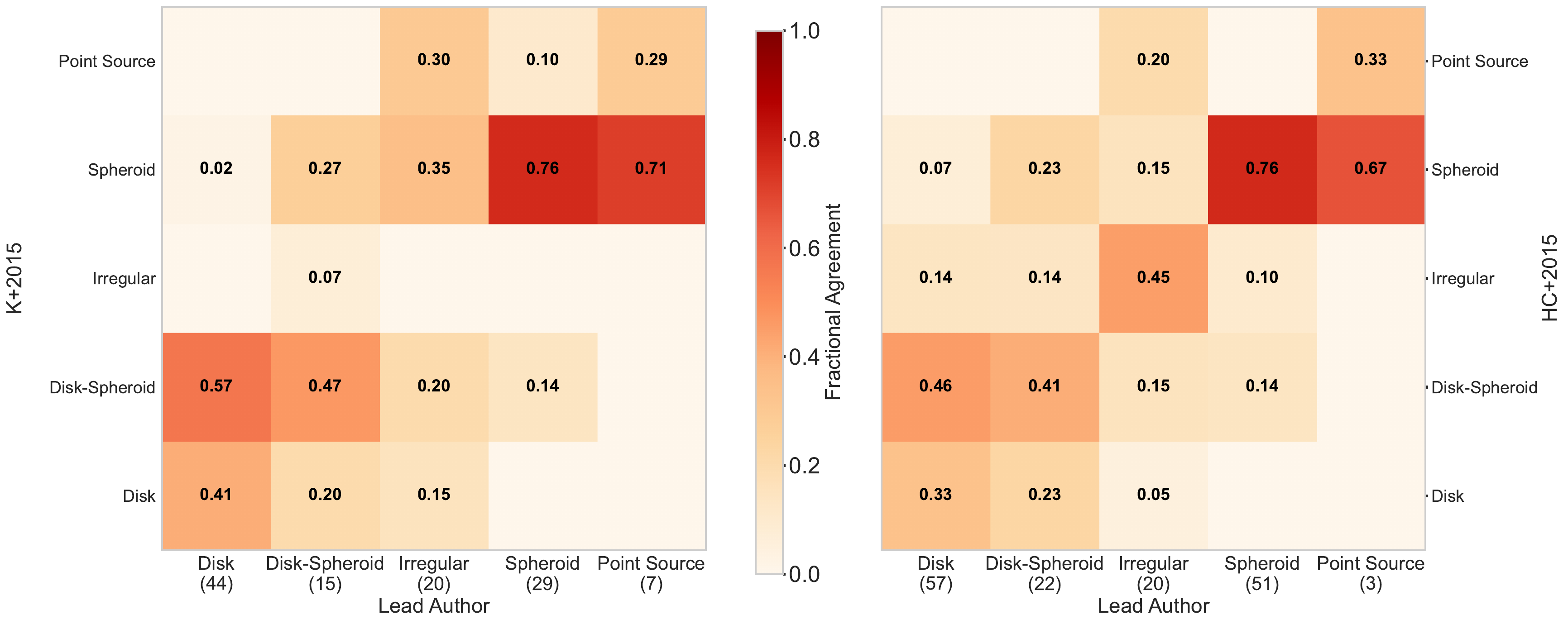}
    \caption{2D-histogram comparing visual-like morphological classifications from (1) \citet{huertascompany} and (2) \citet{Kartaltepe2015morph} with visual classifications from the lead author. Note: 67/194 of the lead author's classifications are from JWST, with the remainder from HST. Both classifiers agree very well for bulge-dominated (spheroid and PS) galaxies. We generally agree whether a source is disk-dominated, but do not agree specifically about whether a spheroidal or halo component is prominent beside the disk. This is not surprising, as the disk-spheroid category is the most ``subject to interpretation" category within our classification scheme. Note also that irregulars show low agreement (45\% with \citet{huertascompany}, 0\% with \citet{Kartaltepe2015morph}). There are several reasons for this: irregular galaxies constitute a wide-range of morphological properties from small with large asymmetries to those undergoing major mergers. Irregular indicators such as tidal streams are also faint, and if they are not sought out, they can remain hidden from studies that do not explicitly look for them. We note that 100\% of this work's visually-classified irregular galaxies show evidence of disturbances or major mergers, improving our confidence in our classification of these sources.}
    \label{fig: wj_hc_jk_comparison}
\end{figure*}

In their paper, K+2015b provided a table of fractional classifications. From this, we constructed our own scheme influenced by that of HC+2015. We mostly rely on primary fractional likelihood values, but include flags to catch disk or bulge-dominated sources and major mergers if applicable. 

To attempt to smooth over differences in classification methodology, while we show direct comparisons for all morphologies in Figure \ref{fig: wj_hc_jk_comparison}, we will quote agreements in terms of whether a source is disk-dominated (disk or disk-spheroid), irregular, or bulge-dominated (spheroid or point source). As K+2015b and HC+2015 both use slightly different definitions of what constitutes each galaxy morphology, grouping in this way allows us to simply compare the primary morphological characteristics without wading into minor methodological differences. 

Figure \ref{fig: wj_hc_jk_comparison} compares our classifications with those of K+2015b and HC+2015. For both catalogs, we find strong agreement ($\sim$75\%) for bulge-dominated sources (spheroids and point sources). Agreement is also high ($>$70\%) for disk-dominated systems (disk or disk–spheroid), though we differ on whether a galaxy should be labeled a pure disk or a disk–spheroid. Our scheme appears stricter in assigning disk–spheroid classifications than either catalog. Foreknowledge that each galaxy hosts an AGN may also have influenced our classifications, whereas K+2015b made no such assumption. Excluding the JWST-based classifications yields nearly identical results, consistent with the good HST–JWST concordance shown in Figure \ref{fig: jeana_agreement}.

Agreement is lowest for irregular galaxies (40\% with HC+2015, 5\% with K+2015b). This is expected: our visual classifications deliberately emphasized faint tidal features and disturbances by using individualized, non-standardized image stretches, whereas neither K+2015b nor HC+2015 were designed to systematically capture such features. Importantly, 100\% of galaxies we classify as irregular show clear evidence of disturbances or mergers, reinforcing that our irregular category specifically traces disturbed systems with recent interactions.

Overall, we find good agreement with K+2015b and HC+2015 for bulge-dominated ($\sim$75\%) and disk-dominated ($\sim$70\%) systems, and lower agreement for irregulars, as anticipated. Grouping all sources into disk-like, bulge-like, or irregular, we achieve an overall concordance of 83\%. While the small number of classifiers makes us cautious about over-relying on visual morphology, especially in ambiguous cases, visual inspection remains essential for identifying faint merger signatures and tidal features that are not captured by automated or profile-based methods.

\section{Analysis} \label{sec: analysis}
In this section, we test whether our structural measurements show systematic trends that could bias our interpretation of host galaxy morphologies. Specifically, we examine whether bulge-to-total (B/T) ratios or point-source contributions evolve with redshift, compare the distribution of B/T ratios across visual morphologies and SED shapes, and analyze the special case of galaxies classified as point sources. Together, these checks allow us to evaluate the robustness of our results and to identify where AGN contamination or observational effects may influence the inferred properties of the host galaxies and will allow us to discuss the physical interpretations of our results in \S \ref{sec: discussion}.

\subsection{Redshift Evolution (or Not) of Visual and Quantifiable Classifications}
\label{sec: z_evolution}

\citet{simmons_urry_serc} showed through simulations of Hubble Space Telescope ACS observations that it is common, and indeed expected, to misclassify galaxies as spheroids as redshift increases. At higher redshifts, more of a galaxy is shifted to fewer pixels and the galaxy appears more bulge-like. Under this model, a flat redshift distribution of B/T ratios is indicative that higher redshift sources may have their B/T ratios overestimated. 

\citet{simmons_urry_serc} also note that B/T ratios are often overestimated by $\sim 10\%$ for AGN host galaxies. Our fitting methodology might account for this overestimation. Galaxies with a bright point source are most likely to contribute to this overestimation. By using the PS ratio to decide whether or not to subtract a point source from a given galaxy, we subtract the point source where necessary without artificially suppressing all B/T ratios. For sources without a large PS ratio, we do not expect the B/T ratio to be significantly overestimated. 

\begin{figure}
    \centering
    \includegraphics[width=\linewidth]{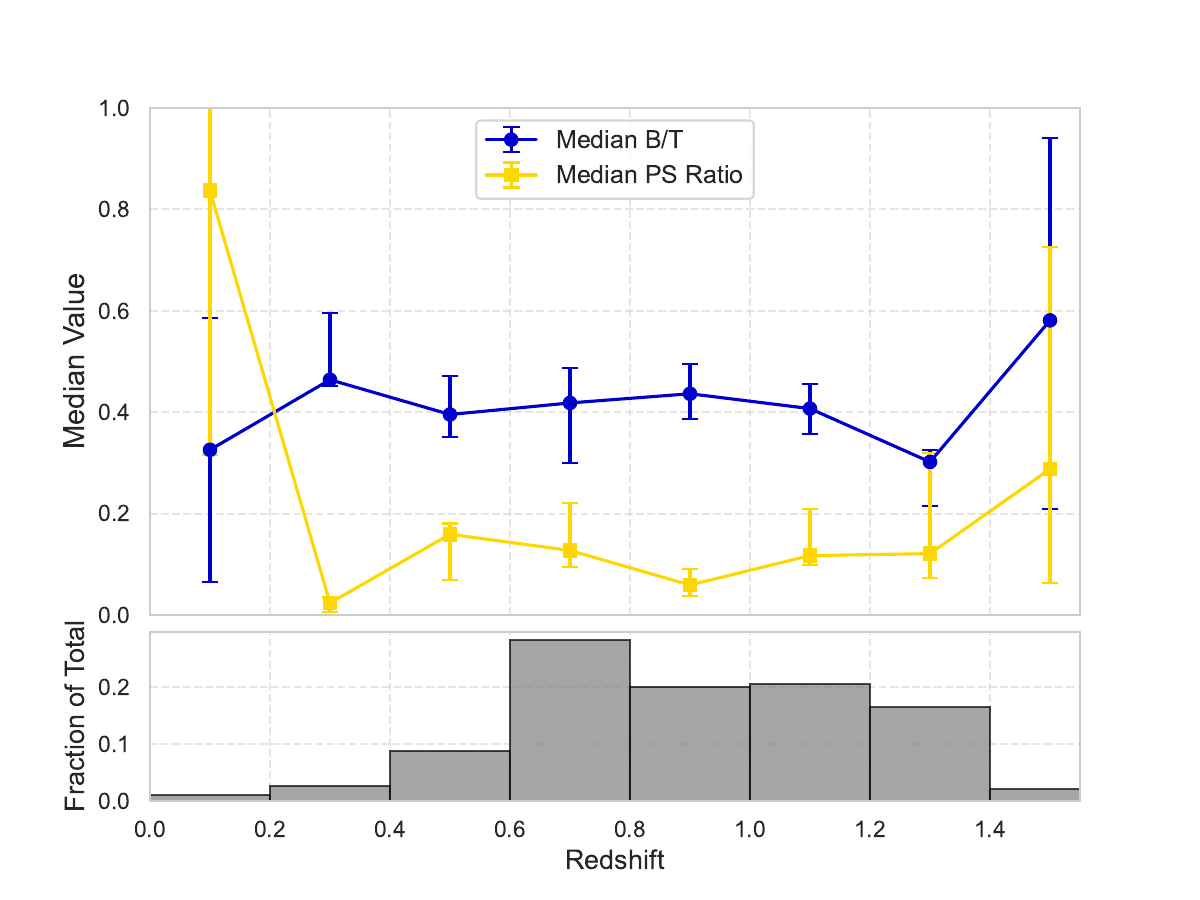}
    \caption{Median B/T and median PS Ratio plotted as a function of redshift. Error bars from the 68\% confidence interval of a thousand-iteration Bayesian Bootstrap. Neither metric significantly increases with redshift. Within the error bars, the overall distributions remain steady or slightly decrease at the high-$z$ end of our redshift range.}
    \label{fig: bt_psratio_z}
\end{figure}

Indeed, as seen in Figure \ref{fig: bt_psratio_z}, within the credible intervals from a Bayesian bootstrap, the bulge/total ratios and PS ratios are broadly constant across the sample, consistent with the picture from \citet{simmons_urry_serc}. 

We also checked for evidence of redshift evolution in the visual classifications and did not find clear evidence of additional spheroid classifications as redshift increased. 

Ultimately, we do not see a clear redshift evolution across the morphologies. That the median B/T and PS ratios are flat may indicate that we are marginally overestimating the B/T ratios for galaxies at higher redshifts, but this effect will be small on our overall statistics since the majority of sources in our sample have $\mathrm{z}<1$.

\begin{figure*}[t]
    \centering
    \includegraphics[width=\textwidth]{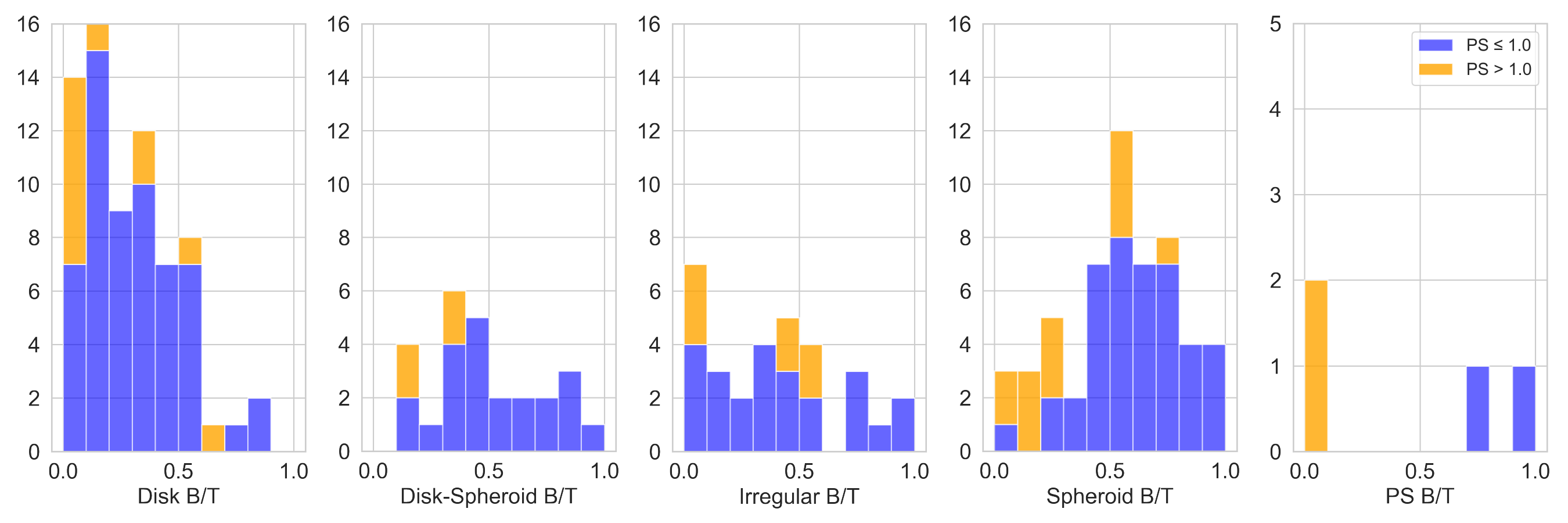}
    \caption{Stacked B/T ratio histograms sorted according to host galaxy morphology. Yellow colors indicate galaxies where a point source has been subtracted, while blue indicates that no point source was removed. The trends show the majority of disks having B/T ratios less than 0.5 and disk-spheroids having a similar distribution. Irregulars show a nearly flat distribution. Spheroids show peaks at both 0.5 and 0.9. All visually-classified point sources have indeed had a point source subtracted.}
    \label{fig: b/t morph comp}
\end{figure*}

\subsection{Analysis of Bulge/Total Light Ratios}
\label{sec: b/t analysis}

Figure \ref{fig: b/t morph comp} compares the visual classification with the distribution of B/T light ratios for each galaxy. Disks and Disk-Spheroids tend to have low B/T ratios with the majority with $\mathrm{PS} \ < \ 0.5$. Their median B/T ratios lie at $0.29$ and $0.31$, respectively, though disk-spheroids entirely lack sources with $\mathrm{B/T} < 0.1$. In contrast, spheroids and point sources tend toward higher median ratios B/T of $0.54$ and $0.75$, respectively. 

Irregular galaxies show a median B/T of $0.42$. Since the majority of galaxies in our sample are disk-dominated with a minority spheroid or bulge-dominated, it is \textit{prima facie} plausible that irregulars should appear representative of the broader sample.

In general, the B/T ratios agree with our visual classifications. None of the classifications diverge dramatically from what we expect and their distributions are similar to those presented in past studies of these galaxies \citep[e.g.,][]{simmons_urry_serc}.

\begin{figure}
    \centering
    \includegraphics[width=0.85\linewidth]{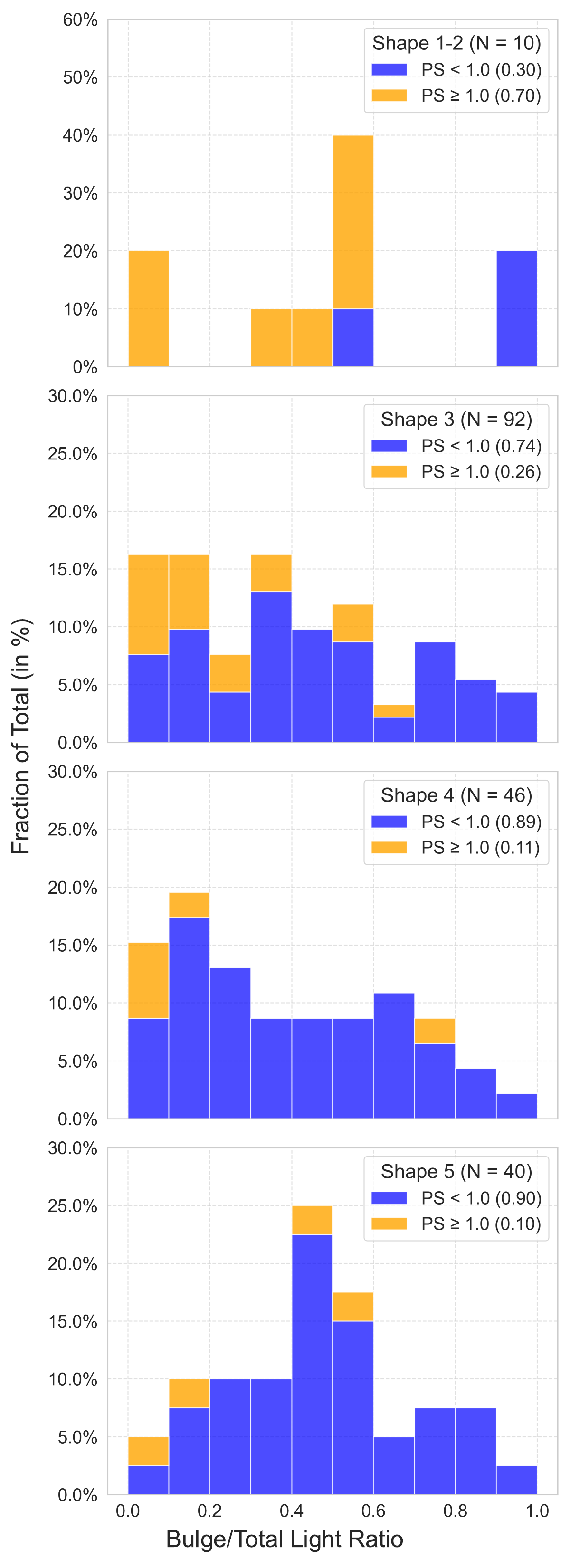}
    \caption{Bulge/total light ratios measured with \texttt{galight}, grouped by SED shape as defined in Table~\ref{tab: panel_lims} and Figure~\ref{fig: goods5p}. The fraction of sources above or below the point source ratio cut, PS = 1, is shown in parentheses. Shape 5 appears visually and statistically distinct from shapes 3 and 4 as discussed in \S \ref{sec: b/t analysis}.}
    \label{fig: b/t_frac5p}
\end{figure}

Figure \ref{fig: b/t_frac5p} shows the B/T ratios for the galaxies in each SED shape. Shape~1-2 shows a broad distribution of B/T ratios. 10/11 have been point source subtracted and there are a broad range of B/T ratios.

Shapes 3 and 4 are similar, with the majority of their galaxies having $\mathrm{B/T} < 0.5$, though shape~4 shows a slightly higher fraction of galaxies with very low B/T ratios ($0 < \mathrm{B/T}<0.2)$. Shape~3 does include more galaxies that have been point source subtracted. 

Finally, Shape~5 is clearly distinct from all other shapes with a sharp peak around $\mathrm{B/T} = 0.5$. It also shows the lowest fraction of point source subtracted galaxies at 17\%. 

To quantify the difference between the SED shapes, we utilize the K-S test \citep{Massey_1951}. The K-S test is a nonparametric test used to test whether two samples originate from the same underlying distribution. In our case, we posit a null hypothesis: for any combination of SED shapes, the B/T distributions originate from the same underlying distribution. Rejecting the null hypothesis indicates that the samples are drawn from differing distributions, or in physical terms, populations of galaxies. 

\begin{deluxetable}{lcc}
\tablenum{5}
\tablecaption{KS Test Results Between Shapes \label{tab: ks_shapes}}
\tablehead{
\colhead{Comparison} & \colhead{K-S Statistic} & \colhead{p-value}}
\startdata
\textbf{Shape 3 vs Shape 4} & \textbf{0.0978} & \textbf{0.9244} \\
\textbf{Shapes 3+4 vs Shape 5} & \textbf{0.2514} & \textbf{0.0320} \\
Shape 3 vs Shape 5 & 0.2478 & 0.0533 \\
Shape 4 vs Shape 5 & 0.3033 & 0.0295 \\
\enddata
\end{deluxetable}

Table \ref{tab: ks_shapes} shows the results of the K-S test. Note that we did not run any tests on shape~1 due to the small sample size (N=11) of that shape. Shapes 3 and 4 appear to be drawn from a similar underlying distribution ($\mathrm{p=0.9244}$). Assuming that shapes 3 and 4 represent a similar underlying population of galaxies, combining and comparing them against shape~5 yields a p-value of $p = 0.032$, implying a frequentist confidence level of 96.8\%. Thus, we conclude that the host galaxies of obscured and faint shape~5 AGN are distinct from those in the shapes 3 and 4. We will further support and discuss the implications and context of this difference in \S \ref{sec: discussion}.

\subsection{The Host Galaxies of Visually-Classified Point Sources}
\begin{figure*}
    \centering
    \includegraphics[width=\linewidth]{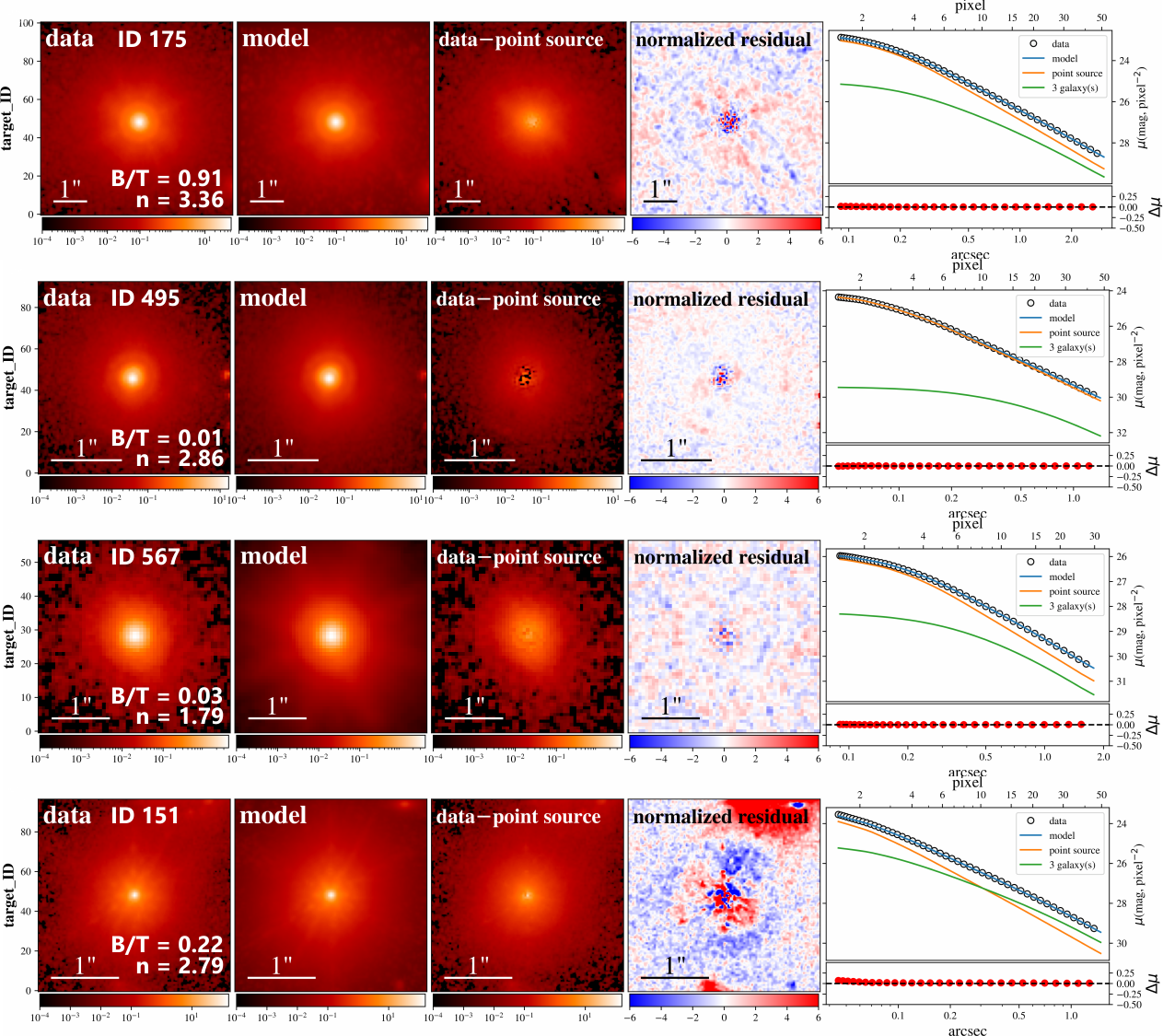}
    \caption{Resulting \texttt{galight} fits for the four visually-classified point sources in our sample. Each galaxy has had a point source subtracted in an attempt to detect each AGN's host galaxy.}
    \label{fig: ps_galight_hosts}
\end{figure*}

\texttt{galight} has previously been used to study the host galaxies of quasars, successfully identifying light from the galaxies and allowing an estimate of some morphological properties \citep{ding_2023_galight_nature}. Thus, utilizing \texttt{galight}, we subtracted point sources from the four visually-classified point source galaxies, GOODS-S ID 175, 495, 567, and GOODS-N ID 151. Note that we do not change the visual classification of the galaxy based on the point source subtracted result. The tentative classifications discussed in this section are only for a discussion of these unique sources. 

Figure \ref{fig: ps_galight_hosts} shows the results for these four galaxies. The galaxies are, in general, of intermediate Sersic Index and B/T ratio after PS subtraction. 
ID 175 appears to be a spheroidal galaxy with a bright central AGN (log $L_{\mathrm{x}} \ =  \ 44.11 \ \mathrm{erg \ s^{-1}}$. Its Sersic Index is n = 3.38 and B/T ratio = 0.90, even after point source subtraction. This is all consistent with an identification of spheroid. 

ID 495 poorly detects any host galaxy light. The subtraction of the point source completely minimized the B/T ratio. Although some host galaxy light may be present, it is difficult to conclusively state anything about this source.

ID 567 shows one of the most interesting subtractions. The Sersic Index n = 1.79, indicating a disk-dominated morphology. Knowing that the top is facing North, the host galaxy is extended from the NW to SE of the image. The B/T ratio (B/T = 0.03) is also consistent with a disk-dominated host galaxy. The galaxy is at moderate redshift (z = 0.74). It shows one of the most extreme point source ratios in the entire sample at 92.85, meaning the point source contributes almost 93 times the light as the host galaxy. This is unsurprising for an unobscured AGN, though it is dramatic considering the X-ray luminosity is only $10^{43.63} \ \mathrm{erg \ s^{-1}}$. The SED is largely unobscured, showing bright UV emission with minor damping of the IR. While difficult to ascertain information about the angle of the system, the unobscured SED inside a disk-dominated morphology indicates that we may be viewing this galaxy directly head-on, allowing us to peer directly to the accretion disk where much of the UV emission is generated.

ID 151 is one of the subset of galaxies imaged with JWST F115W. The galaxy has an intermediate Sersic Index (n=2.79) with a disk-like B/T ratio (B/T = 0.22). The residual image is poor, so it is difficult to determine much about the galaxy’s morphological properties.

Taken together, it is difficult to make any firm conclusions about these galaxies. Two are of indeterminate morphology, one appears bulge-dominated, and one disk-dominated. The only salient point that can be drawn from them is that bright, quasar-like AGN may appear in any galaxy morphology so long as the conditions – for example, available material to accrete or an appropriate viewing angle – exist. 

\section{Discussion}
\label{sec: discussion}

\begin{figure}
    \centering
    \includegraphics[width=\linewidth]{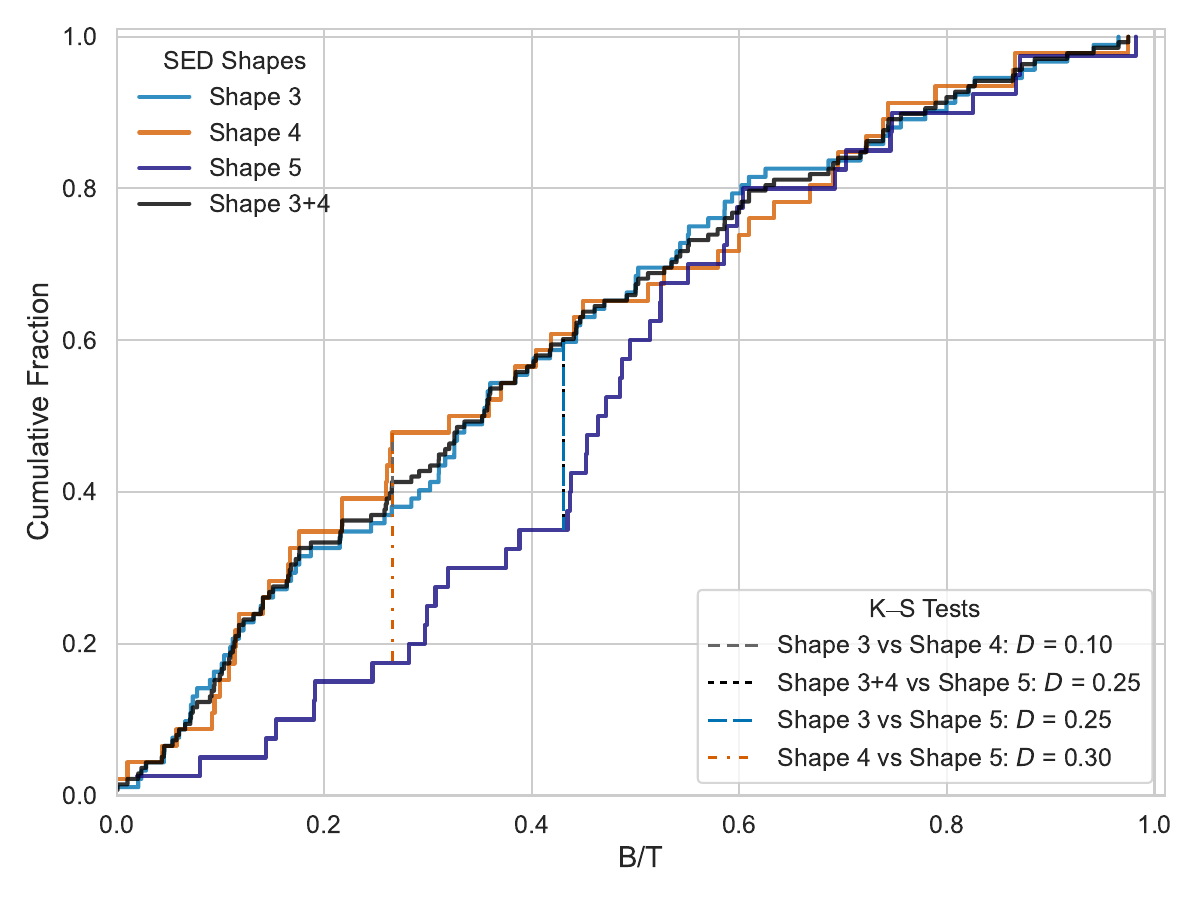}
    \caption{B/T ratio cumulative distribution functions for SED shapes 3, 4, 5, and the combination of shapes 3 and 4 (shape~3+4). Shapes 3 and 4 follow similar growth patterns, where as shape~5 shows a dearth of low B/T ratio values, with a sharp increase between $\mathrm{B/T} = 0.4$ and $\mathrm{B/T} = 0.6$.}
    \label{fig: bt_cdf}
\end{figure}

\begin{figure}
    \centering
    \includegraphics[width=\linewidth]{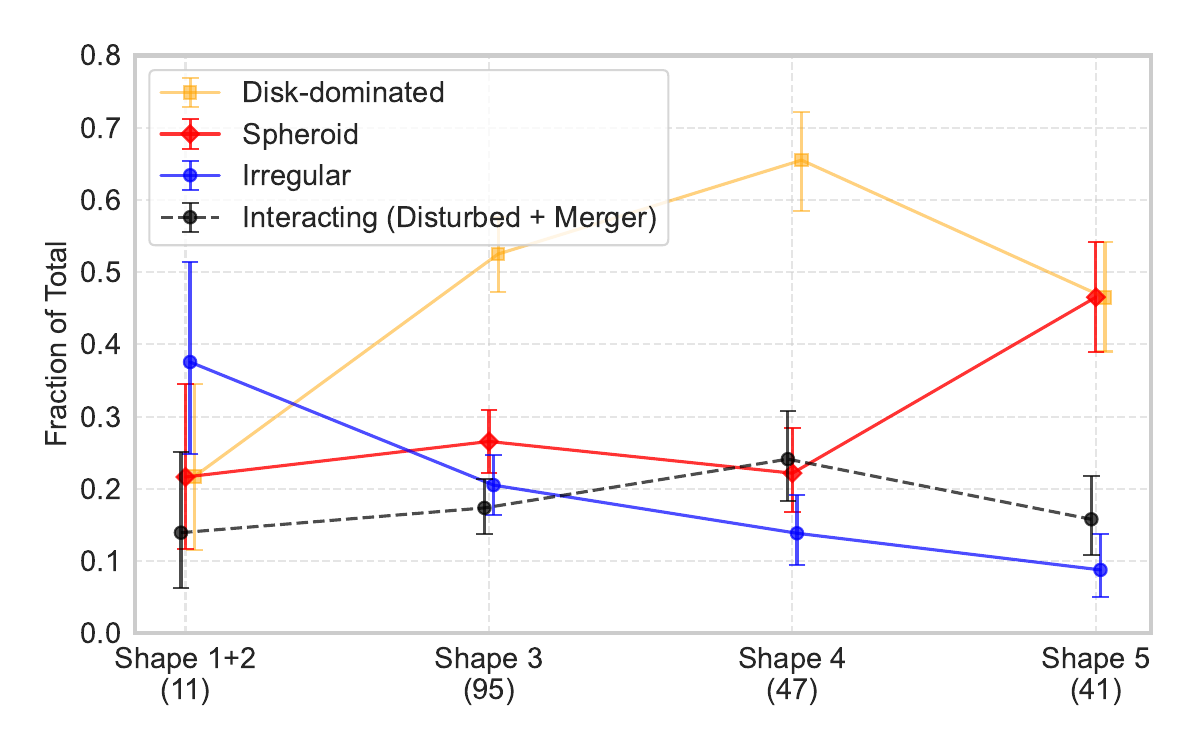}
    \caption{Fraction of galaxies visually classified as irregular (blue), spheroid (red), disk-dominated (orange, disk or disk-spheroid), or interacting (black dashed,  secondary classification of disturbed or major merger). To avoid double counting, the irregular galaxies were not counted towards the "interacting" fraction, meaning only visual secondary classifications for non-irregular galaxies contributed. Bars show the Bayesian credible interval for each SED shape. Irregular galaxies monotonically decrease from Shape~1/2 to Shape~5. Interacting galaxies are flat within the credible intervals. The significantly increased proportion of spheroid galaxies in shape 5 supports the contention that shape 5 emphasizes galaxies in a different evolutionary stage than shapes 3 and 4.}
    \label{fig: sed_shape_irregular}
\end{figure}

Analyzing the distributions of galaxy properties according to the SED shapes discussed in \S \ref{sec: methodology_seds} and \S \ref{sec: b/t analysis} reveals several interesting trends.

\subsection{Interactions Do Not Explain the SED Shape Evolution}

First, we want to begin by discussing the influence of galaxy interactions on our sample. 34\% of AGN in our sample show evidence of interactions. It is noteworthy that the other 66\% do not show such evidence. We posit that this in agreement with \citet{treister_2012_majormerger_agn}, who suggested that major mergers drive accretion for only the most luminous AGN. To test this in our sample, we analyzed the 28 AGN with X-ray luminosities above  $\mathrm{L_x} > 10^{44} \mathrm{erg \ s^{-1}}$. We find that half are undergoing interactions with other galaxies.  Despite a higher prevalence of interactions, that only half of these luminous AGN are undergoing interactions suggests that secular processes are important in driving the AGN. Indeed, many secular mechanisms may combine to explain the observed properties of the AGN SEDs. Thus, let us discuss more distinctly the properties of the galaxies of SED shapes~3, 4, and 5. Because shape~1/2 only contains 9 galaxies with constrained B/T ratios, we will not discuss it at depth. 

The population of galaxies that fall into SED shapes 3 and 4 are statistically similar in their global structures. Although, as seen in Figure \ref{fig: b/t_frac5p}, shape~3 hosts have more than double the fraction of point-source--subtracted systems (0.26 vs.\ 0.11), their bulge-to-total (B/T) distributions are consistent with being from the same parent distribution (K--S test $p = 0.9244$), with median B/T values differing by only $\Delta \mathrm{B/T} \approx 0.013$. The interacting fractions (including disturbed galaxies or major mergers) remain below 30\% for both shapes. This agrees with results indicating that active galaxies are quite similar to their inactive counterparts \citep[e.g.,][]{Rosario2015_bulge_enhancement, bruce2016_active_inactive_galaxies}. This is an important note, as it implies that AGN may activate within any morphology of host galaxy and are not necessarily biased towards systems with ongoing interactions.

Despite these similarities in the B/T ratio distributions of shapes 3 and 4, they are identified as different SED shapes due to differing levels of MIR and X-ray emission. Indeed, \citet{Auge2023} found that shape~3 AGN are on average more X-ray luminous than shape~4 AGN in the full AHA sample by $\sim 0.5$ dex. They also showed that MIR emission is largely powered by the AGN itself, not intrinsic properties of the host galaxy (e.g., star formation). Thus, nuclear-scale factors---such as dust distribution or covering factor---may be more determinative of SED shape than global morphology.  If the host galaxies are morphologically similar, then the systematic difference in SEDs can be plausibly explained by variations in the nuclear environment without invoking differences in the large-scale structure of the host. This view aligns with recent work emphasizing that sub-galactic gas and dust physics can strongly influence observed AGN SEDs \citep[e.g.,][]{Alonso-herrero_clumps_agn_2011, harrison_almedida_agn_review2024}. Thus, we do not need morphological differences to explain the SED. This reinforces the view that AGN arise across morphologically diverse galaxies \citep[e.g.,][]{hickox_alexander_obscuredagn_review2018, harrison_almedida_agn_review2024}. 

Simulations of AGN have shown that different AGN are likely to be fed in different ways. For example, only the most luminous AGN with bolometric luminosities greater than $10^{46} \ \mathrm{erg \ s^{-1}}$ are likely to be substantially fueled via major mergers (e.g.,  \citealt{steinborn_2018}). Lower-mass ($\mathrm{\log M_\star \leq 9.5 \ M_\odot}$) galaxies may be impacted by ram pressure stripping during mergers, with black hole accretion rates being amplified by as much as a factor of two \citep{ricarte_2020_rps_agn}. This could plausibly lead to an increase in the AGN's X-ray luminosity, but considering at most 20\% of AGN host galaxies in the same simulations show clear merger signals, this again supports our results suggesting that the majority of AGN are not triggered by mergers and instead rely upon secular feeding processes.

Our findings are also consistent with several observational results on X-ray AGN hosts in the same redshift regime.  Studies of X-ray--selected AGN at $z\leq 2$ find at most moderate excesses of strong merger signatures compared to controls \citep[e.g.,][]{cisternas2011_weak_agn_merger_link, 
kocevski_candels_agn_merger_2012, povic2012_weak_agn_merger_link, villforth_2014_merger_candels}. The lack of clear connections between mergers and AGN activity except in the most luminous AGN support scenarios where the nuclear SED differences reflect variations in fueling and obscuration on sub-galactic scales within most AGN host galaxies, rather than wholesale transformations of global galaxy morphology on short timescales. 

\subsection{The Identification of Nuclear-obscured Spheroid Galaxies}

In contrast to shapes 3 and 4, which are morphologically similar but likely differ slightly in nuclear properties or geometry, shape 5 shows a coherent host-scale distinction that may indicate a different evolutionary state. Disk fractions fall from $\sim 60$\% in shape~4 to $<30$\% in shape~5 (see Fig. \ref{fig: sed_shape_irregular}), while spheroids become the largest plurality. Viewing the \texttt{galight} results,  shape~5 hosts show a dearth of low B/T galaxies relative to shapes~3--4. This is noteworthy given that \citet{Auge2023} reported similar UV/MIR/FIR-to-intrinsic X-ray luminosity ratios for shapes~4 and~5, and earlier work in \citet{hao_2013_quasar_seds} did not distinguish our shape~4 and 5. To clearly ascertain if this population of galaxies is truly distinct, we will require larger samples. However, we can posit a physical explanation for these galaxies based on our unique sample. 

We interpret shape~5 as a heavily UV-optical-IR obscured phase of post-merger evolution, where galaxies are generally embedded in more bulge-dominated systems that have already undergone galaxy mergers. These galaxies must have a different assembly history compared to the disk-dominated samples in SED shapes 3 and 4. No shape 5 galaxies are undergoing major mergers, and at most 22\% (both the raw fraction, and the top of the Bayesian credible interval after removing double-counted irregulars with disturbances) have evidence of recent interactions. Thus, these spheroids must have undergone major mergers at an earlier evolutionary stage than the galaxies in the other SED shapes. 

We also do not see strong evidence for MIR reprocessing, implying that what obscuring dust is present must be located inside the nucleus. We put forward several suggestions.

A compact, nuclear dust screen (or polar dust) could attenuate the UV while yielding only moderate MIR if the covering factor or dust temperature distribution is low; alternatively, a decline from a prior high-accretion (quasar) phase could reduce the intrinsic UV/X-ray output while leaving residual obscuration. Host-scale interstellar media can also contribute appreciably to X-ray obscuration in some systems \citep{buchner2017_agn_gas_torus}, providing a natural pathway to high obscuration in moderate–high B/T galaxies without invoking ongoing major mergers. 

In this view, shape~5 may represent a late or intermittently fueled stage in which nuclear geometry and host ISM jointly set the observed SED. The host galaxies themselves are likely the post-merger remnants of major mergers in the early universe. We suspect that these galaxies will show suppressed star formation rates consistent with the red sequence. To better understand whether this is truly a population of quiescent galaxies formed through mergers in the early universe, we suggest future studies investigate their colors, star formation rates, and dust properties through SED modeling. However, this in-depth look at these galaxies lies beyond the scope of this paper.

\section{Summary and Conclusions}
In this paper, we have analyzed the $1 \ \mathrm{\mu m}$-normalized SEDs of 194 X-ray luminous AGN in the GOODS fields and examined their connections with their host galaxies' morphologies. We sorted the AGN host galaxies according to the slope of their SEDs in the UV and MIR using the methodology of \citet{Auge2023}. We then visually-classified all the host galaxies and modeled their morphologies using \texttt{galight} to understand if different morphologies correlate with different SED shapes.

We find that the GOODS X-ray-selected AGN live in a wide variety of host galaxies, with $52\%$ being visually disk-dominated and $\sim 31\%$ visually bulge-dominated. The remainder are irregular.  

We note four primary results below:

\begin{itemize}
\item \textbf{Galaxy interactions may not be the primary trigger for obscured AGN:} 34\% of AGN in our sample show signs of interactions. Only Secular processes appear capable of driving AGN to high luminosities, with mergers contributing to only a subset of the population. 
\item \textbf{Post-merger spheroid galaxies appear heavily-obscured:} The most UV- and MIR-obscured AGN are preferentially found in galaxies with higher bulge-to-total ratios, reduced disk fractions, and fewer signs of recent interactions. These systems appear to have assembled earlier in cosmic time and exhibit early-type morphologies while still containing sufficient nuclear dust and gas to fuel and obscure an X-ray–bright AGN.
\item \textbf{Obscured AGN live in diverse hosts:} For the bulk of the sample, spanning lightly to moderately obscured UV/MIR emission, the distributions of host morphologies are statistically similar. Bulge-to-total ratio distributions and interaction fractions remain statistically consistent across these obscuration levels, suggesting that nuclear-scale dust and gas geometry, rather than global host morphology, drives the observed SED differences.
\item \textbf{Most GOODS X-ray AGN exhibit obscured UV/optical SEDs:} Approximately 94\% of the sample exhibit suppressed UV emission, and about half show rising MIR–FIR emission. The high obscured fraction reflects both intrinsic AGN properties and the sensitivity of GOODS to faint dust-reprocessed emission.
\end{itemize}

Using \texttt{galight}, we characterized an interesting population of UV/optical/MIR obscured AGN in spheroid galaxies. This population still contains sufficient dust (and, presumably, gas) to power and obscure an X-ray bright AGN, yet is beginning to show the properties of early-type galaxies. We recommend that future work study these heavily-obscured spheroids to better understand their dust masses and kinematics. Preliminary work in the AHA Collaboration suggests that this population is present in AGN in the wider COSMOS fields, though much more work is necessary to characterize this population's properties. Rigorous SED modeling should elucidate whether these galaxies are quiescent and may hint at their assembly histories.

This study provides additional evidence that a mixture of secular and interaction-driven processes power AGN activity. We find evidence that some of the brightest AGN appear to have host galaxies undergoing or which recently underwent merger episodes, but this is not true of all the most luminous AGN. As larger datasets become available, for example through COSMOS-Web, we expect more clarity as we move from samples sizes in the hundreds to thousands of X-ray selected AGN.

\begin{acknowledgments}
We would like to thank the reviewer for their kind and illustrative feedback. Their comments improved the readability and precision of the argument.

Will Jarvis (WJ) acknowledges support from Research Experience for Undergraduate program at the Institute for Astronomy, University of Hawaii-Manoa funded through NSF grant \#2050710.  WJ acknowledges support from NSF award \#2206844. We thank the Wisconsin Space Grant Consortium through grant $\mathrm{\#UGR23\_6\_0}$ and the National Space Grant College and Fellowship Program: NASA Education Cooperative Agreement \#80NSSC20M0123. WJ thanks the Astronaut Scholarship Foundation for their funding, supporting the purchase of a new research laptop, housing, and academic activities that benefited this project. WJ thanks the University of Wisconsin - Madison Physics Department for their funding through the 2024 Summer Thaxton Fellowship. WJ would like to thank the Institute for Astronomy and the University of Hawai'i  for their hospitality and support during the course of this project. WJ and the authors acknowledge the generous support of the Peter Livingston Scholars Program, whose contributions to undergraduate research have played a key role in the development of this work. Finally, WJ would like to thank L\&L Hawaiian BBQ, La Crepe Cafe, and Pho Viet Manoa Restaurant for showing me a taste of Hawai'i (and Paris!) during his time there. 

This material is based upon work supported by the National Science Foundation Graduate Research Fellowship under Grant No. DGE-2039655. Any opinion, findings, and conclusions or recommendations expressed in this material are those of the authors(s) and do not necessarily reflect the views of the National Science Foundation.

C.A. acknowledge support from NASA through ADAP award 80NSSC22K1126. DS acknowledges support from the NSF under grant No. 1716994, and from NASA through ADAP award No. 80NSSC19K1022. Tonima Tasnim Ananna acknowledges support from ADAP grant 80NSSC24K0692. 

This research made use of Astropy\footnote{http://www.astropy.org}, a community-developed core Python package for Astronomy \citep{astropy2013, astropy2018}. This work has made use of the Rainbow Cosmological Surveys Database, which is operated by the Centro de Astrobiología (CAB/INTA), partnered with the University of California Observatories at Santa Cruz (UCO/Lick,UCSC). We acknowledge the use of the NumPy Package and would like to thank the NumPy team for their hardwork in creating such an invaluable tool \citep{harris2020numpy}. We acknowledge the usage of Matplotlib for plotting and would like to thank the Matplotlib team for their efforts in creating such a wonderful package \citep{Hunter2007matplotlib}. We utilized the \texttt{galight} package which is based on \texttt{lenstronomy} \citep{lenstronomy_software}\footnote{https://github.com/lenstronomy/lenstronomy} \citep{birrer_lenstronomy_2018, birrer_lenstronomy_ii_2021, galight_ding, galight_ding_old, ding_2023_galight_nature}.

Finally, Will acknowledges the hard work of members of the University of Wisconsin - Madison Astronomy Club and surrounding high schools in classifying the morphologies of AGN host galaxies. Although not all of their classifications were utilized due to changes in sample selection throughout the project, the following students (in alphabetical order) demonstrated exemplary dedication to this project: Hojoon Cha, Miles Gamble, Scott Huddleston, Alyssa Jankowski, Jenna Karcheski, Alexander Kiner, Brooke Kotten, Elliot Koziol, Adam Miller, Alex Tellez.

\end{acknowledgments}
%

\newpage
\appendix
\section{Additional Visual Classification Comparisons}
\label{app: additional visual classifications}

\begin{figure}[t!]
    \centering
    \includegraphics[width=\linewidth]{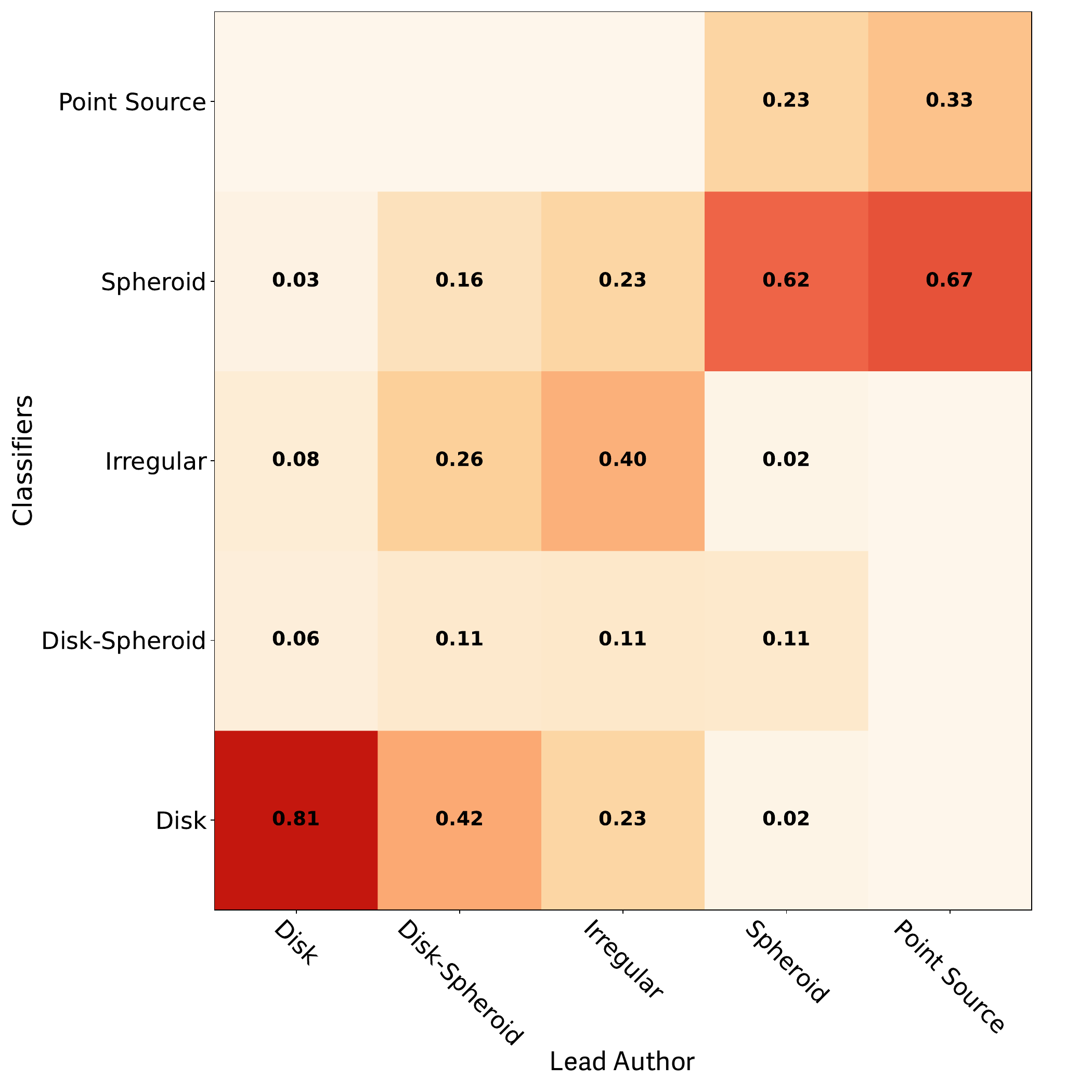}
    \caption{Heat map showing the classification agreement between 7 classifiers and the lead author. The colorbar is the same as Figure \ref{fig: jeana_agreement}. Note that the fractions do not all add to one. If more than half the classifiers failed to reach a consensus on the primary morphology, that source was marked as 'Unclear' and not sorted into any category. In general, we agree whether a source is disk-dominated or spheroidal in nature. We often struggle to agree on the nature of disk-spheroids and irregulars, an unsurprising consequence of these being defined by features that are often difficult to categorize. The broad agreement seen here is encouraging. The classifiers have spent a fraction of the total time studying these sources and have access to less filters (namely, JADES had not been released while they were conducting their classifications) than the author, yet broadly find the same broad results.}
    \label{fig: uw_astro_matches}
\end{figure}

In \S \ref{sec: morph_intro}, we described comparisons between the author's visual classifications and those in the literature from \citet{Kartaltepe2015morph} and \citet{huertascompany}. We also made reference to additional visual classifications which we compared against. In this appendix, we present comparisons with additional classifications carried out by the UW-Madison Astronomy Club and selected additional classifiers. This allowed us to compare classifications between the author and a wider group utilizing the same classification methodology.

We prepared a presentation on galaxy evolution and morphology before giving them a training sample of 35 non-AGN galaxies in the field. These galaxies are identical in redshift range, but have X-ray luminosities less than $10^{42.5}$ $\mathrm{erg\ s^{-1}}$. This ensured students would be trained on galaxies representative of those they would see in the sample without accidentally including any real targets. We also encouraged the students to practice identifying galaxy features using the popular citizen-scientist Galaxy Zoo program \citep{galzoo}. 

The classifiers utilized the Smithsonian Astrophysical Observatory DS9 software to classify the galaxies \citep{sao_ds9}. Our study relies on classifying faint details from mergers or galaxy interactions which does not allow a uniform stretch to be performed on all galaxies. Instead, each student was responsible for determining their own stretch. We instructed them to avoid stretching too far and taught them about common data artifacts. The most common source of confusion was the HST diffraction pattern around point sources, which some classifiers originally believed to be a ring. We corrected the issue by giving the classifiers a basic introduction to the HST point spread function.

Ultimately, seven additional classifiers reviewed the AGN host galaxies. Figure \ref{fig: uw_astro_matches} shows a heat map comparing the author’s classifications with the students'. Note that some of the fractions do not add to 1, as galaxies which did not have a majority agreement were marked as unclear and not included in the table.

In general, the classifiers agree with the author on whether a source is disk or bulge-dominated. We agree $81\%$ of the time on disk galaxies, $40\%$ for irregulars, and $62\%$ of the time on spheroid galaxies. That we disagree most about disk-spheroids is unsurprising; the primary features, as noted in \S \ref{sec: morph_intro}, are often difficult to analyze. 

We conclude that we broadly agree with trained classifiers across the sample, though with some disagreements in the irregular category.

\vspace{5mm}




\newpage

\newpage
\bibliography{reu2021.bib}

@ARTICLE{Kormendy2013BHGal,
       author = {{Kormendy}, John and {Ho}, Luis C.},
        title = "{Coevolution (Or Not) of Supermassive Black Holes and Host Galaxies}",
      journal = {\araa},
         year = 2013,
        month = aug,
       volume = {51},
       number = {1},
        pages = {511-653},
          doi = {10.1146/annurev-astro-082708-101811},
archivePrefix = {arXiv},
       eprint = {1304.7762},
 primaryClass = {astro-ph.CO},
       adsurl = {https://ui.adsabs.harvard.edu/abs/2013ARA&A..51..511K}
}

@ARTICLE{Li2020CT,
       author = {{Li}, Junyao and {Xue}, Yongquan and {Sun}, Mouyuan and {Brandt}, William N. and {Yang}, Guang and {Vito}, Fabio and {Tozzi}, Paolo and {Vignali}, Cristian and {Comastri}, Andrea and {Shu}, Xinwen and {Fang}, Guanwen and {Fan}, Lulu and {Luo}, Bin and {Chen}, Chien-Ting and {Zheng}, Xuechen},
        title = "{Piercing through Highly Obscured and Compton-thick AGNs in the Chandra Deep Fields. II. Are Highly Obscured AGNs the Missing Link in the Merger-triggered AGN-Galaxy Coevolution Models?}",
      journal = {\apj},
         year = 2020,
        month = nov,
       volume = {903},
       number = {1},
          eid = {49},
        pages = {49},
          doi = {10.3847/1538-4357/abb6e7},
archivePrefix = {arXiv},
       eprint = {2008.05863},
 primaryClass = {astro-ph.GA},
       adsurl = {https://ui.adsabs.harvard.edu/abs/2020ApJ...903...49L}
}

@ARTICLE{PersicS2004SFR,
       author = {{Persic}, M. and {Rephaeli}, Y. and {Braito}, V. and {Cappi}, M. and {Della Ceca}, R. and {Franceschini}, A. and {Gruber}, D.~E.},
        title = "{2-10 keV luminosity of high-mass binaries as a gauge of ongoing star-formation rate}",
      journal = {\aap},
         year = 2004,
        month = jun,
       volume = {419},
        pages = {849-862},
          doi = {10.1051/0004-6361:20034500},
archivePrefix = {arXiv},
       eprint = {astro-ph/0402568},
 primaryClass = {astro-ph},
       adsurl = {https://ui.adsabs.harvard.edu/abs/2004A&A...419..849P}
}

@ARTICLE{Suh2019COSMOS,
       author = {{Suh}, Hyewon and {Civano}, Francesca and {Hasinger}, G{\"u}nther and {Lusso}, Elisabeta and {Marchesi}, Stefano and {Schulze}, Andreas and {Onodera}, Masato and {Rosario}, David J. and {Sanders}, David B.},
        title = "{Multi-wavelength Properties of Type 1 and Type 2 AGN Host Galaxies in the Chandra-COSMOS Legacy Survey}",
      journal = {\apj},
         year = 2019,
        month = feb,
       volume = {872},
       number = {2},
          eid = {168},
        pages = {168},
          doi = {10.3847/1538-4357/ab01fb},
archivePrefix = {arXiv},
       eprint = {1902.03244},
 primaryClass = {astro-ph.GA},
       adsurl = {https://ui.adsabs.harvard.edu/abs/2019ApJ...872..168S}
}

@ARTICLE{GUO2013GOODSS,
       author = {{Guo}, Yicheng and {Ferguson}, Henry C. and {Giavalisco}, Mauro and {Barro}, Guillermo and {Willner}, S.~P. and {Ashby}, Matthew L.~N. and {Dahlen}, Tomas and {Donley}, Jennifer L. and {Faber}, Sandra M. and {Fontana}, Adriano and {Galametz}, Audrey and {Grazian}, Andrea and {Huang}, Kuang-Han and {Kocevski}, Dale D. and {Koekemoer}, Anton M. and {Koo}, David C. and {McGrath}, Elizabeth J. and {Peth}, Michael and {Salvato}, Mara and {Wuyts}, Stijn and {Castellano}, Marco and {Cooray}, Asantha R. and {Dickinson}, Mark E. and {Dunlop}, James S. and {Fazio}, G.~G. and {Gardner}, Jonathan P. and {Gawiser}, Eric and {Grogin}, Norman A. and {Hathi}, Nimish P. and {Hsu}, Li-Ting and {Lee}, Kyoung-Soo and {Lucas}, Ray A. and {Mobasher}, Bahram and {Nandra}, Kirpal and {Newman}, Jeffery A. and {van der Wel}, Arjen},
        title = "{CANDELS Multi-wavelength Catalogs: Source Detection and Photometry in the GOODS-South Field}",
      journal = {\apjs},
         year = 2013,
        month = aug,
       volume = {207},
       number = {2},
          eid = {24},
        pages = {24},
          doi = {10.1088/0067-0049/207/2/24},
archivePrefix = {arXiv},
       eprint = {1308.4405},
 primaryClass = {astro-ph.CO},
       adsurl = {https://ui.adsabs.harvard.edu/abs/2013ApJS..207...24G}
}

@ARTICLE{Luo2017CDFS,
       author = {{Luo}, B. and {Brandt}, W.~N. and {Xue}, Y.~Q. and {Lehmer}, B. and {Alexander}, D.~M. and {Bauer}, F.~E. and {Vito}, F. and {Yang}, G. and {Basu-Zych}, A.~R. and {Comastri}, A. and {Gilli}, R. and {Gu}, Q. -S. and {Hornschemeier}, A.~E. and {Koekemoer}, A. and {Liu}, T. and {Mainieri}, V. and {Paolillo}, M. and {Ranalli}, P. and {Rosati}, P. and {Schneider}, D.~P. and {Shemmer}, O. and {Smail}, I. and {Sun}, M. and {Tozzi}, P. and {Vignali}, C. and {Wang}, J. -X.},
        title = "{The Chandra Deep Field-South Survey: 7 Ms Source Catalogs}",
      journal = {\apjs},
         year = 2017,
        month = jan,
       volume = {228},
       number = {1},
          eid = {2},
        pages = {2},
          doi = {10.3847/1538-4365/228/1/2},
archivePrefix = {arXiv},
       eprint = {1611.03501},
 primaryClass = {astro-ph.GA},
       adsurl = {https://ui.adsabs.harvard.edu/abs/2017ApJS..228....2L}
}

@ARTICLE{Xue2016CDFN,
       author = {{Xue}, Y.~Q. and {Luo}, B. and {Brandt}, W.~N. and {Alexander}, D.~M. and {Bauer}, F.~E. and {Lehmer}, B.~D. and {Yang}, G.},
        title = "{The 2 Ms Chandra Deep Field-North Survey and the 250 ks Extended Chandra Deep Field-South Survey: Improved Point-source Catalogs}",
      journal = {\apjs},
         year = 2016,
        month = jun,
       volume = {224},
       number = {2},
          eid = {15},
        pages = {15},
          doi = {10.3847/0067-0049/224/2/15},
archivePrefix = {arXiv},
       eprint = {1602.06299},
 primaryClass = {astro-ph.GA},
       adsurl = {https://ui.adsabs.harvard.edu/abs/2016ApJS..224...15X}
}

@ARTICLE{bandara2009,
       author = {{Bandara}, Kaushala and {Crampton}, David and {Simard}, Luc},
        title = "{A Relationship Between Supermassive Black Hole Mass and the Total Gravitational Mass of the Host Galaxy}",
      journal = {\apj},
         year = 2009,
        month = oct,
       volume = {704},
       number = {2},
        pages = {1135-1145},
          doi = {10.1088/0004-637X/704/2/1135},
archivePrefix = {arXiv},
       eprint = {0909.0269},
 primaryClass = {astro-ph.GA},
       adsurl = {https://ui.adsabs.harvard.edu/abs/2009ApJ...704.1135B}
}

@ARTICLE{Grogin2011GOODSN,
       author = {{Grogin}, Norman A. and {Kocevski}, Dale D. and {Faber}, S.~M. and {Ferguson}, Henry C. and {Koekemoer}, Anton M. and {Riess}, Adam G. and {Acquaviva}, Viviana and {Alexander}, David M. and {Almaini}, Omar and {Ashby}, Matthew L.~N. and {Barden}, Marco and {Bell}, Eric F. and {Bournaud}, Fr{\'e}d{\'e}ric and {Brown}, Thomas M. and {Caputi}, Karina I. and {Casertano}, Stefano and {Cassata}, Paolo and {Castellano}, Marco and {Challis}, Peter and {Chary}, Ranga-Ram and {Cheung}, Edmond and {Cirasuolo}, Michele and {Conselice}, Christopher J. and {Roshan Cooray}, Asantha and {Croton}, Darren J. and {Daddi}, Emanuele and {Dahlen}, Tomas and {Dav{\'e}}, Romeel and {de Mello}, Du{\'\i}lia F. and {Dekel}, Avishai and {Dickinson}, Mark and {Dolch}, Timothy and {Donley}, Jennifer L. and {Dunlop}, James S. and {Dutton}, Aaron A. and {Elbaz}, David and {Fazio}, Giovanni G. and {Filippenko}, Alexei V. and {Finkelstein}, Steven L. and {Fontana}, Adriano and {Gardner}, Jonathan P. and {Garnavich}, Peter M. and {Gawiser}, Eric and {Giavalisco}, Mauro and {Grazian}, Andrea and {Guo}, Yicheng and {Hathi}, Nimish P. and {H{\"a}ussler}, Boris and {Hopkins}, Philip F. and {Huang}, Jia-Sheng and {Huang}, Kuang-Han and {Jha}, Saurabh W. and {Kartaltepe}, Jeyhan S. and {Kirshner}, Robert P. and {Koo}, David C. and {Lai}, Kamson and {Lee}, Kyoung-Soo and {Li}, Weidong and {Lotz}, Jennifer M. and {Lucas}, Ray A. and {Madau}, Piero and {McCarthy}, Patrick J. and {McGrath}, Elizabeth J. and {McIntosh}, Daniel H. and {McLure}, Ross J. and {Mobasher}, Bahram and {Moustakas}, Leonidas A. and {Mozena}, Mark and {Nandra}, Kirpal and {Newman}, Jeffrey A. and {Niemi}, Sami-Matias and {Noeske}, Kai G. and {Papovich}, Casey J. and {Pentericci}, Laura and {Pope}, Alexandra and {Primack}, Joel R. and {Rajan}, Abhijith and {Ravindranath}, Swara and {Reddy}, Naveen A. and {Renzini}, Alvio and {Rix}, Hans-Walter and {Robaina}, Aday R. and {Rodney}, Steven A. and {Rosario}, David J. and {Rosati}, Piero and {Salimbeni}, Sara and {Scarlata}, Claudia and {Siana}, Brian and {Simard}, Luc and {Smidt}, Joseph and {Somerville}, Rachel S. and {Spinrad}, Hyron and {Straughn}, Amber N. and {Strolger}, Louis-Gregory and {Telford}, Olivia and {Teplitz}, Harry I. and {Trump}, Jonathan R. and {van der Wel}, Arjen and {Villforth}, Carolin and {Wechsler}, Risa H. and {Weiner}, Benjamin J. and {Wiklind}, Tommy and {Wild}, Vivienne and {Wilson}, Grant and {Wuyts}, Stijn and {Yan}, Hao-Jing and {Yun}, Min S.},
        title = "{CANDELS: The Cosmic Assembly Near-infrared Deep Extragalactic Legacy Survey}",
      journal = {\apjs},
         year = 2011,
        month = dec,
       volume = {197},
       number = {2},
          eid = {35},
        pages = {35},
          doi = {10.1088/0067-0049/197/2/35},
archivePrefix = {arXiv},
       eprint = {1105.3753},
 primaryClass = {astro-ph.CO},
       adsurl = {https://ui.adsabs.harvard.edu/abs/2011ApJS..197...35G}
}

@ARTICLE{Koekemoer2011CANDELS,
       author = {{Koekemoer}, Anton M. and {Faber}, S.~M. and {Ferguson}, Henry C. and {Grogin}, Norman A. and {Kocevski}, Dale D. and {Koo}, David C. and {Lai}, Kamson and {Lotz}, Jennifer M. and {Lucas}, Ray A. and {McGrath}, Elizabeth J. and {Ogaz}, Sara and {Rajan}, Abhijith and {Riess}, Adam G. and {Rodney}, Steve A. and {Strolger}, Louis and {Casertano}, Stefano and {Castellano}, Marco and {Dahlen}, Tomas and {Dickinson}, Mark and {Dolch}, Timothy and {Fontana}, Adriano and {Giavalisco}, Mauro and {Grazian}, Andrea and {Guo}, Yicheng and {Hathi}, Nimish P. and {Huang}, Kuang-Han and {van der Wel}, Arjen and {Yan}, Hao-Jing and {Acquaviva}, Viviana and {Alexander}, David M. and {Almaini}, Omar and {Ashby}, Matthew L.~N. and {Barden}, Marco and {Bell}, Eric F. and {Bournaud}, Fr{\'e}d{\'e}ric and {Brown}, Thomas M. and {Caputi}, Karina I. and {Cassata}, Paolo and {Challis}, Peter J. and {Chary}, Ranga-Ram and {Cheung}, Edmond and {Cirasuolo}, Michele and {Conselice}, Christopher J. and {Roshan Cooray}, Asantha and {Croton}, Darren J. and {Daddi}, Emanuele and {Dav{\'e}}, Romeel and {de Mello}, Duilia F. and {de Ravel}, Loic and {Dekel}, Avishai and {Donley}, Jennifer L. and {Dunlop}, James S. and {Dutton}, Aaron A. and {Elbaz}, David and {Fazio}, Giovanni G. and {Filippenko}, Alexei V. and {Finkelstein}, Steven L. and {Frazer}, Chris and {Gardner}, Jonathan P. and {Garnavich}, Peter M. and {Gawiser}, Eric and {Gruetzbauch}, Ruth and {Hartley}, Will G. and {H{\"a}ussler}, Boris and {Herrington}, Jessica and {Hopkins}, Philip F. and {Huang}, Jia-Sheng and {Jha}, Saurabh W. and {Johnson}, Andrew and {Kartaltepe}, Jeyhan S. and {Khostovan}, Ali A. and {Kirshner}, Robert P. and {Lani}, Caterina and {Lee}, Kyoung-Soo and {Li}, Weidong and {Madau}, Piero and {McCarthy}, Patrick J. and {McIntosh}, Daniel H. and {McLure}, Ross J. and {McPartland}, Conor and {Mobasher}, Bahram and {Moreira}, Heidi and {Mortlock}, Alice and {Moustakas}, Leonidas A. and {Mozena}, Mark and {Nandra}, Kirpal and {Newman}, Jeffrey A. and {Nielsen}, Jennifer L. and {Niemi}, Sami and {Noeske}, Kai G. and {Papovich}, Casey J. and {Pentericci}, Laura and {Pope}, Alexandra and {Primack}, Joel R. and {Ravindranath}, Swara and {Reddy}, Naveen A. and {Renzini}, Alvio and {Rix}, Hans-Walter and {Robaina}, Aday R. and {Rosario}, David J. and {Rosati}, Piero and {Salimbeni}, Sara and {Scarlata}, Claudia and {Siana}, Brian and {Simard}, Luc and {Smidt}, Joseph and {Snyder}, Diana and {Somerville}, Rachel S. and {Spinrad}, Hyron and {Straughn}, Amber N. and {Telford}, Olivia and {Teplitz}, Harry I. and {Trump}, Jonathan R. and {Vargas}, Carlos and {Villforth}, Carolin and {Wagner}, Cory R. and {Wandro}, Pat and {Wechsler}, Risa H. and {Weiner}, Benjamin J. and {Wiklind}, Tommy and {Wild}, Vivienne and {Wilson}, Grant and {Wuyts}, Stijn and {Yun}, Min S.},
        title = "{CANDELS: The Cosmic Assembly Near-infrared Deep Extragalactic Legacy Survey{\textemdash}The Hubble Space Telescope Observations, Imaging Data Products, and Mosaics}",
      journal = {\apjs},
         year = 2011,
        month = dec,
       volume = {197},
       number = {2},
          eid = {36},
        pages = {36},
          doi = {10.1088/0067-0049/197/2/36},
archivePrefix = {arXiv},
       eprint = {1105.3754},
 primaryClass = {astro-ph.CO},
       adsurl = {https://ui.adsabs.harvard.edu/abs/2011ApJS..197...36K}
}

@ARTICLE{Giavalisco2004GOODS,
       author = {{Giavalisco}, M. and {Ferguson}, H.~C. and {Koekemoer}, A.~M. and {Dickinson}, M. and {Alexander}, D.~M. and {Bauer}, F.~E. and {Bergeron}, J. and {Biagetti}, C. and {Brandt}, W.~N. and {Casertano}, S. and {Cesarsky}, C. and {Chatzichristou}, E. and {Conselice}, C. and {Cristiani}, S. and {Da Costa}, L. and {Dahlen}, T. and {de Mello}, D. and {Eisenhardt}, P. and {Erben}, T. and {Fall}, S.~M. and {Fassnacht}, C. and {Fosbury}, R. and {Fruchter}, A. and {Gardner}, J.~P. and {Grogin}, N. and {Hook}, R.~N. and {Hornschemeier}, A.~E. and {Idzi}, R. and {Jogee}, S. and {Kretchmer}, C. and {Laidler}, V. and {Lee}, K.~S. and {Livio}, M. and {Lucas}, R. and {Madau}, P. and {Mobasher}, B. and {Moustakas}, L.~A. and {Nonino}, M. and {Padovani}, P. and {Papovich}, C. and {Park}, Y. and {Ravindranath}, S. and {Renzini}, A. and {Richardson}, M. and {Riess}, A. and {Rosati}, P. and {Schirmer}, M. and {Schreier}, E. and {Somerville}, R.~S. and {Spinrad}, H. and {Stern}, D. and {Stiavelli}, M. and {Strolger}, L. and {Urry}, C.~M. and {Vandame}, B. and {Williams}, R. and {Wolf}, C.},
        title = "{The Great Observatories Origins Deep Survey: Initial Results from Optical and Near-Infrared Imaging}",
      journal = {\apjl},
         year = 2004,
        month = jan,
       volume = {600},
       number = {2},
        pages = {L93-L98},
          doi = {10.1086/379232},
archivePrefix = {arXiv},
       eprint = {astro-ph/0309105},
 primaryClass = {astro-ph},
       adsurl = {https://ui.adsabs.harvard.edu/abs/2004ApJ...600L..93G}
}

@ARTICLE{Kartaltepe2015morph,
       author = {{Kartaltepe}, Jeyhan S. and {Mozena}, Mark and {Kocevski}, Dale and {McIntosh}, Daniel H. and {Lotz}, Jennifer and {Bell}, Eric F. and {Faber}, Sandy and {Ferguson}, Harry and {Koo}, David and {Bassett}, Robert and {Bernyk}, Maksym and {Blancato}, Kirsten and {Bournaud}, Frederic and {Cassata}, Paolo and {Castellano}, Marco and {Cheung}, Edmond and {Conselice}, Christopher J. and {Croton}, Darren and {Dahlen}, Tomas and {de Mello}, Duilia F. and {DeGroot}, Laura and {Donley}, Jennifer and {Guedes}, Javiera and {Grogin}, Norman and {Hathi}, Nimish and {Hilton}, Matt and {Hollon}, Brett and {Koekemoer}, Anton and {Liu}, Nick and {Lucas}, Ray A. and {Martig}, Marie and {McGrath}, Elizabeth and {McPartland}, Conor and {Mobasher}, Bahram and {Morlock}, Alice and {O'Leary}, Erin and {Peth}, Mike and {Pforr}, Janine and {Pillepich}, Annalisa and {Rosario}, David and {Soto}, Emmaris and {Straughn}, Amber and {Telford}, Olivia and {Sunnquist}, Ben and {Trump}, Jonathan and {Weiner}, Benjamin and {Wuyts}, Stijn and {Inami}, Hanae and {Kassin}, Susan and {Lani}, Caterina and {Poole}, Gregory B. and {Rizer}, Zachary},
        title = "{CANDELS Visual Classifications: Scheme, Data Release, and First Results}",
      journal = {\apjs},
         year = 2015,
        month = nov,
       volume = {221},
       number = {1},
          eid = {11},
        pages = {11},
          doi = {10.1088/0067-0049/221/1/11},
archivePrefix = {arXiv},
       eprint = {1401.2455},
 primaryClass = {astro-ph.GA},
       adsurl = {https://ui.adsabs.harvard.edu/abs/2015ApJS..221...11K}
}

@ARTICLE{shakura1973,
       author = {{Shakura}, N.~I. and {Sunyaev}, R.~A.},
        title = "{Reprint of 1973A\&A....24..337S. Black holes in binary systems. Observational appearance.}",
      journal = {\aap},
         year = 1973,
        month = jun,
       volume = {500},
        pages = {33-51},
       adsurl = {https://ui.adsabs.harvard.edu/abs/1973A&A....24..337S}
}

@ARTICLE{Koratkar1999,
       author = {{Koratkar}, Anuradha and {Blaes}, Omer},
        title = "{The Ultraviolet and Optical Continuum Emission in Active Galactic Nuclei: The Status of Accretion Disks}",
      journal = {\pasp},
         year = 1999,
        month = jan,
       volume = {111},
       number = {755},
        pages = {1-30},
          doi = {10.1086/316294},
       adsurl = {https://ui.adsabs.harvard.edu/abs/1999PASP..111....1K}
}

@ARTICLE{Hickox2017,
       author = {{Hickox}, Ryan C. and {Myers}, Adam D. and {Greene}, Jenny E. and {Hainline}, Kevin N. and {Zakamska}, Nadia L. and {DiPompeo}, Michael A.},
        title = "{Composite Spectral Energy Distributions and Infrared-Optical Colors of Type 1 and Type 2 Quasars}",
      journal = {\apj},
         year = 2017,
        month = nov,
       volume = {849},
       number = {1},
          eid = {53},
        pages = {53},
          doi = {10.3847/1538-4357/aa8c77},
archivePrefix = {arXiv},
       eprint = {1709.04468},
 primaryClass = {astro-ph.GA},
       adsurl = {https://ui.adsabs.harvard.edu/abs/2017ApJ...849...53H}
}

@INPROCEEDINGS{sanders_agn_ulirg_connection,
       author = {{Sanders}, D.~B. and {Kartaltepe}, J.~S. and {Kewley}, L.~J. and {U}, Vivian and {Yuan}, T. and {Evans}, A.~S. and {Armus}, L. and {Mazzarella}, J.~M.},
        title = "{Luminous Infrared Galaxies and the ``Starburst-AGN Connection''}",
    booktitle = {The Starburst-AGN Connection},
         year = 2009,
       editor = {{Wang}, W. and {Yang}, Z. and {Luo}, Z. and {Chen}, Z.},
       series = {Astronomical Society of the Pacific Conference Series},
       volume = {408},
        month = oct,
        pages = {3},
       adsurl = {https://ui.adsabs.harvard.edu/abs/2009ASPC..408....3S}
}

@ARTICLE{Sanders1989,
       author = {{Sanders}, D.~B. and {Phinney}, E.~S. and {Neugebauer}, G. and {Soifer}, B.~T. and {Matthews}, K.},
        title = "{Continuum Energy Distributions of Quasars: Shapes and Origins}",
      journal = {\apj},
         year = 1989,
        month = dec,
       volume = {347},
        pages = {29},
          doi = {10.1086/168094},
       adsurl = {https://ui.adsabs.harvard.edu/abs/1989ApJ...347...29S}
}

@ARTICLE{Elvis1995,
       author = {{Elvis}, Martin and {Wilkes}, Belinda J. and {McDowell}, Jonathan C. and {Green}, Richard F. and {Bechtold}, Jill and {Willner}, S.~P. and {Oey}, M.~S. and {Polomski}, Elisha and {Cutri}, Roc},
        title = "{Atlas of Quasar Energy Distributions}",
      journal = {\apjs},
         year = 1994,
        month = nov,
       volume = {95},
        pages = {1},
          doi = {10.1086/192093},
       adsurl = {https://ui.adsabs.harvard.edu/abs/1994ApJS...95....1E}
}

@ARTICLE{Richards2006,
       author = {{Richards}, Gordon T. and {Lacy}, Mark and {Storrie-Lombardi}, Lisa J. and {Hall}, Patrick B. and {Gallagher}, S.~C. and {Hines}, Dean C. and {Fan}, Xiaohui and {Papovich}, Casey and {Vanden Berk}, Daniel E. and {Trammell}, George B. and {Schneider}, Donald P. and {Vestergaard}, Marianne and {York}, Donald G. and {Jester}, Sebastian and {Anderson}, Scott F. and {Budav{\'a}ri}, Tam{\'a}s and {Szalay}, Alexander S.},
        title = "{Spectral Energy Distributions and Multiwavelength Selection of Type 1 Quasars}",
      journal = {\apjs},
         year = 2006,
        month = oct,
       volume = {166},
       number = {2},
        pages = {470-497},
          doi = {10.1086/506525},
archivePrefix = {arXiv},
       eprint = {astro-ph/0601558},
 primaryClass = {astro-ph},
       adsurl = {https://ui.adsabs.harvard.edu/abs/2006ApJS..166..470R}
}

@ARTICLE{Scoville2007cosmos,
       author = {{Scoville}, N. and {Aussel}, H. and {Brusa}, M. and {Capak}, P. and {Carollo}, C.~M. and {Elvis}, M. and {Giavalisco}, M. and {Guzzo}, L. and {Hasinger}, G. and {Impey}, C. and {Kneib}, J. -P. and {LeFevre}, O. and {Lilly}, S.~J. and {Mobasher}, B. and {Renzini}, A. and {Rich}, R.~M. and {Sanders}, D.~B. and {Schinnerer}, E. and {Schminovich}, D. and {Shopbell}, P. and {Taniguchi}, Y. and {Tyson}, N.~D.},
        title = "{The Cosmic Evolution Survey (COSMOS): Overview}",
      journal = {\apjs},
         year = 2007,
        month = sep,
       volume = {172},
       number = {1},
        pages = {1-8},
          doi = {10.1086/516585},
archivePrefix = {arXiv},
       eprint = {astro-ph/0612305},
 primaryClass = {astro-ph},
       adsurl = {https://ui.adsabs.harvard.edu/abs/2007ApJS..172....1S}
}

@ARTICLE{LaMassa2013stripe,
       author = {{LaMassa}, Stephanie M. and {Urry}, C. Megan and {Cappelluti}, Nico and {Civano}, Francesca and {Ranalli}, Piero and {Glikman}, Eilat and {Treister}, Ezequiel and {Richards}, Gordon and {Ballantyne}, David and {Stern}, Daniel and {Comastri}, Andrea and {Cardamone}, Carie and {Schawinski}, Kevin and {B{\"o}hringer}, Hans and {Chon}, Gayoung and {Murray}, Stephen S. and {Green}, Paul and {Nandra}, Kirpal},
        title = "{Finding rare AGN: XMM-Newton and Chandra observations of SDSS Stripe 82}",
      journal = {\mnras},
         year = 2013,
        month = dec,
       volume = {436},
       number = {4},
        pages = {3581-3601},
          doi = {10.1093/mnras/stt1837},
archivePrefix = {arXiv},
       eprint = {1309.7048},
 primaryClass = {astro-ph.CO},
       adsurl = {https://ui.adsabs.harvard.edu/abs/2013MNRAS.436.3581L}
}

@article{astropy2013,
Adsurl = {http://adsabs.harvard.edu/abs/2013A%26A...558A..33A},
Archiveprefix = {arXiv},
Author = {{Astropy Collaboration} and {Robitaille}, T.~P. and {Tollerud}, E.~J. and {Greenfield}, P. and {Droettboom}, M. and {Bray}, E. and {Aldcroft}, T. and {Davis}, M. and {Ginsburg}, A. and {Price-Whelan}, A.~M. and {Kerzendorf}, W.~E. and {Conley}, A. and {Crighton}, N. and {Barbary}, K. and {Muna}, D. and {Ferguson}, H. and {Grollier}, F. and {Parikh}, M.~M. and {Nair}, P.~H. and {Unther}, H.~M. and {Deil}, C. and {Woillez}, J. and {Conseil}, S. and {Kramer}, R. and {Turner}, J.~E.~H. and {Singer}, L. and {Fox}, R. and {Weaver}, B.~A. and {Zabalza}, V. and {Edwards}, Z.~I. and {Azalee Bostroem}, K. and {Burke}, D.~J. and {Casey}, A.~R. and {Crawford}, S.~M. and {Dencheva}, N. and {Ely}, J. and {Jenness}, T. and {Labrie}, K. and {Lim}, P.~L. and {Pierfederici}, F. and {Pontzen}, A. and {Ptak}, A. and {Refsdal}, B. and {Servillat}, M. and {Streicher}, O.},
Doi = {10.1051/0004-6361/201322068},
Eid = {A33},
Eprint = {1307.6212},
Journal = {\aap},
Month = oct,
Pages = {A33},
Primaryclass = {astro-ph.IM},
Title = {{Astropy: A community Python package for astronomy}},
Volume = 558,
Year = 2013}

@ARTICLE{astropy2018,
       author = {{Astropy Collaboration} and {Price-Whelan}, A.~M. and
         {Sip{\H{o}}cz}, B.~M. and {G{\"u}nther}, H.~M. and {Lim}, P.~L. and
         {Crawford}, S.~M. and {Conseil}, S. and {Shupe}, D.~L. and
         {Craig}, M.~W. and {Dencheva}, N. and {Ginsburg}, A. and {Vand
        erPlas}, J.~T. and {Bradley}, L.~D. and {P{\'e}rez-Su{\'a}rez}, D. and
         {de Val-Borro}, M. and {Aldcroft}, T.~L. and {Cruz}, K.~L. and
         {Robitaille}, T.~P. and {Tollerud}, E.~J. and {Ardelean}, C. and
         {Babej}, T. and {Bach}, Y.~P. and {Bachetti}, M. and {Bakanov}, A.~V. and
         {Bamford}, S.~P. and {Barentsen}, G. and {Barmby}, P. and
         {Baumbach}, A. and {Berry}, K.~L. and {Biscani}, F. and {Boquien}, M. and
         {Bostroem}, K.~A. and {Bouma}, L.~G. and {Brammer}, G.~B. and
         {Bray}, E.~M. and {Breytenbach}, H. and {Buddelmeijer}, H. and
         {Burke}, D.~J. and {Calderone}, G. and {Cano Rodr{\'\i}guez}, J.~L. and
         {Cara}, M. and {Cardoso}, J.~V.~M. and {Cheedella}, S. and {Copin}, Y. and
         {Corrales}, L. and {Crichton}, D. and {D'Avella}, D. and {Deil}, C. and
         {Depagne}, {\'E}. and {Dietrich}, J.~P. and {Donath}, A. and
         {Droettboom}, M. and {Earl}, N. and {Erben}, T. and {Fabbro}, S. and
         {Ferreira}, L.~A. and {Finethy}, T. and {Fox}, R.~T. and
         {Garrison}, L.~H. and {Gibbons}, S.~L.~J. and {Goldstein}, D.~A. and
         {Gommers}, R. and {Greco}, J.~P. and {Greenfield}, P. and
         {Groener}, A.~M. and {Grollier}, F. and {Hagen}, A. and {Hirst}, P. and
         {Homeier}, D. and {Horton}, A.~J. and {Hosseinzadeh}, G. and {Hu}, L. and
         {Hunkeler}, J.~S. and {Ivezi{\'c}}, {\v{Z}}. and {Jain}, A. and
         {Jenness}, T. and {Kanarek}, G. and {Kendrew}, S. and {Kern}, N.~S. and
         {Kerzendorf}, W.~E. and {Khvalko}, A. and {King}, J. and {Kirkby}, D. and
         {Kulkarni}, A.~M. and {Kumar}, A. and {Lee}, A. and {Lenz}, D. and
         {Littlefair}, S.~P. and {Ma}, Z. and {Macleod}, D.~M. and
         {Mastropietro}, M. and {McCully}, C. and {Montagnac}, S. and
         {Morris}, B.~M. and {Mueller}, M. and {Mumford}, S.~J. and {Muna}, D. and
         {Murphy}, N.~A. and {Nelson}, S. and {Nguyen}, G.~H. and
         {Ninan}, J.~P. and {N{\"o}the}, M. and {Ogaz}, S. and {Oh}, S. and
         {Parejko}, J.~K. and {Parley}, N. and {Pascual}, S. and {Patil}, R. and
         {Patil}, A.~A. and {Plunkett}, A.~L. and {Prochaska}, J.~X. and
         {Rastogi}, T. and {Reddy Janga}, V. and {Sabater}, J. and
         {Sakurikar}, P. and {Seifert}, M. and {Sherbert}, L.~E. and
         {Sherwood-Taylor}, H. and {Shih}, A.~Y. and {Sick}, J. and
         {Silbiger}, M.~T. and {Singanamalla}, S. and {Singer}, L.~P. and
         {Sladen}, P.~H. and {Sooley}, K.~A. and {Sornarajah}, S. and
         {Streicher}, O. and {Teuben}, P. and {Thomas}, S.~W. and
         {Tremblay}, G.~R. and {Turner}, J.~E.~H. and {Terr{\'o}n}, V. and
         {van Kerkwijk}, M.~H. and {de la Vega}, A. and {Watkins}, L.~L. and
         {Weaver}, B.~A. and {Whitmore}, J.~B. and {Woillez}, J. and
         {Zabalza}, V. and {Astropy Contributors}},
        title = "{The Astropy Project: Building an Open-science Project and Status of the v2.0 Core Package}",
      journal = {\aj},
         year = 2018,
        month = sep,
       volume = {156},
       number = {3},
          eid = {123},
        pages = {123},
          doi = {10.3847/1538-3881/aabc4f},
archivePrefix = {arXiv},
       eprint = {1801.02634},
 primaryClass = {astro-ph.IM},
       adsurl = {https://ui.adsabs.harvard.edu/abs/2018AJ....156..123A}
}

@Article{         harris2020numpy,
 title         = {Array programming with {NumPy}},
 author        = {Charles R. Harris and K. Jarrod Millman and St{\'{e}}fan J.
                 van der Walt and Ralf Gommers and Pauli Virtanen and David
                 Cournapeau and Eric Wieser and Julian Taylor and Sebastian
                 Berg and Nathaniel J. Smith and Robert Kern and Matti Picus
                 and Stephan Hoyer and Marten H. van Kerkwijk and Matthew
                 Brett and Allan Haldane and Jaime Fern{\'{a}}ndez del
                 R{\'{i}}o and Mark Wiebe and Pearu Peterson and Pierre
                 G{\'{e}}rard-Marchant and Kevin Sheppard and Tyler Reddy and
                 Warren Weckesser and Hameer Abbasi and Christoph Gohlke and
                 Travis E. Oliphant},
 year          = {2020},
 month         = sep,
 journal       = {Nature},
 volume        = {585},
 number        = {7825},
 pages         = {357--362},
 doi           = {10.1038/s41586-020-2649-2},
 publisher     = {Springer Science and Business Media {LLC}},
 url           = {https://doi.org/10.1038/s41586-020-2649-2}
}

@Article{Hunter2007matplotlib,
  Author    = {Hunter, J. D.},
  Title     = {Matplotlib: A 2D graphics environment},
  Journal   = {Computing in Science \& Engineering},
  Volume    = {9},
  Number    = {3},
  Pages     = {90--95},
  publisher = {IEEE COMPUTER SOC},
  doi       = {10.1109/MCSE.2007.55},
  year      = 2007
}

@ARTICLE{huertascompany,
       author = {{Huertas-Company}, M. and {Gravet}, R. and {Cabrera-Vives}, G. and {P{\'e}rez-Gonz{\'a}lez}, P.~G. and {Kartaltepe}, J.~S. and {Barro}, G. and {Bernardi}, M. and {Mei}, S. and {Shankar}, F. and {Dimauro}, P. and {Bell}, E.~F. and {Kocevski}, D. and {Koo}, D.~C. and {Faber}, S.~M. and {Mcintosh}, D.~H.},
        title = "{A Catalog of Visual-like Morphologies in the 5 CANDELS Fields Using Deep Learning}",
      journal = {\apjs},
         year = 2015,
        month = nov,
       volume = {221},
       number = {1},
          eid = {8},
        pages = {8},
          doi = {10.1088/0067-0049/221/1/8},
archivePrefix = {arXiv},
       eprint = {1509.05429},
 primaryClass = {astro-ph.GA},
       adsurl = {https://ui.adsabs.harvard.edu/abs/2015ApJS..221....8H}
}

@ARTICLE{Lamassa2016s82x,
       author = {{LaMassa}, Stephanie M. and {Urry}, C. Megan and {Cappelluti}, Nico and {B{\"o}hringer}, Hans and {Comastri}, Andrea and {Glikman}, Eilat and {Richards}, Gordon and {Ananna}, Tonima and {Brusa}, Marcella and {Cardamone}, Carie and {Chon}, Gayoung and {Civano}, Francesca and {Farrah}, Duncan and {Gilfanov}, Marat and {Green}, Paul and {Komossa}, S. and {Lira}, Paulina and {Makler}, Martin and {Marchesi}, Stefano and {Pecoraro}, Robert and {Ranalli}, Piero and {Salvato}, Mara and {Schawinski}, Kevin and {Stern}, Daniel and {Treister}, Ezequiel and {Viero}, Marco},
        title = "{The 31 Deg$^{2}$ Release of the Stripe 82 X-Ray Survey: The Point Source Catalog}",
      journal = {\apj},
         year = 2016,
        month = feb,
       volume = {817},
       number = {2},
          eid = {172},
        pages = {172},
          doi = {10.3847/0004-637X/817/2/172},
archivePrefix = {arXiv},
       eprint = {1510.00852},
 primaryClass = {astro-ph.GA},
       adsurl = {https://ui.adsabs.harvard.edu/abs/2016ApJ...817..172L}
}

@ARTICLE{urrypadovani,
       author = {{Urry}, C. Megan and {Padovani}, Paolo},
        title = "{Unified Schemes for Radio-Loud Active Galactic Nuclei}",
      journal = {\pasp},
         year = 1995,
        month = sep,
       volume = {107},
        pages = {803},
          doi = {10.1086/133630},
archivePrefix = {arXiv},
       eprint = {astro-ph/9506063},
 primaryClass = {astro-ph},
       adsurl = {https://ui.adsabs.harvard.edu/abs/1995PASP..107..803U}
}

@ARTICLE{galzoo,
       author = {{Lintott}, Chris J. and {Schawinski}, Kevin and {Slosar}, An{\v{z}}e and {Land}, Kate and {Bamford}, Steven and {Thomas}, Daniel and {Raddick}, M. Jordan and {Nichol}, Robert C. and {Szalay}, Alex and {Andreescu}, Dan and {Murray}, Phil and {Vandenberg}, Jan},
        title = "{Galaxy Zoo: morphologies derived from visual inspection of galaxies from the Sloan Digital Sky Survey}",
      journal = {\mnras},
         year = 2008,
        month = sep,
       volume = {389},
       number = {3},
        pages = {1179-1189},
          doi = {10.1111/j.1365-2966.2008.13689.x},
archivePrefix = {arXiv},
       eprint = {0804.4483},
 primaryClass = {astro-ph},
       adsurl = {https://ui.adsabs.harvard.edu/abs/2008MNRAS.389.1179L}
}

@ARTICLE{Auge2023,
       author = {{Auge}, Connor and {Sanders}, David and {Treister}, Ezequiel and {Urry}, C. Megan and {Kirkpatrick}, Allison and {Cappelluti}, Nico and {Ananna}, Tonima Tasnim and {Boquien}, M{\'e}d{\'e}ric and {Balokovi{\'c}}, Mislav and {Civano}, Francesca and {Coleman}, Brandon and {Ghosh}, Aritra and {Kartaltepe}, Jeyhan and {Koss}, Michael and {LaMassa}, Stephanie and {Marchesi}, Stefano and {Peca}, Alessandro and {Powell}, Meredith and {Trakhtenbrot}, Benny and {Turner}, Tracey Jane},
        title = "{The Accretion History of AGN: The Spectral Energy Distributions of X-ray Luminous AGN}",
      journal = {arXiv e-prints},
         year = 2023,
        month = aug,
          eid = {arXiv:2308.10710},
        pages = {arXiv:2308.10710},
          doi = {10.48550/arXiv.2308.10710},
archivePrefix = {arXiv},
       eprint = {2308.10710},
 primaryClass = {astro-ph.GA},
       adsurl = {https://ui.adsabs.harvard.edu/abs/2023arXiv230810710A}
}

@ARTICLE{vanderwel_2012_candels,
       author = {{van der Wel}, A. and {Bell}, E.~F. and {H{\"a}ussler}, B. and {McGrath}, E.~J. and {Chang}, Yu-Yen and {Guo}, Yicheng and {McIntosh}, D.~H. and {Rix}, H. -W. and {Barden}, M. and {Cheung}, E. and {Faber}, S.~M. and {Ferguson}, H.~C. and {Galametz}, A. and {Grogin}, N.~A. and {Hartley}, W. and {Kartaltepe}, J.~S. and {Kocevski}, D.~D. and {Koekemoer}, A.~M. and {Lotz}, J. and {Mozena}, M. and {Peth}, M.~A. and {Peng}, Chien Y.},
        title = "{Structural Parameters of Galaxies in CANDELS}",
      journal = {\apjs},
         year = 2012,
        month = dec,
       volume = {203},
       number = {2},
          eid = {24},
        pages = {24},
          doi = {10.1088/0067-0049/203/2/24},
archivePrefix = {arXiv},
       eprint = {1211.6954},
 primaryClass = {astro-ph.CO},
       adsurl = {https://ui.adsabs.harvard.edu/abs/2012ApJS..203...24V}
}

@ARTICLE{simmons_urry_serc,
       author = {{Simmons}, B.~D. and {Urry}, C.~M.},
        title = "{The Accuracy of Morphological Decomposition of Active Galactic Nucleus Host Galaxies}",
      journal = {\apj},
         year = 2008,
        month = aug,
       volume = {683},
       number = {2},
        pages = {644-658},
          doi = {10.1086/589827},
archivePrefix = {arXiv},
       eprint = {0804.1363},
 primaryClass = {astro-ph},
       adsurl = {https://ui.adsabs.harvard.edu/abs/2008ApJ...683..644S}
}

@ARTICLE{candels_visual,
       author = {{Kartaltepe}, Jeyhan S. and {Mozena}, Mark and {Kocevski}, Dale and {McIntosh}, Daniel H. and {Lotz}, Jennifer and {Bell}, Eric F. and {Faber}, Sandy and {Ferguson}, Harry and {Koo}, David and {Bassett}, Robert and {Bernyk}, Maksym and {Blancato}, Kirsten and {Bournaud}, Frederic and {Cassata}, Paolo and {Castellano}, Marco and {Cheung}, Edmond and {Conselice}, Christopher J. and {Croton}, Darren and {Dahlen}, Tomas and {de Mello}, Duilia F. and {DeGroot}, Laura and {Donley}, Jennifer and {Guedes}, Javiera and {Grogin}, Norman and {Hathi}, Nimish and {Hilton}, Matt and {Hollon}, Brett and {Koekemoer}, Anton and {Liu}, Nick and {Lucas}, Ray A. and {Martig}, Marie and {McGrath}, Elizabeth and {McPartland}, Conor and {Mobasher}, Bahram and {Morlock}, Alice and {O'Leary}, Erin and {Peth}, Mike and {Pforr}, Janine and {Pillepich}, Annalisa and {Rosario}, David and {Soto}, Emmaris and {Straughn}, Amber and {Telford}, Olivia and {Sunnquist}, Ben and {Trump}, Jonathan and {Weiner}, Benjamin and {Wuyts}, Stijn and {Inami}, Hanae and {Kassin}, Susan and {Lani}, Caterina and {Poole}, Gregory B. and {Rizer}, Zachary},
        title = "{CANDELS Visual Classifications: Scheme, Data Release, and First Results}",
      journal = {\apjs},
         year = 2015,
        month = nov,
       volume = {221},
       number = {1},
          eid = {11},
        pages = {11},
          doi = {10.1088/0067-0049/221/1/11},
archivePrefix = {arXiv},
       eprint = {1401.2455},
 primaryClass = {astro-ph.GA},
       adsurl = {https://ui.adsabs.harvard.edu/abs/2015ApJS..221...11K}
}

@ARTICLE{rieke_jades_goods,
       author = {{Rieke}, Marcia J. and {Robertson}, Brant and {Tacchella}, Sandro and {Hainline}, Kevin and {Johnson}, Benjamin D. and {Hausen}, Ryan and {Ji}, Zhiyuan and {Willmer}, Christopher N.~A. and {Eisenstein}, Daniel J. and {Pusk{\'a}s}, D{\'a}vid and {Alberts}, Stacey and {Arribas}, Santiago and {Baker}, William M. and {Baum}, Stefi and {Bhatawdekar}, Rachana and {Bonaventura}, Nina and {Boyett}, Kristan and {Bunker}, Andrew J. and {Cameron}, Alex J. and {Carniani}, Stefano and {Charlot}, Stephane and {Chevallard}, Jacopo and {Chen}, Zuyi and {Curti}, Mirko and {Curtis-Lake}, Emma and {Danhaive}, A. Lola and {DeCoursey}, Christa and {Dressler}, Alan and {Egami}, Eiichi and {Endsley}, Ryan and {Helton}, Jakob M. and {Hviding}, Raphael E. and {Kumari}, Nimisha and {Looser}, Tobias J. and {Lyu}, Jianwei and {Maiolino}, Roberto and {Maseda}, Michael V. and {Nelson}, Erica J. and {Rieke}, George and {Rix}, Hans-Walter and {Sandles}, Lester and {Saxena}, Aayush and {Sharpe}, Katherine and {Shivaei}, Irene and {Skarbinski}, Maya and {Smit}, Renske and {Stark}, Daniel P. and {Stone}, Meredith and {Suess}, Katherine A. and {Sun}, Fengwu and {Topping}, Michael and {{\"U}bler}, Hannah and {Villanueva}, Natalia C. and {Wallace}, Imaan E.~B. and {Williams}, Christina C. and {Willott}, Chris and {Whitler}, Lily and {Witstok}, Joris and {Woodrum}, Charity},
        title = "{JADES Initial Data Release for the Hubble Ultra Deep Field: Revealing the Faint Infrared Sky with Deep JWST NIRCam Imaging}",
      journal = {\apjs},
         year = 2023,
        month = nov,
       volume = {269},
       number = {1},
          eid = {16},
        pages = {16},
          doi = {10.3847/1538-4365/acf44d},
archivePrefix = {arXiv},
       eprint = {2306.02466},
 primaryClass = {astro-ph.GA},
       adsurl = {https://ui.adsabs.harvard.edu/abs/2023ApJS..269...16R}
}

@ARTICLE{ananna_2019_aha_seds,
       author = {{Ananna}, Tonima Tasnim and {Treister}, Ezequiel and {Urry}, C. Megan and {Ricci}, C. and {Kirkpatrick}, Allison and {LaMassa}, Stephanie and {Buchner}, Johannes and {Civano}, Francesca and {Tremmel}, Michael and {Marchesi}, Stefano},
        title = "{The Accretion History of AGNs. I. Supermassive Black Hole Population Synthesis Model}",
      journal = {\apj},
         year = 2019,
        month = feb,
       volume = {871},
       number = {2},
          eid = {240},
        pages = {240},
          doi = {10.3847/1538-4357/aafb77},
archivePrefix = {arXiv},
       eprint = {1810.02298},
 primaryClass = {astro-ph.GA},
       adsurl = {https://ui.adsabs.harvard.edu/abs/2019ApJ...871..240A}
}

@article{khrykin+2021_quasar_lifetime,
    author = {Khrykin, Ilya S and Hennawi, Joseph F and Worseck, G√°bor and Davies, Frederick B},
    title = {The first measurement of the quasar lifetime distribution},
    journal = {Monthly Notices of the Royal Astronomical Society},
    volume = {505},
    number = {1},
    pages = {649-662},
    year = {2021},
    month = {05},
    issn = {0035-8711},
    doi = {10.1093/mnras/stab1288},
    url = {https://doi.org/10.1093/mnras/stab1288},
    eprint = {https://academic.oup.com/mnras/article-pdf/505/1/649/38360921/stab1288.pdf},
}

@ARTICLE{galight_ding,
       author = {{Ding}, Xuheng and {Silverman}, John and {Treu}, Tommaso and {Schulze}, Andreas and {Schramm}, Malte and {Birrer}, Simon and {Park}, Daeseong and {Jahnke}, Knud and {Bennert}, Vardha N. and {Kartaltepe}, Jeyhan S. and {Koekemoer}, Anton M. and {Malkan}, Matthew A. and {Sanders}, David},
        title = "{The Mass Relations between Supermassive Black Holes and Their Host Galaxies at 1 < z < 2 HST-WFC3}",
      journal = {\apj},
         year = 2020,
        month = jan,
       volume = {888},
       number = {1},
          eid = {37},
        pages = {37},
          doi = {10.3847/1538-4357/ab5b90},
archivePrefix = {arXiv},
       eprint = {1910.11875},
 primaryClass = {astro-ph.GA},
       adsurl = {https://ui.adsabs.harvard.edu/abs/2020ApJ...888...37D}
}

@ARTICLE{galight_ding_old,
       author = {{Ding}, Xuheng and {Birrer}, Simon and {Treu}, Tommaso and {Silverman}, John D.},
        title = "{Galaxy shapes of Light (GaLight): a 2D modeling of galaxy images}",
      journal = {arXiv e-prints},
         year = 2021,
        month = nov,
          eid = {arXiv:2111.08721},
        pages = {arXiv:2111.08721},
          doi = {10.48550/arXiv.2111.08721},
archivePrefix = {arXiv},
       eprint = {2111.08721},
 primaryClass = {astro-ph.GA},
       adsurl = {https://ui.adsabs.harvard.edu/abs/2021arXiv211108721D}
}

@ARTICLE{sanders_1988_ulirgs,
       author = {{Sanders}, D.~B. and {Soifer}, B.~T. and {Elias}, J.~H. and {Neugebauer}, G. and {Matthews}, K.},
        title = "{Warm Ultraluminous Galaxies in the IRAS Survey: The Transition from Galaxy to Quasar?}",
      journal = {\apjl},
         year = 1988,
        month = may,
       volume = {328},
        pages = {L35},
          doi = {10.1086/185155},
       adsurl = {https://ui.adsabs.harvard.edu/abs/1988ApJ...328L..35S}
}

@ARTICLE{treister_2004_obscuredagn,
       author = {{Treister}, Ezequiel and {Urry}, C. Megan and {Chatzichristou}, Eleni and {Bauer}, Franz and {Alexander}, David M. and {Koekemoer}, Anton and {Van Duyne}, Jeffrey and {Brandt}, William N. and {Bergeron}, Jacqueline and {Stern}, Daniel and {Moustakas}, Leonidas A. and {Chary}, Ranga-Ram and {Conselice}, Christopher and {Cristiani}, Stefano and {Grogin}, Norman},
        title = "{Obscured Active Galactic Nuclei and the X-Ray, Optical, and Far-Infrared Number Counts of Active Galactic Nuclei in the GOODS Fields}",
      journal = {\apj},
         year = 2004,
        month = nov,
       volume = {616},
       number = {1},
        pages = {123-135},
          doi = {10.1086/424891},
archivePrefix = {arXiv},
       eprint = {astro-ph/0408099},
 primaryClass = {astro-ph},
       adsurl = {https://ui.adsabs.harvard.edu/abs/2004ApJ...616..123T}
}

@ARTICLE{elbaz_2011_goods_herschel,
       author = {{Elbaz}, D. and {Dickinson}, M. and {Hwang}, H.~S. and {D{\'\i}az-Santos}, T. and {Magdis}, G. and {Magnelli}, B. and {Le Borgne}, D. and {Galliano}, F. and {Pannella}, M. and {Chanial}, P. and {Armus}, L. and {Charmandaris}, V. and {Daddi}, E. and {Aussel}, H. and {Popesso}, P. and {Kartaltepe}, J. and {Altieri}, B. and {Valtchanov}, I. and {Coia}, D. and {Dannerbauer}, H. and {Dasyra}, K. and {Leiton}, R. and {Mazzarella}, J. and {Alexander}, D.~M. and {Buat}, V. and {Burgarella}, D. and {Chary}, R. -R. and {Gilli}, R. and {Ivison}, R.~J. and {Juneau}, S. and {Le Floc'h}, E. and {Lutz}, D. and {Morrison}, G.~E. and {Mullaney}, J.~R. and {Murphy}, E. and {Pope}, A. and {Scott}, D. and {Brodwin}, M. and {Calzetti}, D. and {Cesarsky}, C. and {Charlot}, S. and {Dole}, H. and {Eisenhardt}, P. and {Ferguson}, H.~C. and {F{\"o}rster Schreiber}, N. and {Frayer}, D. and {Giavalisco}, M. and {Huynh}, M. and {Koekemoer}, A.~M. and {Papovich}, C. and {Reddy}, N. and {Surace}, C. and {Teplitz}, H. and {Yun}, M.~S. and {Wilson}, G.},
        title = "{GOODS-Herschel: an infrared main sequence for star-forming galaxies}",
      journal = {\aap},
         year = 2011,
        month = sep,
       volume = {533},
          eid = {A119},
        pages = {A119},
          doi = {10.1051/0004-6361/201117239},
archivePrefix = {arXiv},
       eprint = {1105.2537},
 primaryClass = {astro-ph.CO},
       adsurl = {https://ui.adsabs.harvard.edu/abs/2011A&A...533A.119E}
}

@ARTICLE{barro2019_goodsn_cat,
       author = {{Barro}, Guillermo and {P{\'e}rez-Gonz{\'a}lez}, Pablo G. and {Cava}, Antonio and {Brammer}, Gabriel and {Pandya}, Viraj and {Eliche Moral}, Carmen and {Esquej}, Pilar and {Dom{\'\i}nguez-S{\'a}nchez}, Helena and {Alcalde Pampliega}, Belen and {Guo}, Yicheng and {Koekemoer}, Anton M. and {Trump}, Jonathan R. and {Ashby}, Matthew L.~N. and {Cardiel}, Nicolas and {Castellano}, Marco and {Conselice}, Christopher J. and {Dickinson}, Mark E. and {Dolch}, Timothy and {Donley}, Jennifer L. and {Espino Briones}, N{\'e}stor and {Faber}, Sandra M. and {Fazio}, Giovanni G. and {Ferguson}, Henry and {Finkelstein}, Steve and {Fontana}, Adriano and {Galametz}, Audrey and {Gardner}, Jonathan P. and {Gawiser}, Eric and {Giavalisco}, Mauro and {Grazian}, Andrea and {Grogin}, Norman A. and {Hathi}, Nimish P. and {Hemmati}, Shoubaneh and {Hern{\'a}n-Caballero}, Antonio and {Kocevski}, Dale and {Koo}, David C. and {Kodra}, Dritan and {Lee}, Kyoung-Soo and {Lin}, Lihwai and {Lucas}, Ray A. and {Mobasher}, Bahram and {McGrath}, Elizabeth J. and {Nandra}, Kirpal and {Nayyeri}, Hooshang and {Newman}, Jeffrey A. and {Pforr}, Janine and {Peth}, Michael and {Rafelski}, Marc and {Rodr{\'\i}guez-Munoz}, Lucia and {Salvato}, Mara and {Stefanon}, Mauro and {van der Wel}, Arjen and {Willner}, Steven P. and {Wiklind}, Tommy and {Wuyts}, Stijn},
        title = "{The CANDELS/SHARDS Multiwavelength Catalog in GOODS-N: Photometry, Photometric Redshifts, Stellar Masses, Emission-line Fluxes, and Star Formation Rates}",
      journal = {\apjs},
         year = 2019,
        month = aug,
       volume = {243},
       number = {2},
          eid = {22},
        pages = {22},
          doi = {10.3847/1538-4365/ab23f2},
archivePrefix = {arXiv},
       eprint = {1908.00569},
 primaryClass = {astro-ph.GA},
       adsurl = {https://ui.adsabs.harvard.edu/abs/2019ApJS..243...22B}
}

@ARTICLE{eisenstein2023_jades_goodss,
       author = {{Eisenstein}, Daniel J. and {Johnson}, Benjamin D. and {Robertson}, Brant and {Tacchella}, Sandro and {Hainline}, Kevin and {Jakobsen}, Peter and {Maiolino}, Roberto and {Bonaventura}, Nina and {Bunker}, Andrew J. and {Cameron}, Alex J. and {Cargile}, Phillip A. and {Curtis-Lake}, Emma and {Hausen}, Ryan and {Pusk{\'a}s}, D{\'a}vid and {Rieke}, Marcia and {Sun}, Fengwu and {Willmer}, Christopher N.~A. and {Willott}, Chris and {Alberts}, Stacey and {Arribas}, Santiago and {Baker}, William M. and {Baum}, Stefi and {Bhatawdekar}, Rachana and {Carniani}, Stefano and {Charlot}, Stephane and {Chen}, Zuyi and {Chevallard}, Jacopo and {Curti}, Mirko and {DeCoursey}, Christa and {D'Eugenio}, Francesco and {de Graaff}, Anna and {Egami}, Eiichi and {Helton}, Jakob M. and {Ji}, Zhiyuan and {Jones}, Gareth C. and {Kumari}, Nimisha and {L{\"u}tzgendorf}, Nora and {Laseter}, Isaac and {Looser}, Tobias J. and {Lyu}, Jianwei and {Maseda}, Michael V. and {Nelson}, Erica and {Parlanti}, Eleonora and {Rauscher}, Bernard J. and {Rawle}, Tim and {Rieke}, George and {Rix}, Hans-Walter and {Rujopakarn}, Wiphu and {Sandles}, Lester and {Saxena}, Aayush and {Scholtz}, Jan and {Sharpe}, Katherine and {Shivaei}, Irene and {Simmonds}, Charlotte and {Smit}, Renske and {Topping}, Michael W. and {{\"U}bler}, Hannah and {Venturi}, Giacomo and {Williams}, Christina C. and {Witstok}, Joris and {Woodrum}, Charity},
        title = "{The JADES Origins Field: A New JWST Deep Field in the JADES Second NIRCam Data Release}",
      journal = {arXiv e-prints},
         year = 2023,
        month = oct,
          eid = {arXiv:2310.12340},
        pages = {arXiv:2310.12340},
          doi = {10.48550/arXiv.2310.12340},
archivePrefix = {arXiv},
       eprint = {2310.12340},
 primaryClass = {astro-ph.GA},
       adsurl = {https://ui.adsabs.harvard.edu/abs/2023arXiv231012340E}
}

@ARTICLE{eisenstein_jades_2023_overview,
       author = {{Eisenstein}, Daniel J. and {Willott}, Chris and {Alberts}, Stacey and {Arribas}, Santiago and {Bonaventura}, Nina and {Bunker}, Andrew J. and {Cameron}, Alex J. and {Carniani}, Stefano and {Charlot}, Stephane and {Curtis-Lake}, Emma and {D'Eugenio}, Francesco and {Endsley}, Ryan and {Ferruit}, Pierre and {Giardino}, Giovanna and {Hainline}, Kevin and {Hausen}, Ryan and {Jakobsen}, Peter and {Johnson}, Benjamin D. and {Maiolino}, Roberto and {Rieke}, Marcia and {Rieke}, George and {Rix}, Hans-Walter and {Robertson}, Brant and {Stark}, Daniel P. and {Tacchella}, Sandro and {Williams}, Christina C. and {Willmer}, Christopher N.~A. and {Baker}, William M. and {Baum}, Stefi and {Bhatawdekar}, Rachana and {Boyett}, Kristan and {Chen}, Zuyi and {Chevallard}, Jacopo and {Circosta}, Chiara and {Curti}, Mirko and {Danhaive}, A. Lola and {DeCoursey}, Christa and {de Graaff}, Anna and {Dressler}, Alan and {Egami}, Eiichi and {Helton}, Jakob M. and {Hviding}, Raphael E. and {Ji}, Zhiyuan and {Jones}, Gareth C. and {Kumari}, Nimisha and {L{\"u}tzgendorf}, Nora and {Laseter}, Isaac and {Looser}, Tobias J. and {Lyu}, Jianwei and {Maseda}, Michael V. and {Nelson}, Erica and {Parlanti}, Eleonora and {Perna}, Michele and {Pusk{\'a}s}, D{\'a}vid and {Rawle}, Tim and {Rodr{\'\i}guez Del Pino}, Bruno and {Sandles}, Lester and {Saxena}, Aayush and {Scholtz}, Jan and {Sharpe}, Katherine and {Shivaei}, Irene and {Silcock}, Maddie S. and {Simmonds}, Charlotte and {Skarbinski}, Maya and {Smit}, Renske and {Stone}, Meredith and {Suess}, Katherine A. and {Sun}, Fengwu and {Tang}, Mengtao and {Topping}, Michael W. and {{\"U}bler}, Hannah and {Villanueva}, Natalia C. and {Wallace}, Imaan E.~B. and {Whitler}, Lily and {Witstok}, Joris and {Woodrum}, Charity},
        title = "{Overview of the JWST Advanced Deep Extragalactic Survey (JADES)}",
      journal = {arXiv e-prints},
         year = 2023,
        month = jun,
          eid = {arXiv:2306.02465},
        pages = {arXiv:2306.02465},
          doi = {10.48550/arXiv.2306.02465},
archivePrefix = {arXiv},
       eprint = {2306.02465},
 primaryClass = {astro-ph.GA},
       adsurl = {https://ui.adsabs.harvard.edu/abs/2023arXiv230602465E}
}

@ARTICLE{hao_2013_quasar_seds,
       author = {{Hao}, Heng and {Elvis}, Martin and {Bongiorno}, Angela and {Zamorani}, Gianni and {Merloni}, Andrea and {Kelly}, Brandon C. and {Civano}, Francesca and {Celotti}, Annalisa and {Ho}, Luis C. and {Jahnke}, Knud and {Comastri}, Andrea and {Trump}, Jonathan R. and {Mainieri}, Vincenzo and {Salvato}, Mara and {Brusa}, Marcella and {Impey}, Chris D. and {Koekemoer}, Anton M. and {Lanzuisi}, Giorgio and {Vignali}, Cristian and {Silverman}, John D. and {Urry}, C. Megan and {Schawinski}, Kevin},
        title = "{A quasar-galaxy mixing diagram: quasar spectral energy distribution shapes in the optical to near-infrared}",
      journal = {\mnras},
         year = 2013,
        month = oct,
       volume = {434},
       number = {4},
        pages = {3104-3121},
          doi = {10.1093/mnras/stt1228},
archivePrefix = {arXiv},
       eprint = {1210.3044},
 primaryClass = {astro-ph.GA},
       adsurl = {https://ui.adsabs.harvard.edu/abs/2013MNRAS.434.3104H}
}

@ARTICLE{hao2014_type1_seds,
       author = {{Hao}, Heng and {Elvis}, Martin and {Civano}, Francesca and {Zamorani}, Gianni and {Ho}, Luis C. and {Comastri}, Andrea and {Brusa}, Marcella and {Bongiorno}, Angela and {Merloni}, Andrea and {Trump}, Jonathan R. and {Salvato}, Mara and {Impey}, Chris D. and {Koekemoer}, Anton M. and {Lanzuisi}, Giorgio and {Celotti}, Annalisa and {Jahnke}, Knud and {Vignali}, Cristian and {Silverman}, John D. and {Urry}, C. Megan and {Schawinski}, Kevin and {Capak}, Peter},
        title = "{Spectral energy distributions of type 1 AGN in XMM-COSMOS - II. Shape evolution}",
      journal = {\mnras},
         year = 2014,
        month = feb,
       volume = {438},
       number = {2},
        pages = {1288-1304},
          doi = {10.1093/mnras/stt2274},
archivePrefix = {arXiv},
       eprint = {1210.3033},
 primaryClass = {astro-ph.GA},
       adsurl = {https://ui.adsabs.harvard.edu/abs/2014MNRAS.438.1288H}
}

@ARTICLE{suh_2020_no_z_relation_evolution,
       author = {{Suh}, Hyewon and {Civano}, Francesca and {Trakhtenbrot}, Benny and {Shankar}, Francesco and {Hasinger}, G{\"u}nther and {Sanders}, David B. and {Allevato}, Viola},
        title = "{No Significant Evolution of Relations between Black Hole Mass and Galaxy Total Stellar Mass Up to z {\ensuremath{\sim}} 2.5}",
      journal = {\apj},
         year = 2020,
        month = jan,
       volume = {889},
       number = {1},
          eid = {32},
        pages = {32},
          doi = {10.3847/1538-4357/ab5f5f},
archivePrefix = {arXiv},
       eprint = {1912.02824},
 primaryClass = {astro-ph.GA},
       adsurl = {https://ui.adsabs.harvard.edu/abs/2020ApJ...889...32S}
}

@ARTICLE{coleman_2022_aha_host_galaxy,
       author = {{Coleman}, Brandon and {Kirkpatrick}, Allison and {Cooke}, Kevin C. and {Glikman}, Eilat and {LaMassa}, Stephanie and {Marchesi}, Stefano and {Peca}, Alessandro and {Treister}, Ezequiel and {Auge}, Connor and {Urry}, C. Megan and {Sanders}, Dave and {Turner}, Tracey Jane and {Ananna}, Tonima Tasnim},
        title = "{Accretion history of AGN: Estimating the host galaxy properties in X-ray luminous AGN from z = 0-3}",
      journal = {\mnras},
         year = 2022,
        month = sep,
       volume = {515},
       number = {1},
        pages = {82-98},
          doi = {10.1093/mnras/stac1679},
archivePrefix = {arXiv},
       eprint = {2206.06398},
 primaryClass = {astro-ph.GA},
       adsurl = {https://ui.adsabs.harvard.edu/abs/2022MNRAS.515...82C}
}

@ARTICLE{gebhardt_2000_msigma,
       author = {{Gebhardt}, Karl and {Bender}, Ralf and {Bower}, Gary and {Dressler}, Alan and {Faber}, S.~M. and {Filippenko}, Alexei V. and {Green}, Richard and {Grillmair}, Carl and {Ho}, Luis C. and {Kormendy}, John and {Lauer}, Tod R. and {Magorrian}, John and {Pinkney}, Jason and {Richstone}, Douglas and {Tremaine}, Scott},
        title = "{A Relationship between Nuclear Black Hole Mass and Galaxy Velocity Dispersion}",
      journal = {\apjl},
         year = 2000,
        month = aug,
       volume = {539},
       number = {1},
        pages = {L13-L16},
          doi = {10.1086/312840},
archivePrefix = {arXiv},
       eprint = {astro-ph/0006289},
 primaryClass = {astro-ph},
       adsurl = {https://ui.adsabs.harvard.edu/abs/2000ApJ...539L..13G}
}

@ARTICLE{conselice_2014_gal_structure,
       author = {{Conselice}, Christopher J.},
        title = "{The Evolution of Galaxy Structure Over Cosmic Time}",
      journal = {\araa},
         year = 2014,
        month = aug,
       volume = {52},
        pages = {291-337},
          doi = {10.1146/annurev-astro-081913-040037},
archivePrefix = {arXiv},
       eprint = {1403.2783},
 primaryClass = {astro-ph.GA},
       adsurl = {https://ui.adsabs.harvard.edu/abs/2014ARA&A..52..291C}
}

@ARTICLE{hopkins_2006_agn_model,
       author = {{Hopkins}, Philip F. and {Hernquist}, Lars and {Cox}, Thomas J. and {Di Matteo}, Tiziana and {Robertson}, Brant and {Springel}, Volker},
        title = "{A Unified, Merger-driven Model of the Origin of Starbursts, Quasars, the Cosmic X-Ray Background, Supermassive Black Holes, and Galaxy Spheroids}",
      journal = {\apjs},
         year = 2006,
        month = mar,
       volume = {163},
       number = {1},
        pages = {1-49},
          doi = {10.1086/499298},
archivePrefix = {arXiv},
       eprint = {astro-ph/0506398},
 primaryClass = {astro-ph},
       adsurl = {https://ui.adsabs.harvard.edu/abs/2006ApJS..163....1H}
}

@ARTICLE{simmons_2017_galzoo_CANDELS,
       author = {{Simmons}, B.~D. and {Lintott}, Chris and {Willett}, Kyle W. and {Masters}, Karen L. and {Kartaltepe}, Jeyhan S. and {H{\"a}u{\ss}ler}, Boris and {Kaviraj}, Sugata and {Krawczyk}, Coleman and {Kruk}, S.~J. and {McIntosh}, Daniel H. and {Smethurst}, R.~J. and {Nichol}, Robert C. and {Scarlata}, Claudia and {Schawinski}, Kevin and {Conselice}, Christopher J. and {Almaini}, Omar and {Ferguson}, Henry C. and {Fortson}, Lucy and {Hartley}, William and {Kocevski}, Dale and {Koekemoer}, Anton M. and {Mortlock}, Alice and {Newman}, Jeffrey A. and {Bamford}, Steven P. and {Grogin}, N.~A. and {Lucas}, Ray A. and {Hathi}, Nimish P. and {McGrath}, Elizabeth and {Peth}, Michael and {Pforr}, Janine and {Rizer}, Zachary and {Wuyts}, Stijn and {Barro}, Guillermo and {Bell}, Eric F. and {Castellano}, Marco and {Dahlen}, Tomas and {Dekel}, Avishai and {Ownsworth}, Jamie and {Faber}, Sandra M. and {Finkelstein}, Steven L. and {Fontana}, Adriano and {Galametz}, Audrey and {Gr{\"u}tzbauch}, Ruth and {Koo}, David and {Lotz}, Jennifer and {Mobasher}, Bahram and {Mozena}, Mark and {Salvato}, Mara and {Wiklind}, Tommy},
        title = "{Galaxy Zoo: quantitative visual morphological classifications for 48 000 galaxies from CANDELS}",
      journal = {\mnras},
         year = 2017,
        month = feb,
       volume = {464},
       number = {4},
        pages = {4420-4447},
          doi = {10.1093/mnras/stw2587},
archivePrefix = {arXiv},
       eprint = {1610.03070},
 primaryClass = {astro-ph.GA},
       adsurl = {https://ui.adsabs.harvard.edu/abs/2017MNRAS.464.4420S}
}

@ARTICLE{renalli_2003_lx_sfr,
       author = {{Ranalli}, Piero and {Comastri}, Andrea and {Setti}, Giancarlo},
        title = "{The 2-10 keV luminosity as a Star Formation Rate indicator}",
      journal = {arXiv e-prints},
         year = 2003,
        month = feb,
          eid = {astro-ph/0202241},
        pages = {astro-ph/0202241},
          doi = {10.48550/arXiv.astro-ph/0202241},
archivePrefix = {arXiv},
       eprint = {astro-ph/0202241},
 primaryClass = {astro-ph},
       adsurl = {https://ui.adsabs.harvard.edu/abs/2002astro.ph..2241R}
}

@ARTICLE{fabbiano_1989_normal_xrays,
       author = {{Fabbiano}, G.},
        title = "{X-rays from normal galaxies.}",
      journal = {\araa},
         year = 1989,
        month = jan,
       volume = {27},
        pages = {87-138},
          doi = {10.1146/annurev.aa.27.090189.000511},
       adsurl = {https://ui.adsabs.harvard.edu/abs/1989ARA&A..27...87F}
}

@ARTICLE{neugebauer_quasar_energy,
       author = {{Neugebauer}, G. and {Green}, R.~F. and {Matthews}, K. and {Schmidt}, M. and {Soifer}, B.~T. and {Bennett}, J.},
        title = "{Continuum Energy Distributions of Quasars in the Palomar-Green Survey}",
      journal = {\apjs},
         year = 1987,
        month = mar,
       volume = {63},
        pages = {615},
          doi = {10.1086/191175},
       adsurl = {https://ui.adsabs.harvard.edu/abs/1987ApJS...63..615N}
}

@ARTICLE{sanders_quasar_continuum,
       author = {{Sanders}, D.~B. and {Phinney}, E.~S. and {Neugebauer}, G. and {Soifer}, B.~T. and {Matthews}, K.},
        title = "{Continuum Energy Distributions of Quasars: Shapes and Origins}",
      journal = {\apj},
         year = 1989,
        month = dec,
       volume = {347},
        pages = {29},
          doi = {10.1086/168094},
       adsurl = {https://ui.adsabs.harvard.edu/abs/1989ApJ...347...29S}
}

@MISC{lallo_2005_hst_orbital_breathing,
       author = {{Lallo}, M. and {Makidon}, R.~B. and {Casertano}, S. and {Gilliland}, R. and {Stys}, J.},
        title = "{HST Temporal Optical Behavior \& Current Focus Status}",
 howpublished = {Instrument Science Report TEL 2005-03, 20 pages},
         year = 2005,
        month = oct,
        pages = {3},
       adsurl = {https://ui.adsabs.harvard.edu/abs/2005tel..rept....3L}
}

@ARTICLE{ding_2023_galight_nature,
       author = {{Ding}, Xuheng and {Onoue}, Masafusa and {Silverman}, John D. and {Matsuoka}, Yoshiki and {Izumi}, Takuma and {Strauss}, Michael A. and {Jahnke}, Knud and {Phillips}, Camryn L. and {Li}, Junyao and {Volonteri}, Marta and {Haiman}, Zoltan and {Andika}, Irham Taufik and {Aoki}, Kentaro and {Baba}, Shunsuke and {Bieri}, Rebekka and {Bosman}, Sarah E.~I. and {Bottrell}, Connor and {Eilers}, Anna-Christina and {Fujimoto}, Seiji and {Habouzit}, Melanie and {Imanishi}, Masatoshi and {Inayoshi}, Kohei and {Iwasawa}, Kazushi and {Kashikawa}, Nobunari and {Kawaguchi}, Toshihiro and {Kohno}, Kotaro and {Lee}, Chien-Hsiu and {Lupi}, Alessandro and {Lyu}, Jianwei and {Nagao}, Tohru and {Overzier}, Roderik and {Schindler}, Jan-Torge and {Schramm}, Malte and {Shimasaku}, Kazuhiro and {Toba}, Yoshiki and {Trakhtenbrot}, Benny and {Trebitsch}, Maxime and {Treu}, Tommaso and {Umehata}, Hideki and {Venemans}, Bram P. and {Vestergaard}, Marianne and {Walter}, Fabian and {Wang}, Feige and {Yang}, Jinyi},
        title = "{Detection of stellar light from quasar host galaxies at redshifts above 6}",
      journal = {\nat},
         year = 2023,
        month = sep,
       volume = {621},
       number = {7977},
        pages = {51-55},
          doi = {10.1038/s41586-023-06345-5},
archivePrefix = {arXiv},
       eprint = {2211.14329},
 primaryClass = {astro-ph.GA},
       adsurl = {https://ui.adsabs.harvard.edu/abs/2023Natur.621...51D}
}

@MISC{lajoie_2023_jwst_psf,
       author = {{Lajoie}, Charles-Philippe and {Lallo}, Matthew and {Mel{\'e}ndez}, Marcio and {Flagey}, Nicolas and {Telfer}, Randal and {Comeau}, Thomas M. and {Kulp}, Bernard A. and {Beck}, Tracy and {Brady}, Gregory R. and {Perrin}, Marshall D.},
        title = "{OTE Science Performance Memo 2 - A Year of Wavefront Sensing with JWST in Flight: Cycle 1 Telescope Monitoring \& Maintenance Summary}",
 howpublished = {Technical Report JWST-STScI-008497},
         year = 2023,
        month = jul,
        pages = {8497},
          doi = {10.48550/arXiv.2307.11179},
archivePrefix = {arXiv},
       eprint = {2307.11179},
 primaryClass = {astro-ph.IM},
       adsurl = {https://ui.adsabs.harvard.edu/abs/2023jwst.rept.8497L}
}

@software{sao_ds9,
       author = {{Smithsonian Astrophysical Observatory}},
        title = "{SAOImage DS9: A utility for displaying astronomical images in the X11 window environment}",
 howpublished = {Astrophysics Source Code Library, record ascl:0003.002},
         year = 2000,
        month = mar,
          eid = {ascl:0003.002},
       adsurl = {https://ui.adsabs.harvard.edu/abs/2000ascl.soft03002S}
}

@ARTICLE{Merritt_Ferrarese_01_msigma,
       author = {{Merritt}, David and {Ferrarese}, Laura},
        title = "{The M$_{{\textbullet}}$-{\ensuremath{\sigma}} Relation for Supermassive Black Holes}",
      journal = {\apj},
         year = 2001,
        month = jan,
       volume = {547},
       number = {1},
        pages = {140-145},
          doi = {10.1086/318372},
archivePrefix = {arXiv},
       eprint = {astro-ph/0008310},
 primaryClass = {astro-ph},
       adsurl = {https://ui.adsabs.harvard.edu/abs/2001ApJ...547..140M}
}

@ARTICLE{fotopoulou_2016_5-10_xlf,
       author = {{Fotopoulou}, S. and {Buchner}, J. and {Georgantopoulos}, I. and {Hasinger}, G. and {Salvato}, M. and {Georgakakis}, A. and {Cappelluti}, N. and {Ranalli}, P. and {Hsu}, L.~T. and {Brusa}, M. and {Comastri}, A. and {Miyaji}, T. and {Nandra}, K.  and  {Aird}, J. and {Paltani}, S.},
        title = "{The 5-10 keV AGN luminosity function at 0.01 < z < 4.0}",
      journal = {\aap},
         year = 2016,
        month = mar,
       volume = {587},
          eid = {A142},
        pages = {A142},
          doi = {10.1051/0004-6361/201424763},
archivePrefix = {arXiv},
       eprint = {1601.06002},
 primaryClass = {astro-ph.GA},
       adsurl = {https://ui.adsabs.harvard.edu/abs/2016A&A...587A.142F}
}

@ARTICLE{lenstronomy_software,
       author = {{Birrer}, Simon and {Shajib}, Anowar and {Gilman}, Daniel and {Galan}, Aymeric and {Aalbers}, Jelle and {Millon}, Martin and {Morgan}, Robert and {Pagano}, Giulia and {Park}, Ji and {Teodori}, Luca and {Tessore}, Nicolas and {Ueland}, Madison and {Van de Vyvere}, Lyne and {Wagner-Carena}, Sebastian and {Wempe}, Ewoud and {Yang}, Lilan and {Ding}, Xuheng and {Schmidt}, Thomas and {Sluse}, Dominique and {Zhang}, Ming and {Amara}, Adam},
        title = "{lenstronomy II: A gravitational lensing software ecosystem}",
      journal = {The Journal of Open Source Software},
         year = 2021,
        month = jun,
       volume = {6},
       number = {62},
          eid = {3283},
        pages = {3283},
          doi = {10.21105/joss.03283},
archivePrefix = {arXiv},
       eprint = {2106.05976},
 primaryClass = {astro-ph.CO},
       adsurl = {https://ui.adsabs.harvard.edu/abs/2021JOSS....6.3283B}
}

@article{Ghosh2023,
   author = {Aritra Ghosh and C. Megan Urry and Aayush Mishra and Laurence Perreault-Levasseur and Priyamvada Natarajan and David B. Sanders and Daisuke Nagai and Chuan Tian and Nico Cappelluti and Jeyhan S. Kartaltepe and Meredith C. Powell and Amrit Rau and Ezequiel Treister},
   doi = {10.3847/1538-4357/acd546},
   issn = {0004-637X},
   issue = {2},
   journal = {The Astrophysical Journal},
   month = {8},
   pages = {134},
   title = {Morphological Parameters and Associated Uncertainties for 8 Million Galaxies in the Hyper Suprime-Cam Wide Survey},
   volume = {953},
   year = {2023}
}

@article{Ghosh2024,
   author = {Aritra Ghosh and C. Megan Urry and Meredith C. Powell and Rhythm Shimakawa and Frank C. van den Bosch and Daisuke Nagai and Kaustav Mitra and Andrew J. Connolly},
   doi = {10.3847/1538-4357/ad596f},
   issn = {0004-637X},
   issue = {2},
   journal = {The Astrophysical Journal},
   month = {8},
   pages = {142},
   title = {Denser Environments Cultivate Larger Galaxies: A Comprehensive Study beyond the Local Universe with 3 Million Hyper Suprime-Cam Galaxies},
   volume = {971},
   year = {2024}
}

@ARTICLE{peca_2023_xray_s82x,
       author = {{Peca}, Alessandro and {Cappelluti}, Nico and {Urry}, C. Megan and {LaMassa}, Stephanie and {Marchesi}, Stefano and {Ananna}, Tonima Tasnim and {Balokovi{\'c}}, Mislav and {Sanders}, David and {Auge}, Connor and {Treister}, Ezequiel and {Powell}, Meredith and {Turner}, Tracey Jane and {Kirkpatrick}, Allison and {Tian}, Chuan},
        title = "{On the Cosmic Evolution of AGN Obscuration and the X-Ray Luminosity Function: XMM-Newton and Chandra Spectral Analysis of the 31.3 deg$^{2}$ Stripe 82X}",
      journal = {\apj},
         year = 2023,
        month = feb,
       volume = {943},
       number = {2},
          eid = {162},
        pages = {162},
          doi = {10.3847/1538-4357/acac28},
archivePrefix = {arXiv},
       eprint = {2210.08030},
 primaryClass = {astro-ph.GA},
       adsurl = {https://ui.adsabs.harvard.edu/abs/2023ApJ...943..162P}
}

@ARTICLE{peca_2024_s82xl,
       author = {{Peca}, Alessandro and {Cappelluti}, Nico and {LaMassa}, Stephanie and {Urry}, C. Megan and {Moscetti}, Massimo and {Marchesi}, Stefano and {Sanders}, David and {Auge}, Connor and {Ghosh}, Aritra and {Ananna}, Tonima Tasnim and {Torres-Alb{\`a}}, N{\'u}ria and {Treister}, Ezequiel},
        title = "{Stripe 82-XL: The {\ensuremath{\sim}}54.8 deg$^{2}$ and {\ensuremath{\sim}}18.8 Ms Chandra and XMM-Newton Point-source Catalog and Number of Counts}",
      journal = {\apj},
         year = 2024,
        month = oct,
       volume = {974},
       number = {2},
          eid = {156},
        pages = {156},
          doi = {10.3847/1538-4357/ad6df4},
archivePrefix = {arXiv},
       eprint = {2407.09617},
 primaryClass = {astro-ph.GA},
       adsurl = {https://ui.adsabs.harvard.edu/abs/2024ApJ...974..156P}
}

@ARTICLE{treister_2012_majormerger_agn,
       author = {{Treister}, E. and {Schawinski}, K. and {Urry}, C.~M. and {Simmons}, B.~D.},
        title = "{Major Galaxy Mergers Only Trigger the Most Luminous Active Galactic Nuclei}",
      journal = {\apjl},
         year = 2012,
        month = oct,
       volume = {758},
       number = {2},
          eid = {L39},
        pages = {L39},
          doi = {10.1088/2041-8205/758/2/L39},
archivePrefix = {arXiv},
       eprint = {1209.5393},
 primaryClass = {astro-ph.CO},
       adsurl = {https://ui.adsabs.harvard.edu/abs/2012ApJ...758L..39T}
}

@ARTICLE{simmons2011_obscured_goods_agn,
       author = {{Simmons}, B.~D. and {Van Duyne}, J. and {Urry}, C.~M. and {Treister}, E. and {Koekemoer}, A.~M. and {Grogin}, N.~A. and {GOODS Team}},
        title = "{Obscured GOODS Active Galactic Nuclei and Their Host Galaxies at z < 1.25: The Slow Black Hole Growth Phase}",
      journal = {\apj},
         year = 2011,
        month = jun,
       volume = {734},
       number = {2},
          eid = {121},
        pages = {121},
          doi = {10.1088/0004-637X/734/2/121},
archivePrefix = {arXiv},
       eprint = {1104.2619},
 primaryClass = {astro-ph.CO},
       adsurl = {https://ui.adsabs.harvard.edu/abs/2011ApJ...734..121S}
}

@article{Massey_1951, title={The Kolmogorov-Smirnov test for goodness of fit}, volume={46}, DOI={10.2307/2280095}, number={253}, journal={Journal of the American Statistical Association}, author={Massey, Frank J.}, year={1951}, month={Mar}, pages={68}}

@ARTICLE{harrison_almedida_agn_review2024,
       author = {{Harrison}, Chris M. and {Ramos Almeida}, Cristina},
        title = "{Observational Tests of Active Galactic Nuclei Feedback: An Overview of Approaches and Interpretation}",
      journal = {Galaxies},
         year = 2024,
        month = apr,
       volume = {12},
       number = {2},
          eid = {17},
        pages = {17},
          doi = {10.3390/galaxies12020017},
archivePrefix = {arXiv},
       eprint = {2404.08050},
 primaryClass = {astro-ph.GA},
       adsurl = {https://ui.adsabs.harvard.edu/abs/2024Galax..12...17H}
}

@ARTICLE{buchner2017_agn_gas_torus,
       author = {{Buchner}, Johannes and {Bauer}, Franz E.},
        title = "{Galaxy gas as obscurer - II. Separating the galaxy-scale and nuclear obscurers of active galactic nuclei}",
      journal = {\mnras},
         year = 2017,
        month = mar,
       volume = {465},
       number = {4},
        pages = {4348-4362},
          doi = {10.1093/mnras/stw2955},
archivePrefix = {arXiv},
       eprint = {1610.09380},
 primaryClass = {astro-ph.HE},
       adsurl = {https://ui.adsabs.harvard.edu/abs/2017MNRAS.465.4348B}
}

@ARTICLE{cisternas2011_weak_agn_merger_link,
       author = {{Cisternas}, Mauricio and {Jahnke}, Knud and {Inskip}, Katherine J. and {Kartaltepe}, Jeyhan and {Koekemoer}, Anton M. and {Lisker}, Thorsten and {Robaina}, Aday R. and {Scodeggio}, Marco and {Sheth}, Kartik and {Trump}, Jonathan R. and {Andrae}, Ren{\'e} and {Miyaji}, Takamitsu and {Lusso}, Elisabeta and {Brusa}, Marcella and {Capak}, Peter and {Cappelluti}, Nico and {Civano}, Francesca and {Ilbert}, Olivier and {Impey}, Chris D. and {Leauthaud}, Alexie and {Lilly}, Simon J. and {Salvato}, Mara and {Scoville}, Nick Z. and {Taniguchi}, Yoshi},
        title = "{The Bulk of the Black Hole Growth Since z \raisebox{-0.5ex}\textasciitilde 1 Occurs in a Secular Universe: No Major Merger-AGN Connection}",
      journal = {\apj},
         year = 2011,
        month = jan,
       volume = {726},
       number = {2},
          eid = {57},
        pages = {57},
          doi = {10.1088/0004-637X/726/2/57},
archivePrefix = {arXiv},
       eprint = {1009.3265},
 primaryClass = {astro-ph.CO},
       adsurl = {https://ui.adsabs.harvard.edu/abs/2011ApJ...726...57C}
}

@ARTICLE{povic2012_weak_agn_merger_link,
       author = {{Povi{\'c}}, M. and {S{\'a}nchez-Portal}, M. and {P{\'e}rez Garc{\'\i}a}, A.~M. and {Bongiovanni}, A. and {Cepa}, J. and {Huertas-Company}, M. and {Lara-L{\'o}pez}, M.~A. and {Fern{\'a}ndez Lorenzo}, M. and {Ederoclite}, A. and {Alfaro}, E. and {Casta{\~n}eda}, H. and {Gallego}, J. and {Gonz{\'a}lez-Serrano}, J.~I. and {Gonz{\'a}lez}, J.~J.},
        title = "{AGN-host galaxy connection: morphology and colours of X-ray selected AGN at z {\ensuremath{\leq}} 2}",
      journal = {\aap},
         year = 2012,
        month = may,
       volume = {541},
          eid = {A118},
        pages = {A118},
          doi = {10.1051/0004-6361/201117314},
archivePrefix = {arXiv},
       eprint = {1202.1662},
 primaryClass = {astro-ph.CO},
       adsurl = {https://ui.adsabs.harvard.edu/abs/2012A&A...541A.118P}
}

@ARTICLE{villforth_2014_merger_candels,
       author = {{Villforth}, C. and {Hamann}, F. and {Rosario}, D.~J. and {Santini}, P. and {McGrath}, E.~J. and {van der Wel}, A. and {Chang}, Y.~Y. and {Guo}, Y. and {Dahlen}, T. and {Bell}, E.~F. and {Conselice}, C.~J. and {Croton}, D. and {Dekel}, A. and {Faber}, S.~M. and {Grogin}, N. and {Hamilton}, T. and {Hopkins}, P.~F. and {Juneau}, S. and {Kartaltepe}, J. and {Kocevski}, D. and {Koekemoer}, A. and {Koo}, D.~C. and {Lotz}, J. and {McIntosh}, D. and {Mozena}, M. and {Somerville}, R. and {Wild}, V.},
        title = "{Morphologies of z {\ensuremath{\sim}} 0.7 AGN host galaxies in CANDELS: no trend of merger incidence with AGN luminosity}",
      journal = {\mnras},
         year = 2014,
        month = apr,
       volume = {439},
       number = {4},
        pages = {3342-3356},
          doi = {10.1093/mnras/stu173},
archivePrefix = {arXiv},
       eprint = {1401.5477},
 primaryClass = {astro-ph.GA},
       adsurl = {https://ui.adsabs.harvard.edu/abs/2014MNRAS.439.3342V}
}

@ARTICLE{rosario2015_bulge_enhancement,
       author = {{Rosario}, D.~J. and {McIntosh}, D.~H. and {van der Wel}, A. and {Kartaltepe}, J. and {Lang}, P. and {Santini}, P. and {Wuyts}, S. and {Lutz}, D. and {Rafelski}, M. and {Villforth}, C. and {Alexander}, D.~M. and {Bauer}, F.~E. and {Bell}, E.~F. and {Berta}, S. and {Brandt}, W.~N. and {Conselice}, C.~J. and {Dekel}, A. and {Faber}, S.~M. and {Ferguson}, H.~C. and {Genzel}, R. and {Grogin}, N.~A. and {Kocevski}, D.~D. and {Koekemoer}, A.~M. and {Koo}, D.~C. and {Lotz}, J.~M. and {Magnelli}, B. and {Maiolino}, R. and {Mozena}, M. and {Mullaney}, J.~R. and {Papovich}, C.~J. and {Popesso}, P. and {Tacconi}, L.~J. and {Trump}, J.~R. and {Avadhuta}, S. and {Bassett}, R. and {Bell}, A. and {Bernyk}, M. and {Bournaud}, F. and {Cassata}, P. and {Cheung}, E. and {Croton}, D. and {Donley}, J. and {DeGroot}, L. and {Guedes}, J. and {Hathi}, N. and {Herrington}, J. and {Hilton}, M. and {Lai}, K. and {Lani}, C. and {Martig}, M. and {McGrath}, E. and {Mutch}, S. and {Mortlock}, A. and {McPartland}, C. and {O'Leary}, E. and {Peth}, M. and {Pillepich}, A. and {Poole}, G. and {Snyder}, D. and {Straughn}, A. and {Telford}, O. and {Tonini}, C. and {Wandro}, P.},
        title = "{The host galaxies of X-ray selected active galactic nuclei to z = 2.5: Structure, star formation, and their relationships from CANDELS and Herschel/PACS}",
      journal = {\aap},
         year = 2015,
        month = jan,
       volume = {573},
          eid = {A85},
        pages = {A85},
          doi = {10.1051/0004-6361/201423782},
archivePrefix = {arXiv},
       eprint = {1409.5122},
 primaryClass = {astro-ph.GA},
       adsurl = {https://ui.adsabs.harvard.edu/abs/2015A&A...573A..85R}
}

@ARTICLE{bruce2016_active_inactive_galaxies,
       author = {{Bruce}, V.~A. and {Dunlop}, J.~S. and {Mortlock}, A. and {Kocevski}, D.~D. and {McGrath}, E.~J. and {Rosario}, D.~J.},
        title = "{The bulge-disc decomposition of AGN host galaxies}",
      journal = {\mnras},
         year = 2016,
        month = may,
       volume = {458},
       number = {3},
        pages = {2391-2404},
          doi = {10.1093/mnras/stw467},
archivePrefix = {arXiv},
       eprint = {1510.03870},
 primaryClass = {astro-ph.GA},
       adsurl = {https://ui.adsabs.harvard.edu/abs/2016MNRAS.458.2391B}
}

@ARTICLE{haring_smbh_bulge_mass,
       author = {{H{\"a}ring}, Nadine and {Rix}, Hans-Walter},
        title = "{On the Black Hole Mass-Bulge Mass Relation}",
      journal = {\apjl},
         year = 2004,
        month = apr,
       volume = {604},
       number = {2},
        pages = {L89-L92},
          doi = {10.1086/383567},
archivePrefix = {arXiv},
       eprint = {astro-ph/0402376},
 primaryClass = {astro-ph},
       adsurl = {https://ui.adsabs.harvard.edu/abs/2004ApJ...604L..89H}
}

@ARTICLE{Haehnelt_quasar_duty_cycle,
       author = {{Haehnelt}, Martin G. and {Rees}, Martin J.},
        title = "{The formation of nuclei in newly formed galaxies and the evolution of the quasar population}",
      journal = {\mnras},
         year = 1993,
        month = jul,
       volume = {263},
       number = {1},
        pages = {168-178},
          doi = {10.1093/mnras/263.1.168},
       adsurl = {https://ui.adsabs.harvard.edu/abs/1993MNRAS.263..168H}
}

@ARTICLE{hickox_alexander_obscuredagn_review2018,
       author = {{Hickox}, Ryan C. and {Alexander}, David M.},
        title = "{Obscured Active Galactic Nuclei}",
      journal = {\araa},
         year = 2018,
        month = sep,
       volume = {56},
        pages = {625-671},
          doi = {10.1146/annurev-astro-081817-051803},
archivePrefix = {arXiv},
       eprint = {1806.04680},
 primaryClass = {astro-ph.GA},
       adsurl = {https://ui.adsabs.harvard.edu/abs/2018ARA&A..56..625H}
}

@ARTICLE{kocevski_candels_agn_merger_2012,
       author = {{Kocevski}, Dale D. and {Faber}, S.~M. and {Mozena}, Mark and {Koekemoer}, Anton M. and {Nandra}, Kirpal and {Rangel}, Cyprian and {Laird}, Elise S. and {Brusa}, Marcella and {Wuyts}, Stijn and {Trump}, Jonathan R. and {Koo}, David C. and {Somerville}, Rachel S. and {Bell}, Eric F. and {Lotz}, Jennifer M. and {Alexander}, David M. and {Bournaud}, Frederic and {Conselice}, Christopher J. and {Dahlen}, Tomas and {Dekel}, Avishai and {Donley}, Jennifer L. and {Dunlop}, James S. and {Finoguenov}, Alexis and {Georgakakis}, Antonis and {Giavalisco}, Mauro and {Guo}, Yicheng and {Grogin}, Norman A. and {Hathi}, Nimish P. and {Juneau}, St{\'e}phanie and {Kartaltepe}, Jeyhan S. and {Lucas}, Ray A. and {McGrath}, Elizabeth J. and {McIntosh}, Daniel H. and {Mobasher}, Bahram and {Robaina}, Aday R. and {Rosario}, David and {Straughn}, Amber N. and {van der Wel}, Arjen and {Villforth}, Carolin},
        title = "{CANDELS: Constraining the AGN-Merger Connection with Host Morphologies at z \raisebox{-0.5ex}\textasciitilde 2}",
      journal = {\apj},
         year = 2012,
        month = jan,
       volume = {744},
       number = {2},
          eid = {148},
        pages = {148},
          doi = {10.1088/0004-637X/744/2/148},
archivePrefix = {arXiv},
       eprint = {1109.2588},
 primaryClass = {astro-ph.CO},
       adsurl = {https://ui.adsabs.harvard.edu/abs/2012ApJ...744..148K}
}

@ARTICLE{lyu_2022_completeness,
       author = {{Lyu}, Jianwei and {Alberts}, Stacey and {Rieke}, George H. and {Rujopakarn}, Wiphu},
        title = "{AGN Selection and Demographics in GOODS-S/HUDF from X-Ray to Radio}",
      journal = {\apj},
         year = 2022,
        month = dec,
       volume = {941},
       number = {2},
          eid = {191},
        pages = {191},
          doi = {10.3847/1538-4357/ac9e5d},
archivePrefix = {arXiv},
       eprint = {2209.06219},
 primaryClass = {astro-ph.GA},
       adsurl = {https://ui.adsabs.harvard.edu/abs/2022ApJ...941..191L}
}

@ARTICLE{ciesla_2015_agn_seds,
       author = {{Ciesla}, L. and {Charmandaris}, V. and {Georgakakis}, A. and {Bernhard}, E. and {Mitchell}, P.~D. and {Buat}, V. and {Elbaz}, D. and {LeFloc'h}, E. and {Lacey}, C.~G. and {Magdis}, G.~E. and {Xilouris}, M.},
        title = "{Constraining the properties of AGN host galaxies with spectral energy distribution modelling}",
      journal = {\aap},
         year = 2015,
        month = apr,
       volume = {576},
          eid = {A10},
        pages = {A10},
          doi = {10.1051/0004-6361/201425252},
archivePrefix = {arXiv},
       eprint = {1501.03672},
 primaryClass = {astro-ph.GA},
       adsurl = {https://ui.adsabs.harvard.edu/abs/2015A&A...576A..10C}
}

@ARTICLE{smethurst_non_merger_2024,
       author = {{Smethurst}, R.~J. and {Beckmann}, R.~S. and {Simmons}, B.~D. and {Coil}, A. and {Devriendt}, J. and {Dubois}, Y. and {Garland}, I.~L. and {Lintott}, C.~J. and {Martin}, G. and {Peirani}, S.},
        title = "{Evidence for non-merger co-evolution of galaxies and their supermassive black holes}",
      journal = {\mnras},
         year = 2024,
        month = feb,
       volume = {527},
       number = {4},
        pages = {10855-10866},
          doi = {10.1093/mnras/stad1794},
       adsurl = {https://ui.adsabs.harvard.edu/abs/2024MNRAS.52710855S}
}

@ARTICLE{baldassare_2020_low_mass_m_sigma,
       author = {{Baldassare}, Vivienne F. and {Dickey}, Claire and {Geha}, Marla and {Reines}, Amy E.},
        title = "{Populating the Low-mass End of the M$_{BH}$- \{\textbackslashsigma \}\_\{* \} Relation}",
      journal = {\apjl},
         year = 2020,
        month = jul,
       volume = {898},
       number = {1},
          eid = {L3},
        pages = {L3},
          doi = {10.3847/2041-8213/aba0c1},
archivePrefix = {arXiv},
       eprint = {2006.15150},
 primaryClass = {astro-ph.GA},
       adsurl = {https://ui.adsabs.harvard.edu/abs/2020ApJ...898L...3B}
}

@ARTICLE{greene_2020_imbh,
       author = {{Greene}, Jenny E. and {Strader}, Jay and {Ho}, Luis C.},
        title = "{Intermediate-Mass Black Holes}",
      journal = {\araa},
         year = 2020,
        month = aug,
       volume = {58},
        pages = {257-312},
          doi = {10.1146/annurev-astro-032620-021835},
archivePrefix = {arXiv},
       eprint = {1911.09678},
 primaryClass = {astro-ph.GA},
       adsurl = {https://ui.adsabs.harvard.edu/abs/2020ARA&A..58..257G}
}

@ARTICLE{tonima_aha_2019_pop_synthesis,
       author = {{Ananna}, Tonima Tasnim and {Treister}, Ezequiel and {Urry}, C. Megan and {Ricci}, C. and {Kirkpatrick}, Allison and {LaMassa}, Stephanie and {Buchner}, Johannes and {Civano}, Francesca and {Tremmel}, Michael and {Marchesi}, Stefano},
        title = "{The Accretion History of AGNs. I. Supermassive Black Hole Population Synthesis Model}",
      journal = {\apj},
         year = 2019,
        month = feb,
       volume = {871},
       number = {2},
          eid = {240},
        pages = {240},
          doi = {10.3847/1538-4357/aafb77},
archivePrefix = {arXiv},
       eprint = {1810.02298},
 primaryClass = {astro-ph.GA},
       adsurl = {https://ui.adsabs.harvard.edu/abs/2019ApJ...871..240A}
}

@ARTICLE{ueda_2014_xlf,
       author = {{Ueda}, Yoshihiro and {Akiyama}, Masayuki and {Hasinger}, G{\"u}nther and {Miyaji}, Takamitsu and {Watson}, Michael G.},
        title = "{Toward the Standard Population Synthesis Model of the X-Ray Background: Evolution of X-Ray Luminosity and Absorption Functions of Active Galactic Nuclei Including Compton-thick Populations}",
      journal = {\apj},
         year = 2014,
        month = may,
       volume = {786},
       number = {2},
          eid = {104},
        pages = {104},
          doi = {10.1088/0004-637X/786/2/104},
archivePrefix = {arXiv},
       eprint = {1402.1836},
 primaryClass = {astro-ph.CO},
       adsurl = {https://ui.adsabs.harvard.edu/abs/2014ApJ...786..104U}
}

@ARTICLE{weaver_2022_cosmos2020_cat,
       author = {{Weaver}, J.~R. and {Kauffmann}, O.~B. and {Ilbert}, O. and {McCracken}, H.~J. and {Moneti}, A. and {Toft}, S. and {Brammer}, G. and {Shuntov}, M. and {Davidzon}, I. and {Hsieh}, B.~C. and {Laigle}, C. and {Anastasiou}, A. and {Jespersen}, C.~K. and {Vinther}, J. and {Capak}, P. and {Casey}, C.~M. and {McPartland}, C.~J.~R. and {Milvang-Jensen}, B. and {Mobasher}, B. and {Sanders}, D.~B. and {Zalesky}, L. and {Arnouts}, S. and {Aussel}, H. and {Dunlop}, J.~S. and {Faisst}, A. and {Franx}, M. and {Furtak}, L.~J. and {Fynbo}, J.~P.~U. and {Gould}, K.~M.~L. and {Greve}, T.~R. and {Gwyn}, S. and {Kartaltepe}, J.~S. and {Kashino}, D. and {Koekemoer}, A.~M. and {Kokorev}, V. and {Le F{\`e}vre}, O. and {Lilly}, S. and {Masters}, D. and {Magdis}, G. and {Mehta}, V. and {Peng}, Y. and {Riechers}, D.~A. and {Salvato}, M. and {Sawicki}, M. and {Scarlata}, C. and {Scoville}, N. and {Shirley}, R. and {Silverman}, J.~D. and {Sneppen}, A. and {Smolc̆i{\'c}}, V. and {Steinhardt}, C. and {Stern}, D. and {Tanaka}, M. and {Taniguchi}, Y. and {Teplitz}, H.~I. and {Vaccari}, M. and {Wang}, W. -H. and {Zamorani}, G.},
        title = "{COSMOS2020: A Panchromatic View of the Universe to z{\ensuremath{\sim}}10 from Two Complementary Catalogs}",
      journal = {\apjs},
         year = 2022,
        month = jan,
       volume = {258},
       number = {1},
          eid = {11},
        pages = {11},
          doi = {10.3847/1538-4365/ac3078},
archivePrefix = {arXiv},
       eprint = {2110.13923},
 primaryClass = {astro-ph.GA},
       adsurl = {https://ui.adsabs.harvard.edu/abs/2022ApJS..258...11W}
}

@ARTICLE{tonima_max_likelihood_2017,
       author = {{Ananna}, Tonima Tasnim and {Salvato}, Mara and {LaMassa}, Stephanie and {Urry}, C. Megan and {Cappelluti}, Nico and {Cardamone}, Carolin and {Civano}, Francesca and {Farrah}, Duncan and {Gilfanov}, Marat and {Glikman}, Eilat and {Hamilton}, Mark and {Kirkpatrick}, Allison and {Lanzuisi}, Giorgio and {Marchesi}, Stefano and {Merloni}, Andrea and {Nandra}, Kirpal and {Natarajan}, Priyamvada and {Richards}, Gordon T. and {Timlin}, John},
        title = "{AGN Populations in Large-volume X-Ray Surveys: Photometric Redshifts and Population Types Found in the Stripe 82X Survey}",
      journal = {\apj},
         year = 2017,
        month = nov,
       volume = {850},
       number = {1},
          eid = {66},
        pages = {66},
          doi = {10.3847/1538-4357/aa937d},
archivePrefix = {arXiv},
       eprint = {1710.01296},
 primaryClass = {astro-ph.GA},
       adsurl = {https://ui.adsabs.harvard.edu/abs/2017ApJ...850...66A}
}

@ARTICLE{Schwartz_BIC_1978,
       author = {{Schwarz}, Gideon},
        title = "{Estimating the Dimension of a Model}",
      journal = {Annals of Statistics},
         year = 1978,
        month = jul,
       volume = {6},
       number = {2},
        pages = {461-464},
       adsurl = {https://ui.adsabs.harvard.edu/abs/1978AnSta...6..461S}
}

@ARTICLE{birrer_lenstronomy_2018,
       author = {{Birrer}, Simon and {Amara}, Adam},
        title = "{lenstronomy: Multi-purpose gravitational lens modelling software package}",
      journal = {Physics of the Dark Universe},
         year = 2018,
        month = dec,
       volume = {22},
        pages = {189-201},
          doi = {10.1016/j.dark.2018.11.002},
archivePrefix = {arXiv},
       eprint = {1803.09746},
 primaryClass = {astro-ph.CO},
       adsurl = {https://ui.adsabs.harvard.edu/abs/2018PDU....22..189B}
}

@ARTICLE{birrer_lenstronomy_ii_2021,
       author = {{Birrer}, Simon and {Shajib}, Anowar and {Gilman}, Daniel and {Galan}, Aymeric and {Aalbers}, Jelle and {Millon}, Martin and {Morgan}, Robert and {Pagano}, Giulia and {Park}, Ji and {Teodori}, Luca and {Tessore}, Nicolas and {Ueland}, Madison and {Van de Vyvere}, Lyne and {Wagner-Carena}, Sebastian and {Wempe}, Ewoud and {Yang}, Lilan and {Ding}, Xuheng and {Schmidt}, Thomas and {Sluse}, Dominique and {Zhang}, Ming and {Amara}, Adam},
        title = "{lenstronomy II: A gravitational lensing software ecosystem}",
      journal = {The Journal of Open Source Software},
         year = 2021,
        month = jun,
       volume = {6},
       number = {62},
          eid = {3283},
        pages = {3283},
          doi = {10.21105/joss.03283},
archivePrefix = {arXiv},
       eprint = {2106.05976},
 primaryClass = {astro-ph.CO},
       adsurl = {https://ui.adsabs.harvard.edu/abs/2021JOSS....6.3283B}
}

@ARTICLE{steinborn_2018,
       author = {{Steinborn}, Lisa K. and {Hirschmann}, Michaela and {Dolag}, Klaus and {Shankar}, Francesco and {Juneau}, St{\'e}phanie and {Krumpe}, Mirko and {Remus}, Rhea-Silvia and {Teklu}, Adelheid F.},
        title = "{Cosmological simulations of black hole growth II: how (in)significant are merger events for fuelling nuclear activity?}",
      journal = {\mnras},
         year = 2018,
        month = nov,
       volume = {481},
       number = {1},
        pages = {341-360},
          doi = {10.1093/mnras/sty2288},
archivePrefix = {arXiv},
       eprint = {1805.06902},
 primaryClass = {astro-ph.GA},
       adsurl = {https://ui.adsabs.harvard.edu/abs/2018MNRAS.481..341S}
}

@ARTICLE{ricarte_2020_rps_agn,
       author = {{Ricarte}, Angelo and {Tremmel}, Michael and {Natarajan}, Priyamvada and {Quinn}, Thomas},
        title = "{A Link between Ram Pressure Stripping and Active Galactic Nuclei}",
      journal = {\apjl},
         year = 2020,
        month = may,
       volume = {895},
       number = {1},
          eid = {L8},
        pages = {L8},
          doi = {10.3847/2041-8213/ab9022},
archivePrefix = {arXiv},
       eprint = {2003.05950},
 primaryClass = {astro-ph.GA},
       adsurl = {https://ui.adsabs.harvard.edu/abs/2020ApJ...895L...8R}
}

@ARTICLE{Alonso-herrero_clumps_agn_2011,
       author = {{Alonso-Herrero}, Almudena and {Ramos Almeida}, Cristina and {Mason}, Rachel and {Asensio Ramos}, Andr{\'e}s and {Roche}, Patrick F. and {Levenson}, Nancy A. and {Elitzur}, Moshe and {Packham}, Christopher and {Rodr{\'\i}guez Espinosa}, Jos{\'e} Miguel and {Young}, Stuart and {D{\'\i}az-Santos}, Tanio and {P{\'e}rez-Garc{\'\i}a}, Ana M.},
        title = "{Torus and Active Galactic Nucleus Properties of Nearby Seyfert Galaxies: Results from Fitting Infrared Spectral Energy Distributions and Spectroscopy}",
      journal = {\apj},
         year = 2011,
        month = aug,
       volume = {736},
       number = {2},
          eid = {82},
        pages = {82},
          doi = {10.1088/0004-637X/736/2/82},
archivePrefix = {arXiv},
       eprint = {1105.2368},
 primaryClass = {astro-ph.CO},
       adsurl = {https://ui.adsabs.harvard.edu/abs/2011ApJ...736...82A}
}


\end{document}